\documentclass[dvips]{article}
\usepackage{supertabular,lscape,epsfig}
\usepackage{amssymb}
\usepackage{amsmath}

\usepackage[polish]{babel}
\usepackage[T1]{fontenc}
\usepackage[latin2]{inputenc}
\usepackage{pslatex}
\usepackage{graphicx}

\DeclareSymbolFont{ppa}{OT1}{ppl}{m}{it}
\DeclareMathSymbol{\vv}{\mathalpha}{ppa}{'166}
\setkeys{Gin}{bb=18 144 592 718}

\begin{document}

\newcommand{\dd}{\,{\rm d}}
\newcommand{\ie}{{\it i.e.},\,}
\newcommand{\etal}{{\it et al.\ }}
\newcommand{\eg}{{\it e.g.},\,}
\newcommand{\cf}{{\it cf.\ }}
\newcommand{\vs}{{\it vs.\ }}
\newcommand{\zdot}{\makebox[0pt][l]{.}}
\newcommand{\up}[1]{\ifmmode^{\rm #1}\else$^{\rm #1}$\fi}
\newcommand{\dn}[1]{\ifmmode_{\rm #1}\else$_{\rm #1}$\fi}
\newcommand{\upd}{\up{d}}
\newcommand{\uph}{\up{h}}
\newcommand{\upm}{\up{m}}  
\newcommand{\ups}{\up{s}}
\newcommand{\arcd}{\ifmmode^{\circ}\else$^{\circ}$\fi}
\newcommand{\arcm}{\ifmmode{'}\else$'$\fi}
\newcommand{\arcs}{\ifmmode{''}\else$''$\fi}
\newcommand{\MS}{{\rm M}\ifmmode_{\odot}\else$_{\odot}$\fi}
\newcommand{\RS}{{\rm R}\ifmmode_{\odot}\else$_{\odot}$\fi}
\newcommand{\LS}{{\rm L}\ifmmode_{\odot}\else$_{\odot}$\fi}

\newcommand{\Abstract}[2]{{\footnotesize\begin{center}ABSTRACT\end{center}
\vspace{1mm}\par#1\par   
\noindent
{~}{\it #2}}}

\newcommand{\TabCap}[2]{\begin{center}\parbox[t]{#1}{\begin{center}
  \small {\spaceskip 2pt plus 1pt minus 1pt T a b l e}
  \refstepcounter{table}\thetable \\[2mm]
  \footnotesize #2 \end{center}}\end{center}}

\newcommand{\TableSep}[2]{\begin{table}[p]\vspace{#1}
\TabCap{#2}\end{table}}

\newcommand{\FigCap}[1]{\footnotesize\par\noindent Fig.\  %
  \refstepcounter{figure}\thefigure. #1\par}

\newcommand{\TableFont}{\footnotesize}
\newcommand{\TableFontIt}{\ttit}
\newcommand{\SetTableFont}[1]{\renewcommand{\TableFont}{#1}}

\newcommand{\MakeTable}[4]{\begin{table}[htb]\TabCap{#2}{#3}
  \begin{center} \TableFont \begin{tabular}{#1} #4
  \end{tabular}\end{center}\end{table}}

\newcommand{\MakeTableSep}[4]{\begin{table}[p]\TabCap{#2}{#3}
  \begin{center} \TableFont \begin{tabular}{#1} #4
  \end{tabular}\end{center}\end{table}}

\newenvironment{references}%
{
\footnotesize \frenchspacing
\renewcommand{\thesection}{}
\renewcommand{\in}{{\rm in }}
\renewcommand{\AA}{Astron.\ Astrophys.}
\newcommand{\AAS}{Astron.~Astrophys.~Suppl.~Ser.}
\newcommand{\ApJ}{Astrophys.\ J.}
\newcommand{\ApJS}{Astrophys.\ J.~Suppl.~Ser.}
\newcommand{\ApJL}{Astrophys.\ J.~Letters}
\newcommand{\AJ}{Astron.\ J.}
\newcommand{\IBVS}{IBVS}
\newcommand{\PASP}{P.A.S.P.}
\newcommand{\Acta}{Acta Astron.}
\newcommand{\MNRAS}{MNRAS}
\renewcommand{\and}{{\rm and }}
\section{{\rm REFERENCES}}
\sloppy \hyphenpenalty10000
\begin{list}{}{\leftmargin1cm\listparindent-1cm
\itemindent\listparindent\parsep0pt\itemsep0pt}}%
{\end{list}\vspace{2mm}}
 
\def\TYLDA{~}
\newlength{\DW}
\settowidth{\DW}{0}
\newcommand{\dw}{\hspace{\DW}}

\newcommand{\refitem}[5]{\item[]{#1} #2%
\def\REFARG{#3}\ifx\REFARG\TYLDA\else, {\it#3}\fi
\def\REFARG{#4}\ifx\REFARG\TYLDA\else, {\bf#4}\fi
\def\REFARG{#5}\ifx\REFARG\TYLDA\else, {#5}\fi.}

\newcommand{\Section}[1]{\section{#1}}
\newcommand{\Subsection}[1]{\subsection{#1}}
\newcommand{\Acknow}[1]{\par\vspace{5mm}{\bf Acknowledgements.} #1}
\pagestyle{myheadings}

\newfont{\bb}{ptmbi8t at 12pt}
\newcommand{\xrule}{\rule{0pt}{2.5ex}}  
\newcommand{\xxrule}{\rule[-1.8ex]{0pt}{4.5ex}}  
\def\thefootnote{\fnsymbol{footnote}}

\begin{center}

{\Large\bf
Binary Lenses in OGLE-III EWS Database. Season 2004}

\vskip1.7cm
{\bf M.~~ J~a~r~o~s~z~y~\'n~s~k~i, 
~~J.~~ S~k~o~w~r~o~n,
~~A.~~ U~d~a~l~s~k~i,
~~M.~~ K~u~b~i~a~k,
~~M.\,K.~~ S~z~y~m~a~\'n~s~k~i,
~~G.~~ P~i~e~t~r~z~y~\'n~s~k~i,
~~I.~~ S~o~s~z~y~\'n~s~k~i,\\
~~K.~~ \.Z~e~b~r~u~\'n,
~~O.~~ S~z~e~w~c~z~y~k
~~and~~\L.~~ W~y~r~z~y~k~o~w~s~k~i}

{Warsaw University Observatory, Al.~Ujazdowskie~4,~00-478~Warszawa, Poland\\
e-mail:(mj,jskowron,udalski,mk,msz,pietrzyn,soszynsk,zebrun,szewczyk,wyrzykow)
@astrouw.edu.pl}
\end{center}
\vskip1.5cm

\Abstract{We present 19 binary lens candidates from OGLE-III Early
Warning System database for the season of 2004. We have also found five
events interpreted as single mass lensing of double sources. The candidates
have been selected by visual light curves inspection.  Examining the models
of binary lenses of this and our previous studies (10 caustic crossing
events of OGLE-II seasons 1997--1999 and 15 binary lens events of OGLE-III
seasons 2002--2003) we find one case of extreme mass ratio binary
($q\approx0.005$, a known planetary lens OGLE 2003-BLG-235/MOA 2003-BLG-53)
and almost all other models with mass ratios in the range $0.1<q<1.0$,
which may indicate the division between planetary systems and binary
stars.}{Gravitational lensing -- Galaxy: center -- binaries: general}

\vskip9mm
\Section{Introduction}
In this paper we present the results of the search for binary lens 
events among microlensing phenomena discovered by the  Early Warning
System (EWS -- Udalski \etal 1994, Udalski 2003) of the third phase of
the Optical Gravitational Lensing Experiment (OGLE-III) in the season of
2004. This is a continuation of the study of binary lenses in OGLE-II
(Jaroszy\'nski 2002, hereafter Paper I) and OGLE-III databases
(Jaroszyński \etal 2004, hereafter Paper~II). The results of the similar
search for binary lens events in MACHO data have been presented by Alcock
\etal (2000). 

The motivation of the study remains the same -- we are going to obtain a
uniform sample of binary lens events, selected and modeled with the same
methods for all seasons. The sample may be used to study the population of
binary systems in the Galaxy. The method of observation of the binaries
(gravitational lensing) allows to study their mass ratios distribution,
since they are directly given by the models. The binary separations are
more difficult, because only their projection into the sky expressed in
Einstein radius units enters the models. In small number of cases the
estimation of the masses and distances to the lenses may be possible.

Cases of extremely low binary mass ratios ($q\le0.01$) are usually
considered as planetary lensing. Such events have been discovered in
OGLE-III database for season 2003 (Bond \etal 2004) and 2005 (Udalski
\etal 2005, Gould \etal 2006, Beaulieu \etal 2006), but are missing in
season 2004.

Our approach follows that of Papers I and II, where the references to
earlier work on the subject are given. Some basic ideas for binary lens
analysis can be found in the review article by Paczy\'nski (1996).

Paper I presents the analysis of 18 binary lens events found in OGLE-II
data with 10 safe caustic crossing cases. Paper II gives 15 binary lens
events. 

In Section~2 we describe the selection of binary lens candidates. In
Section~3 we describe the procedure of fitting  models to the data. The
results are described in Section~4, and the discussion follows in
Section~5. The extensive graphical material is shown in Appendix. 

\Section{Choice of Candidates} 
The OGLE-III data is routinely reduced with difference photometry ({\sc
dia}, Alard and Lupton 1998, Alard 2000) which gives high quality light
curves of variable objects. The EWS system of OGLE-III (Udalski 2003)
automatically picks up candidate objects with microlensing-like
variability.

There are 608 microlensing event candidates selected by EWS in the 2004
season. We visually inspect all candidate light curves looking for features
characteristic for binary lenses (multiple peaks, U-shapes,
asymmetry). Light curves showing excessive noise are omitted. We select 25
candidate binary events in 2004 data for further study. For these candidate
events we apply our standard procedure of finding binary lens models
(compare Papers I, II and Section~3).

\Section{Fitting Binary Lens Models}
The models of the two point mass lens were investigated by many authors 
(Schneider and Weiss 1986, Mao and DiStefano 1995, DiStefano and Mao
1996, Dominik 1998, to mention only a few). The effective methods applicable
for extended  sources  have recently been described by Mao and Loeb
(2001). While we use mostly the point source approximation, we
extensively employ their efficient numerical schemes for calculating
the binary lens caustic structure and source  magnification. 

We fit binary lens models using the $\chi^2$ minimization  method for
the  light curves. It is convenient to model the flux at the time $t_i$
as: 
$$F_i=F(t_i)=A(t_i)\times F_s+F_b\equiv(A(t_i)-1)\times F_s+F_0\eqno(1)$$
where $F_s$ is the flux of the source being lensed, $F_b$ the blended
flux (from the source close neighbors and possibly the lens), and the
combination $F_b+F_s=F_0$ is the total flux, measured long before or
long after the event. The last parameter can be reasonably well
estimated with observations performed in seasons preceding and following
2004, as a weighted mean:
$$F_0=\frac{\sum\limits_{i^\prime=1}^{N^\prime}
\displaystyle{\frac{F_i}{\sigma_i}}}
{\sum\limits_{i^\prime=1}^{N^\prime}
\displaystyle{\frac{1}{\sigma_i}}}\eqno(2)$$
where $F_i$ are the observed fluxes and $\sigma_i$ their estimated
photometric errors. The summation over $i^\prime$ does not include
observations of 2004, and $N^\prime$ is the number of relevant
observations.

In fitting the models we use rescaled errors (compare Papers I and II).
More detailed analysis (\eg Wyrzykowski 2005) shows that the OGLE
photometric errors are overestimated for very faint sources and
underestimated for bright ones. Error scaling used here, based on the
scatter of the source flux in seasons when it is supposedly invariable,
is the simplest approach. It gives the estimate of the combined effect
of the observational errors and possibly undetectable, low amplitude
internal source variability. We require that constant flux source model
fits well the other seasons data after error rescaling:
$$\chi_{\rm other}^2=\sum\limits_{i^\prime=1}^{N^\prime}
\frac{(F_i-F_0)^2}{(s\sigma_i)^2}=N^\prime-1\eqno(3)$$
where $s$ is the error scaling factor.

The lens magnification (amplification) of the source 
$A(t_i)=A(t_i;p_j)$ depends on the set of model parameters $p_j$. 
Using this notation one has for the $\chi^2$:
$$\chi^2=\sum\limits_{i=1}^N\frac{((A_i-1)F_s+F_0-F_i)^2}{\sigma_i^2}.\eqno(4)$$
The dependence of $\chi^2$ on the binary lens parameters $p_j$ is
complicated, while the dependence on the source flux is quadratic. The
equation $\partial\chi^2/\partial F_s=0$ can be solved algebraically,
giving $F_s=F_s(p_j;\{F_i\})$, thus effectively reducing the dimension
of parameter space. Any method of minimizing $\chi^2$ may (in some
cases) give unphysical solutions with $F_s>F_0$, which would imply a
negative blended flux. To reduce the occurrence of such faulty solutions
we add an extra term to $\chi^2$ which vanishes automatically for
physically correct models with $F_s \le F_0$, but is a fast growing
function of the source flux $F_s$ whenever it exceeds the base flux
$F_0$. 

Our analysis of the models, their fit quality etc. is based on the
$\chi_1^2$ calculated with the rescaled errors:
$$\chi_1^2\equiv\frac{\chi^2}{s^2}\eqno(5)$$
which is displayed in the tables and plots below. For events with multiple
models (representing different local minima of $\chi^2$), we assess the
relevance of each model with the relative weight $w\sim\exp(-\chi_1^2/2)$.

For most of the light curves we investigate the caustic crossings are
not  well sampled and we are forced to use a point source approximation
in majority of our models.  In three cases (events OGLE 2004-BLG-035,
OGLE 2004-BLG-039, and OGLE 2004-BLG-207) the caustic crossings are
resolved, so the extended source models can be fitted. In these cases
the strategy resembling Albrow \etal (1999) for finding binary lens
models can be used. It is based on the fact that some of the parameters
(the source angular size, the strength of the caustic) can be fitted
independently, so for an initial fit one can split the parameter space
into two lower dimensionality sub-manifolds. 

The binary system consists of two masses $m_1$ and $m_2$, where by
convention  ${m_1\le m_2}$. The Einstein radius of the binary lens is
defined as: 
$$r_{\rm E}=\sqrt{\frac{4G(m_1+m_2)}{c^2}\frac{d_{\rm OL}d_{\rm LS}}
{d_{\rm OS}}}\eqno(6)$$
where $G$ is the constant of gravity, $c$ is the speed of light, $d_{\rm
OL}$ is the observer--lens distance, $d_{\rm LS}$ is the lens--source 
distance, and $d_{\rm OS}\equiv d_{\rm OL}+d_{\rm LS}$ is the distance
between the observer and the source. The Einstein radius serves as a
length unit and the Einstein time: ${t_{\rm E}=r_{\rm E}/\vv_\perp}$,
where $\vv_\perp$ is the lens velocity relative to the line joining the
observer with the source, serves as a time unit. The passage of the
source in the lens background is defined by seven parameters: ${q\equiv 
m_1/m_2}$ (${0<q\le1}$) -- the binary mass ratio, $d$ -- binary
separation  expressed in $r_{\rm E}$ units, $\beta$ -- the angle between
the source  trajectory as projected onto the sky and the projection of
the binary axis, $b$ -- the impact parameter relative to the binary
center of mass, $t_0$ -- the time of closest approach of the source to
the binary center of mass, $t_{\rm E}$ -- the Einstein time, and $r_s$ the 
source radius. Thus we are left with the seven or six  dimensional
parameter space, depending on the presence/absence of observations
covering the caustic crossings. 

\renewcommand{\arraystretch}{1.05}
\MakeTable{c@{\hspace{5pt}}c@{\hspace{5pt}}c@{\hspace{5pt}}rccccrcccc}
{12.5cm}{The binary lens models of season 2004 events}
{\hline
\noalign{\vskip 3pt}
 year & event & & $\chi_1^2/{\rm DOF}$ & $s$ & $q$ & $d$ & $\beta$ & 
\multicolumn{1}{c}{$b$} & $t_0$ & $t_{\rm E}$ & $f$ \\
\noalign{\vskip3pt}
\hline
2004 & 035 & b &   216.0/236 &  2.12 & 0.119 & 1.019 &   57.35 & $  0.15$ &  3085.7 &   67.3 &  0.78 & \\
2004  & 039 & b &   399.0/382 &  1.94 & 0.094 & 1.165 &  174.39 & $ -0.11$ &  3082.4 &   38.9 &  0.88 \\
2004 & 207 & b &   501.7/387 &  1.56 & 0.308 & 1.634 &   62.86 & $  0.78$ &  3151.8 &   28.7 &  0.84 & \\
2004 & 226 & b &   361.3/309 &  1.48 & 0.323 & 1.401 &  296.26 & $ -0.40$ &  3147.7 &   32.9 &  0.67 & \\
     &     & b &   363.9/309 &  1.48 & 0.078 & 0.992 &  161.51 & $  0.16$ &  3142.0 &   39.5 &  0.25 & \\
2004 & 250 & b &   276.9/274 &  2.04 & 0.705 & 0.830 &  155.04 & $ -0.12$ &  3160.8 &   75.2 &  0.11 \\
2004 & 273 & b &   241.0/233 &  1.93 & 0.450 & 0.550 &   63.99 & $  0.06$ &  3157.8 &   30.8 &  1.00 \\
2004 & 280 & b &   182.2/241 &  2.34 & 0.455 & 2.052 &  139.49 & $  0.19$ &  3181.0 &   26.3 &  0.51 \\
2004 & 309 & b &   230.6/296 &  2.35 & 0.346 & 1.293 &   14.86 & $ -0.08$ &  3193.1 &   60.4 &  0.12 \\
2004 & 325 & b &   351.0/349 &  1.62 & 0.528 & 1.193 &  224.64 & $  0.06$ &  3180.5 &   57.0 &  0.32 \\
2004 & 347 & d &   293.7/291 &  1.47 & 0.879 & 3.091 &  220.57 & $  0.67$ &  3263.9 &   54.6 &  0.60 \\
     &     & d &   296.4/291 &  1.47 & 0.452 & 0.587 &  122.70 & $ -0.35$ &  3208.8 &   35.0 &  0.93 \\
2004 & 354 & b &   343.4/335 &  3.64 & 0.681 & 1.869 &   74.34 & $  0.85$ &  3192.8 &   79.2 &  0.03 \\
2004 & 362 & b &   399.7/418 &  1.93 & 0.267 & 1.036 &  149.07 & $ -0.27$ &  3195.3 &  126.8 &  0.04 \\
2004 & 366 & b &   213.9/218 &  1.98 & 0.242 & 0.897 &   39.96 & $ -0.27$ &  3198.7 &  187.0 &  0.05 \\
2004 & 367 & ? &  1422.0/295 &  1.36 & 0.052 & 0.633 &   97.64 & $ -0.01$ &  3180.8 &   44.7 &  0.20 \\
2004 & 373 & b &   544.7/550 &  1.52 & 0.708 & 0.955 &  231.36 & $ -0.09$ &  3195.6 &   33.9 &  0.38 \\
2004 & 379 & b &   432.0/339 &  1.41 & 0.687 & 0.821 &  226.94 & $  0.14$ &  3196.7 &   30.7 &  0.65 \\
2004 & 406 & b &   422.7/405 &  1.44 & 0.575 & 0.885 &  214.84 & $ -0.16$ &  3190.0 &   22.7 &  0.23 \\
     &     & b &   423.1/405 &  1.44 & 0.353 & 2.635 &  233.05 & $ -1.28$ &  3120.5 &   69.5 &  0.12 \\
2004 & 444 & d &   209.6/240 &  1.98 & 0.675 & 1.282 &  241.57 & $ -0.19$ &  3186.8 &   41.0 &  0.02 \\
2004 & 451 & b &   295.5/313 &  1.51 & 0.227 & 0.963 &  109.45 & $  0.23$ &  3224.4 &   35.7 &  0.13 \\
2004 & 460 & b &   340.2/343 &  1.59 & 0.396 & 1.178 &  350.94 & $ -0.20$ &  3204.1 &   33.0 &  0.18 \\
     &     & b &   344.8/343 &  1.59 & 0.894 & 1.240 &   17.54 & $  0.02$ &  3200.4 &   31.3 &  0.21 \\
2004 & 480 & b &   363.0/342 &  1.63 & 0.163 & 1.370 &  322.43 & $ -0.15$ &  3225.1 &   10.4 &  0.71 \\
     &     & b &   367.3/342 &  1.63 & 0.038 & 0.797 &  200.47 & $ -0.11$ &  3225.3 &   14.2 &  0.41 \\
2004 & 490 & d &   289.2/313 &  1.80 & 0.111 & 0.722 &   53.63 & $ -0.17$ &  3223.5 &   17.1 &  0.45 \\
     &     & d &   291.2/313 &  1.80 & 0.001 & 1.515 &  169.49 & $  0.17$ &  3223.0 &   18.5 &  0.36 \\
2004 & 559 & ? &   438.2/199 &  1.78 & 0.658 & 0.881 &   91.86 & $  0.06$ &  3269.4 &   17.9 &  0.27 \\
2004 & 572 & b &   259.7/240 &  1.26 & 0.793 & 0.760 &  270.06 & $  0.03$ &  3288.0 &   31.0 &  0.16 \\
2004 & 605 & d &   370.8/410 &  1.57 & 0.227 & 0.671 &   19.94 & $  0.07$ &  3309.9 &   80.6 &  0.16 \\
     &     & d &   374.9/410 &  1.57 & 0.193 & 3.801 &   77.56 & $  2.89$ &  3477.3 &  264.8 &  0.12 \\
\noalign{\vskip3pt}
\hline
\noalign{\vskip3pt}
\multicolumn{13}{p{12.4cm}}{Note: The columns show: the
event year and EWS number, the event classification ("b" for a binary
lens, "d" for a double source event, "?" for low quality fits and/or
uncompelling cases), the rescaled $\chi^2_1$, number of DOF, the scaling
factor $s$, the mass ratio $q$, the binary separation $d$, the source
trajectory direction $\beta$, the impact parameter $b$, the time of the
closest center of mass approach $t_0$, the Einstein time $t_{\rm E}$
and the  blending parameter $f\equiv F_s/F_0$.}}
We begin with a scan of the parameter space using a logarithmic grid of
points in $(q,d)$ plane (${10^{-3}\le q\le 1}$, ${0.1\le d\le 10}$) and
allowing for  continuous variation of other parameters. The choice
of starting points combines systematic and Monte Carlo searching of
regions in parameter space allowing for caustic crossing or cusp
approaching events. The $\chi^2$ minimization is based on downhill
method and uses standard  numerical algorithms. When a local minimum is
found we make a small Monte Carlo jump in the parameter space and repeat
the downhill search. In some cases it allows to find a different
local minimum. If it does not work several times, we stop and try next
starting point.

Only the events with characteristics of caustic crossing (apparent
discontinuities in observed light curves, U-shapes) can be treated as
safe binary lens cases. The double peak events may result from cusps
approaches, but may also be produced by double sources (\eg Gaudi and
Han 2004). In such cases we also check the double source fit of the
event postulating: 
$$F(t)=A(u_1(t))\times F_{s1}+A(u_2(t))\times F_{s2}+F_b\eqno(7)$$
where $F_{s1}$, $F_{s2}$ are the fluxes of the source components, $F_b$
is the blended flux, and $A(u)$ is the single lens amplification
(Paczyński 1986). The dimensionless source -- lens separations are given
as:
$$u_1(t)=\sqrt{{b_1}^2+\frac{(t-t_{01})^2}{{t_{\rm E}}^2}}\qquad
u_2(t)=\sqrt{{b_2}^2+\frac{(t-t_{02})^2}{{t_{\rm E}}^2}}\eqno(8)$$
where $t_{01}$, $t_{02}$ are the closest approach times of the source
components, $b_1$, $b_2$ are the respective impact parameters, and   
$t_{\rm E}$ is the (common) Einstein time.

\Section{Results} 
Our fitting procedures applied to selected 25 candidate events give the
results summarized in Table~1. In a few cases we find concurrent models of
similar fit quality and we give their parameters in the consecutive rows of
Table~1. In the third column of the table we assess the character of the
events. In 19 cases (of 25 investigated) the events are safe binary lens
phenomena in our opinion (designated "b" in Table~1). There are three cases
classified as double source events ("d" in Table~1) and three events with
low quality fits ("?" in Table~1). The source paths and model light
curves are shown in the first part of Appendix. 

The results of double source modeling are summarized in Table~2. The
double source modeling is applied to the majority of the binary lens
candidates and some other non-standard events. While formally the fits
are usually better for binary lenses, in a few cases we prefer double
source models as more natural, giving less complicated light curves. The
comparison of two kinds of fits is given in the second part of Appendix,
and the well separated double source events -- in the third. 

\renewcommand{\arraystretch}{1}
\MakeTable{llrcccccrcc}{12.5cm}{Parameters of double source modeling}
{\hline
\noalign{\vskip3pt}
Year & \multicolumn{1}{c}{Event} & $\chi^2/$DOF & $b_1$ & $b_2$ & $t_{01}$ & $t_{02}$ & $t_{\rm E}$ &  $f_1$ & $f_2$ \\
\noalign{\vskip3pt}
\hline
2004 & 004 & 1054./849 & 0.8803 & 0.0014 & 3048.75 & 3063.12 &    12.1 & 0.996 & 0.004 \\
2004 & 226 &  404./312 & 0.0591 & 0.0067 & 3137.71 & 3148.21 &    71.1 & 0.109 & 0.014 \\
2004 & 280 &  597./242 & 0.0684 & 0.0057 & 3167.15 & 3186.43 &   122.6 & 0.021 & 0.008 \\
2004 & 328 &  244./277 & 0.7078 & 0.0080 & 3177.45 & 3255.48 &    12.9 & 0.935 & 0.065 \\
2004 & 347 &  368./294 & 0.1749 & 0.0790 & 3204.96 & 3219.74 &    45.9 & 0.276 & 0.140 \\
2004 & 354 & 1857./338 & 0.7987 & 0.4766 & 3176.11 & 3192.13 &     3.5 & 0.865 & 0.135 \\
2004 & 362 &  746./419 & 0.0000 & 0.0259 & 3200.43 & 3210.20 &   483.3 & 0.000 & 0.005 \\
2004 & 366 &  650./221 & 1.1336 & 0.0125 & 3169.15 & 3180.90 &    10.1 & 0.979 & 0.021 \\
2004 & 367 & 2480./298 & 0.0151 & 0.0124 & 3179.86 & 3181.00 &    20.1 & 0.108 & 0.461 \\
2004 & 406 &  546./408 & 0.1870 & 0.0000 & 3190.99 & 3195.64 &     8.8 & 1.000 & 0.000 \\
2004 & 444 &  255./243 & 2.0329 & 0.0001 & 3190.11 & 3204.32 &     8.1 & 0.993 & 0.007 \\
2004 & 451 &  405./314 & 0.0976 & 0.0001 & 3205.18 & 3225.93 &    28.1 & 0.079 & 0.007 \\
2004 & 460 &  375./346 & 0.7770 & 0.0000 & 3199.70 & 3210.71 &    19.4 & 1.000 & 0.000 \\
2004 & 480 &  501./343 & 0.0001 & 0.2528 & 3219.32 & 3225.11 &    10.5 & 0.099 & 0.901 \\
2004 & 490 &  291./316 & 0.1745 & 0.0004 & 3222.93 & 3238.96 &    17.7 & 0.373 & 0.009 \\
2004 & 559 & 3489./202 & 0.1343 & 0.1559 & 3258.82 & 3266.48 &    10.2 & 0.699 & 0.301 \\
2004 & 572 &  399./243 & 0.1962 & 0.3326 & 3275.19 & 3301.65 &    27.6 & 0.374 & 0.626 \\
2004 & 605 &  380./413 & 0.0002 & 0.0007 & 3297.45 & 3316.19 &    42.5 & 0.245 & 0.210 \\
\noalign{\vskip3pt}
\hline
\noalign{\vskip3pt}
\multicolumn{11}{p{12.0cm}}{Note: The columns contain: the year and event
number according to EWS, the rescaled $\chi^2$ value and the DOF number,
the impact parameters $b_1$ and $b_2$ for the two source components,
times of the closest approaches $t_{01}$ and $t_{02}$, the Einstein
time $t_{\rm E}$, and the blending parameters $f_1=F_{s1}/(F_{s1}+
F_{s2}+F_b)$ and $f_2=F_{s2}/(F_{s1}+F_{s2}+F_b)$.}}

Our sample of binary lenses includes now $10+15+19=44$ events of Paper~I,
Paper~II, and the present work, some of them with multiple models. Using
the sample we study the  distributions of various binary lens
parameters. In Fig.~1 we show the histograms for the mass ratio and the
binary separation. The mass ratio is practically limited to the range 
$0.1\le q\le1$ with very small probability of finding a model in the
range $0.01\to0.1$ and a single planetary lens with $q<0.01$.
\begin{figure}[htb]

\includegraphics[height=63.0mm,width=62.0mm]{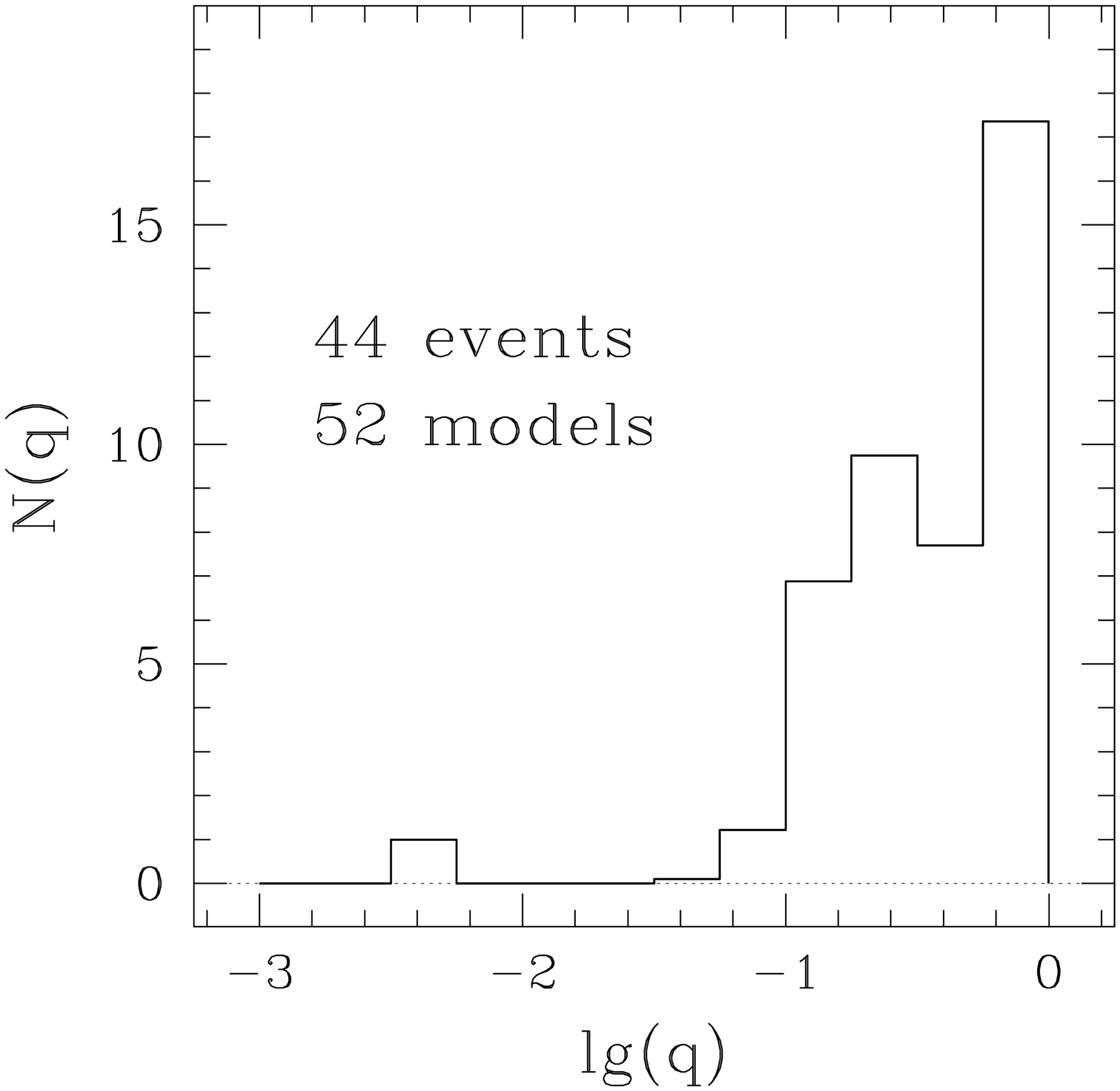} 
\includegraphics[height=63.0mm,width=62.0mm]{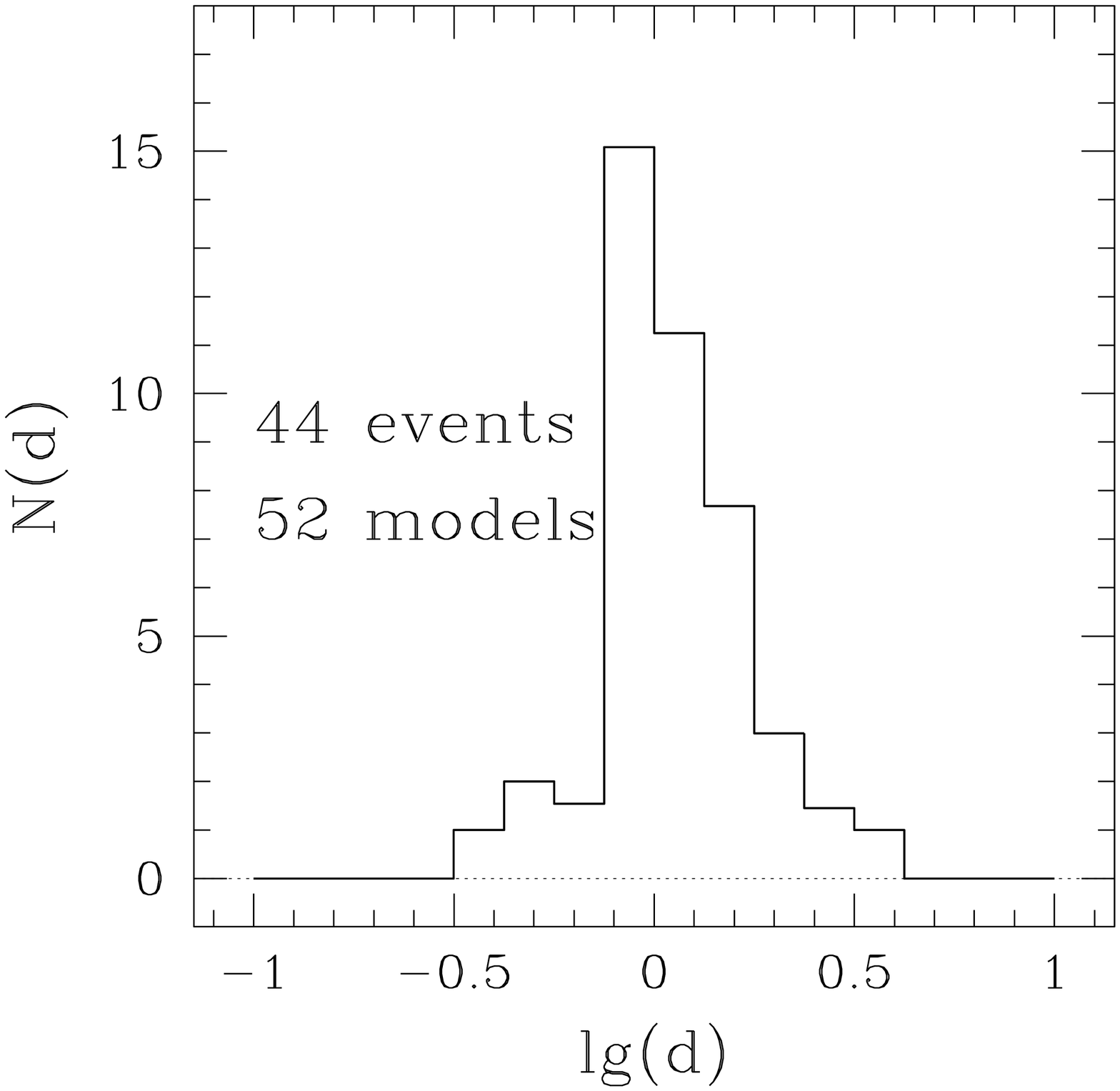}
 
\FigCap{Histogram of mass ratios ({\it left}) and separations ({\it
right}) for binary lens events of OGLE-II (Paper~I) and OGLE-III (Paper
II and this work). The histogram includes 44 events, some of them with
multiple models. The alternative models of any event have been assigned
fractional weights.}
\end{figure}

\Section{Discussion}
Our classification of the investigated events into the binary lens
double source/ unknown categories needs further explanation. The binary
lens model of the event 367 is rather superficial with its four maxima
in the light curve, while there is only one in the data; the double
source model is more natural, but we reject it on the basis of low fit
quality. Both kinds of models for the event 444 are formally acceptable,
but not compelling because of the low flux amplitude. In the case of
559, the binary lens model looks superficial (case similar to 367) and
the double source model gives a very poor fit. We include event 347 in
the double source category despite the better formal quality of the
binary lens fit. We choose the binary lens interpretation for the event
280, since its light curve exhibits three maxima (two of them rather
weak) and the double source model is qualitatively wrong. For the events
490 and 605 we choose the double source interpretation because such
models are simpler as compared to binary lens models, with formally
similar quality. In the case of 226 we arbitrarily choose binary lens
interpretation, despite the existence of the similar quality double
source model.

Our sample of OGLE binary lens events contains now 44 cases. The
bimodality of the mass ratio distribution and the lack of intermediate
$q$ values remains a valid interpretation of the data. We are not trying
a statistical interpretation of mass ratio distribution in this paper
skipping it into a future publication including events of 2005 season
with another three planetary lenses.

We neglect parallax effect in the binary lens models included in this
work. The inclusion of the effect improves some of the presented models,
but the difference is never dramatic. Since simultaneous measurement of the
parallax effect and the source size allows the lens mass estimation (An
\etal 2002) we are going to investigate it with detail in a paper devoted
to events with observations covering caustic crossings.

\Acknow{We thank Bohdan Paczy\'nski for many helpful discussions  and
Shude Mao for the permission of using his binary lens modeling
software. This work was supported in part by the Polish MNiSW grant
2-P03D-016-24, the NSF grant AST-0607070, and NASA grant NNG06GE27G.}

\newpage
\centerline{\bf Appendix}

\noindent 
{\bf Binary Lens Models}

Below we present plots for the 25 events for which  the binary lens
modeling has been applied. Some of the events, especially cases of
heavily blended sources or events without apparent caustic crossing, may
have alternative double source models. In such cases we show the
comparison of the binary lens and double source fits to the data in the
next subsection. 

The events are ordered and named according to their position  in the
OGLE EWS database for the season 2004. Some events have more than one
binary lens model of comparable fit quality (compare Table~1). We always
show the best (first) model, and the second one only if it is not rejected
at 95\% confidence level based on the difference in its $\chi^2$ value.

Each case is illustrated with two panels. The most interesting part of
the source trajectory, the binary and its caustic structure are shown in
the left panel for the case considered. The labels give the $q$ and $d$
values. On the right panels the part of the best fit light curve  is compared
with observations. The labels give  the rescaled $\chi_1^2$ /DOF values.
The source radius (as projected into the lens plane and expressed in
Einstein radius units) is labeled only for three events with resolved
caustic crossings. Below the light curves we show the differences
between the observed and modeled flux in units of rescaled errors. The
dotted lines show the $\pm 3\sigma$ band.

\vskip 0.5cm
\noindent\parbox{12.75cm}{
\leftline {\bf OGLE 2004-BLG-035} 

 \includegraphics[height=63mm,width=62mm]{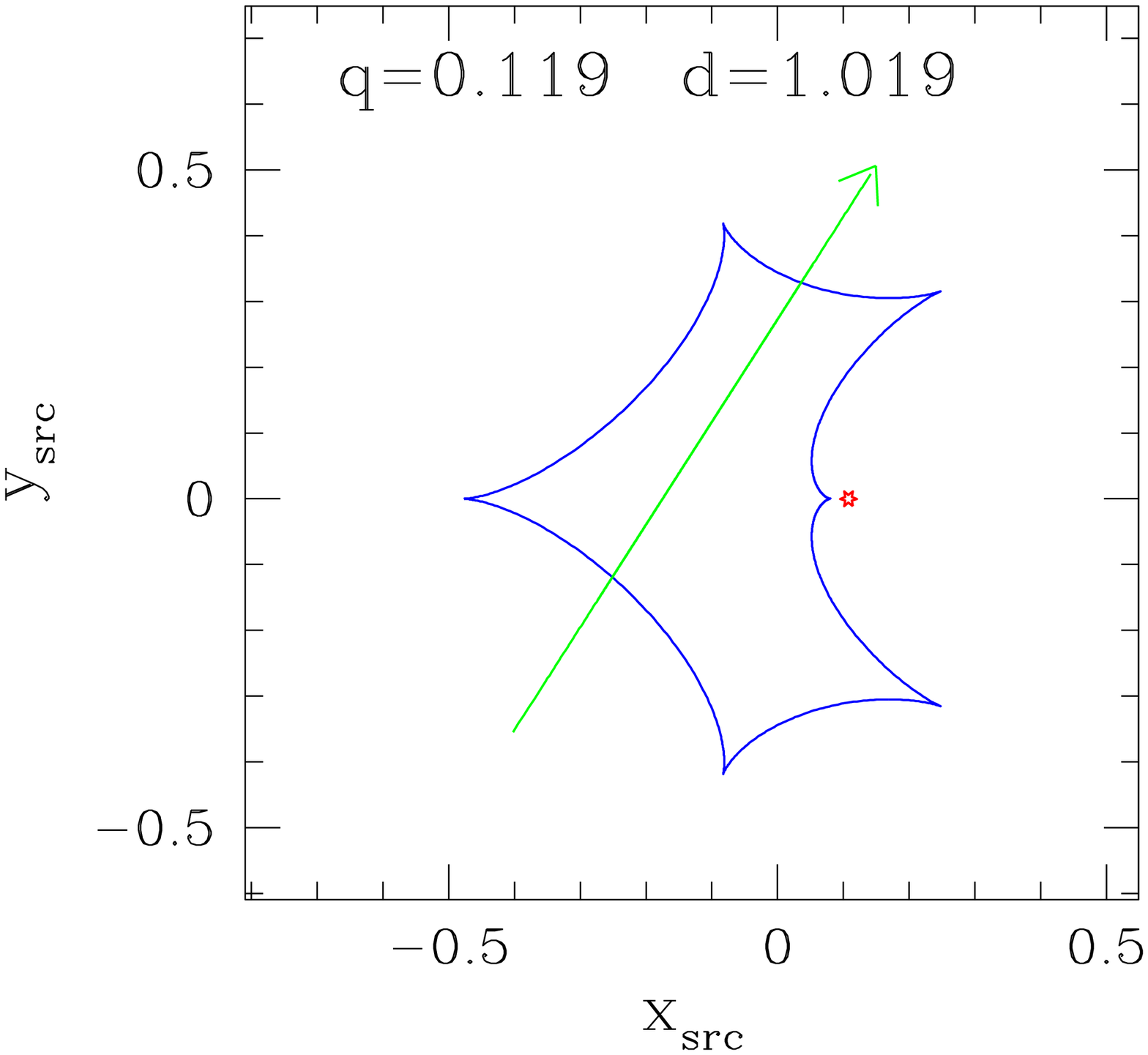}%
 \includegraphics[height=63mm,width=62mm]{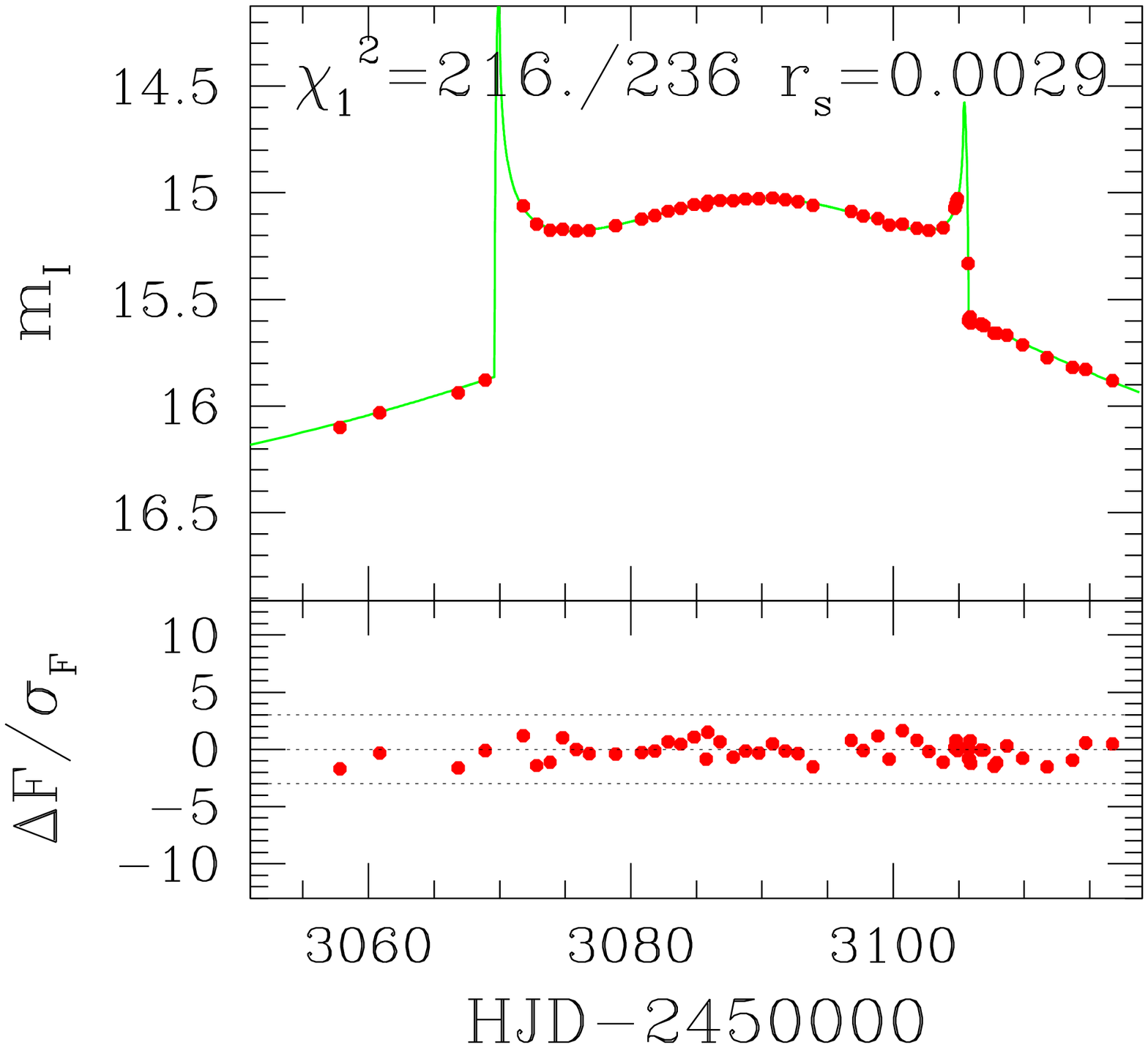}%

}

\noindent\parbox{12.75cm}{
\leftline {\bf OGLE 2004-BLG-039} 

 \includegraphics[height=63mm,width=62mm]{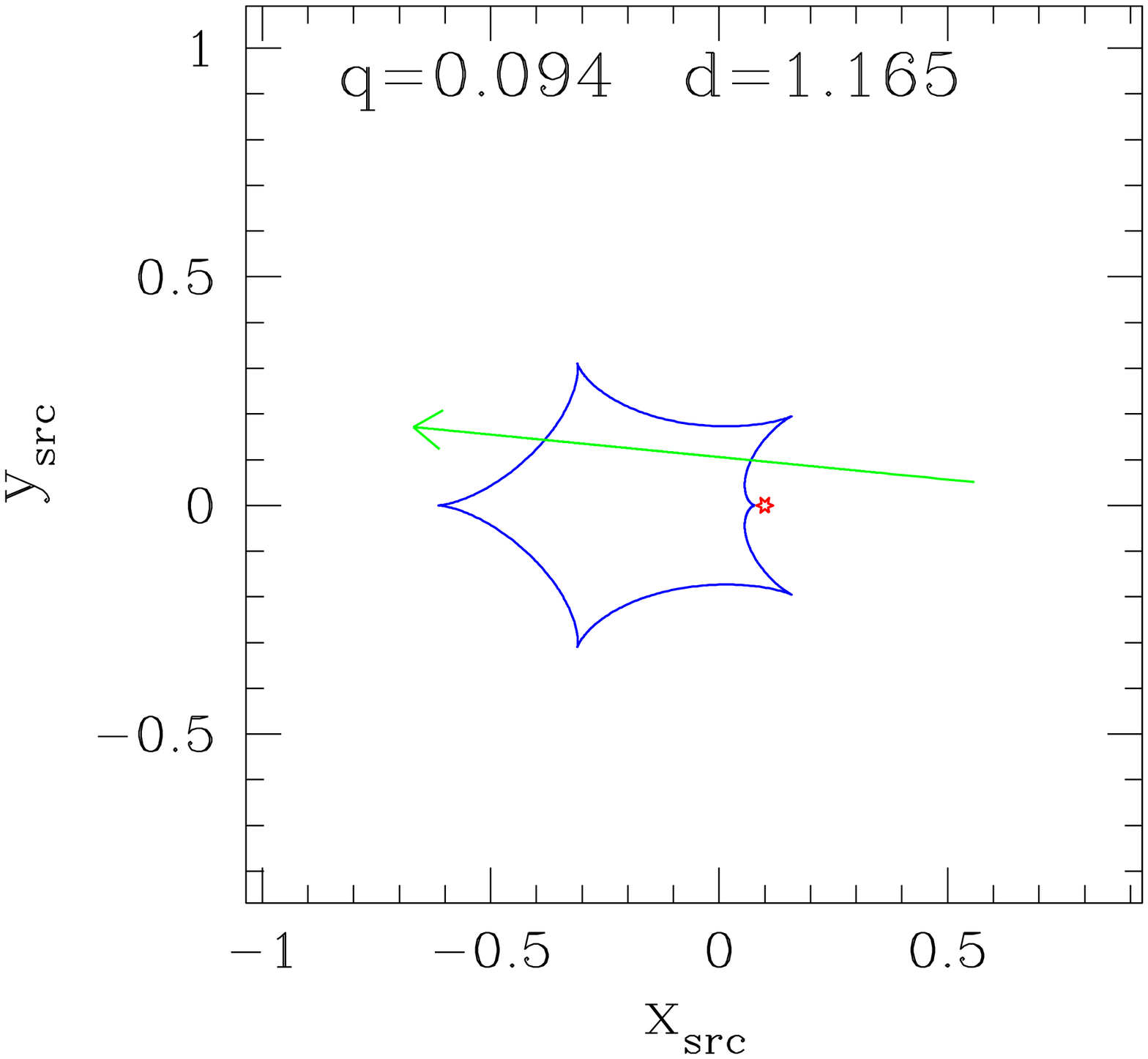}%
 \includegraphics[height=63mm,width=62mm]{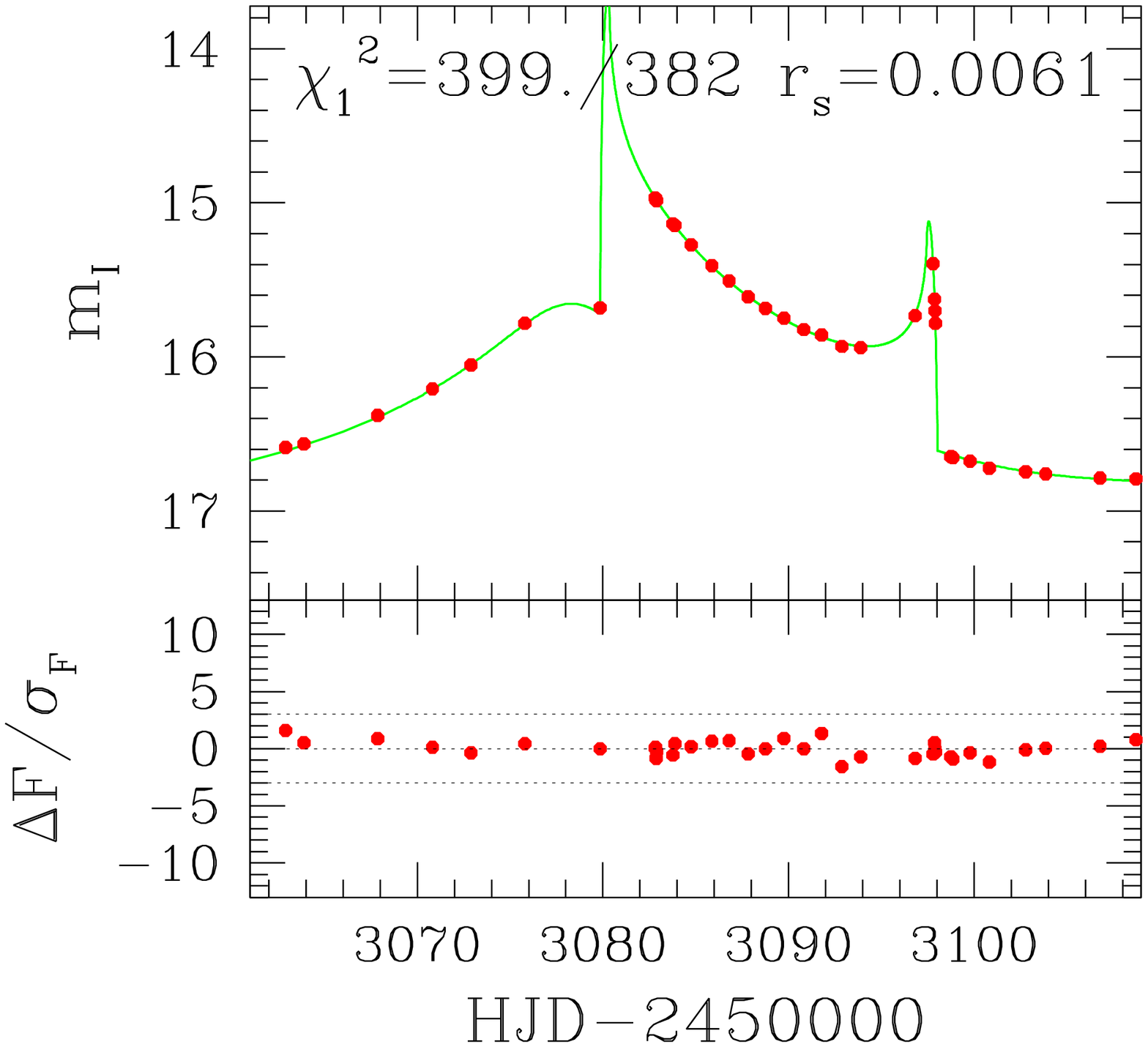}%

}

\noindent\parbox{12.75cm}{
\leftline {\bf OGLE 2004-BLG-207} 

 \includegraphics[height=63mm,width=62mm]{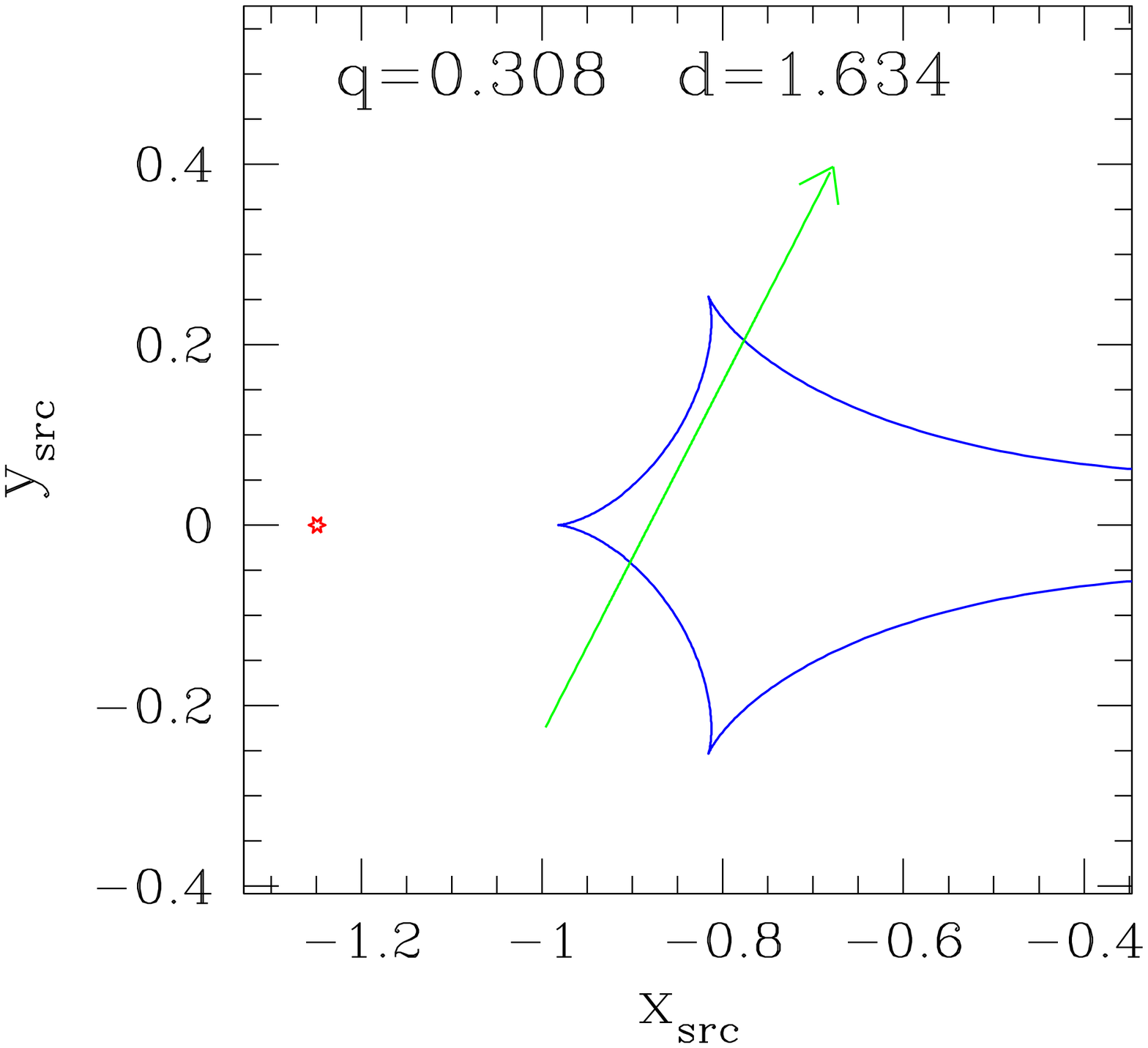}%
 \includegraphics[height=63mm,width=62mm]{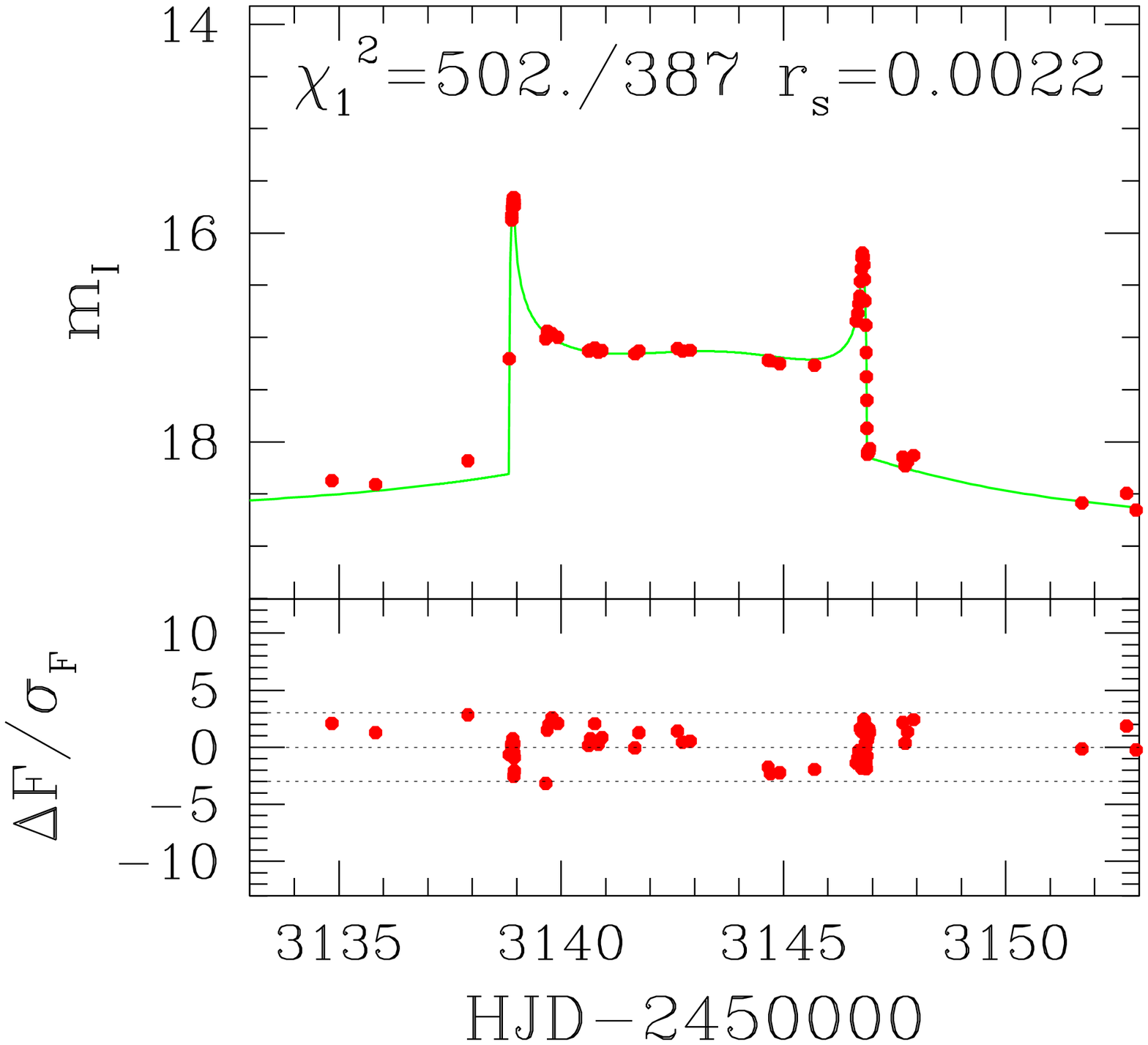}%

}

\noindent\parbox{12.75cm}{
\leftline {\bf OGLE 2004-BLG-226: I} 

\includegraphics[height=63mm,width=62mm]{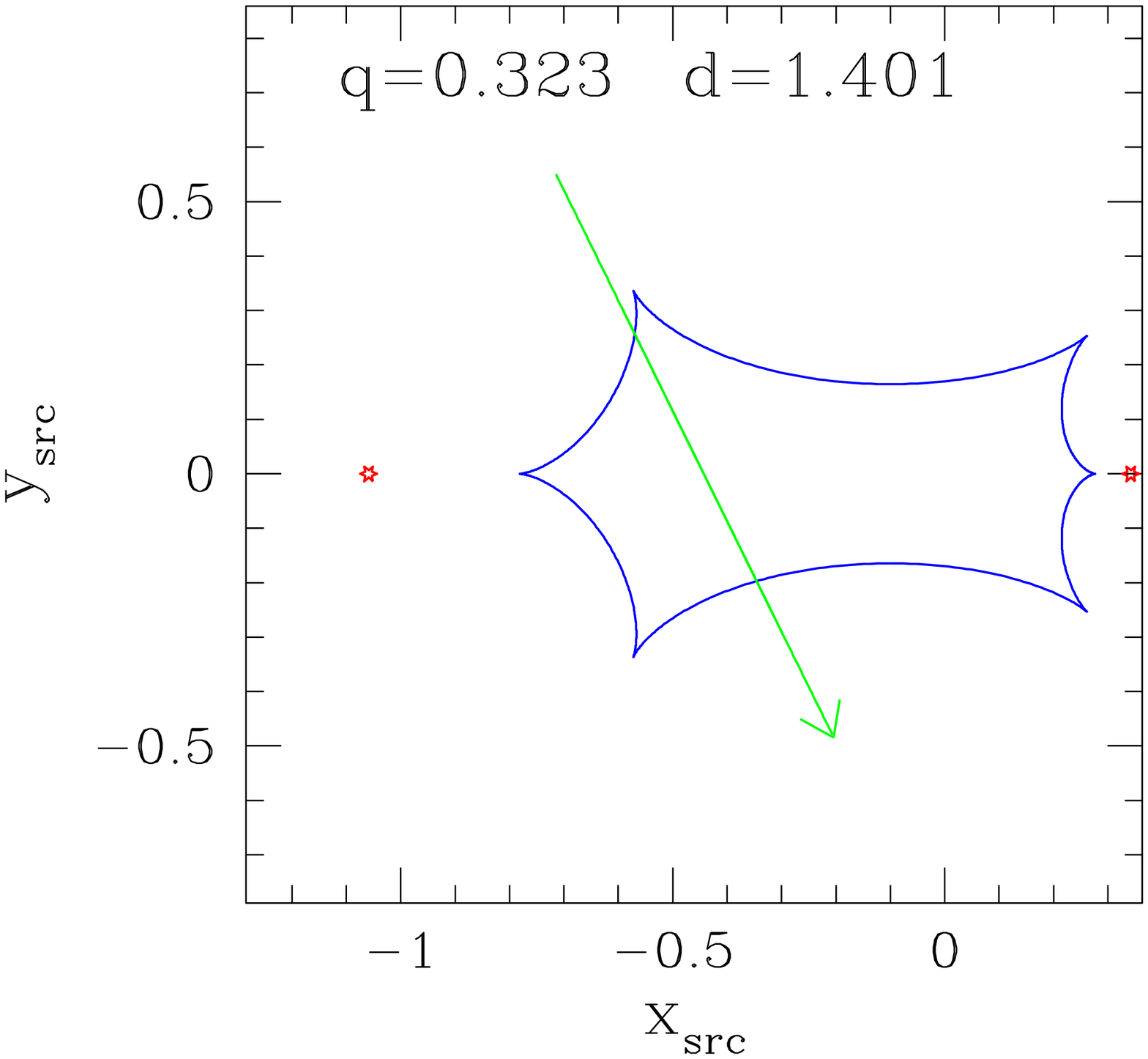}%
\includegraphics[height=63mm,width=62mm]{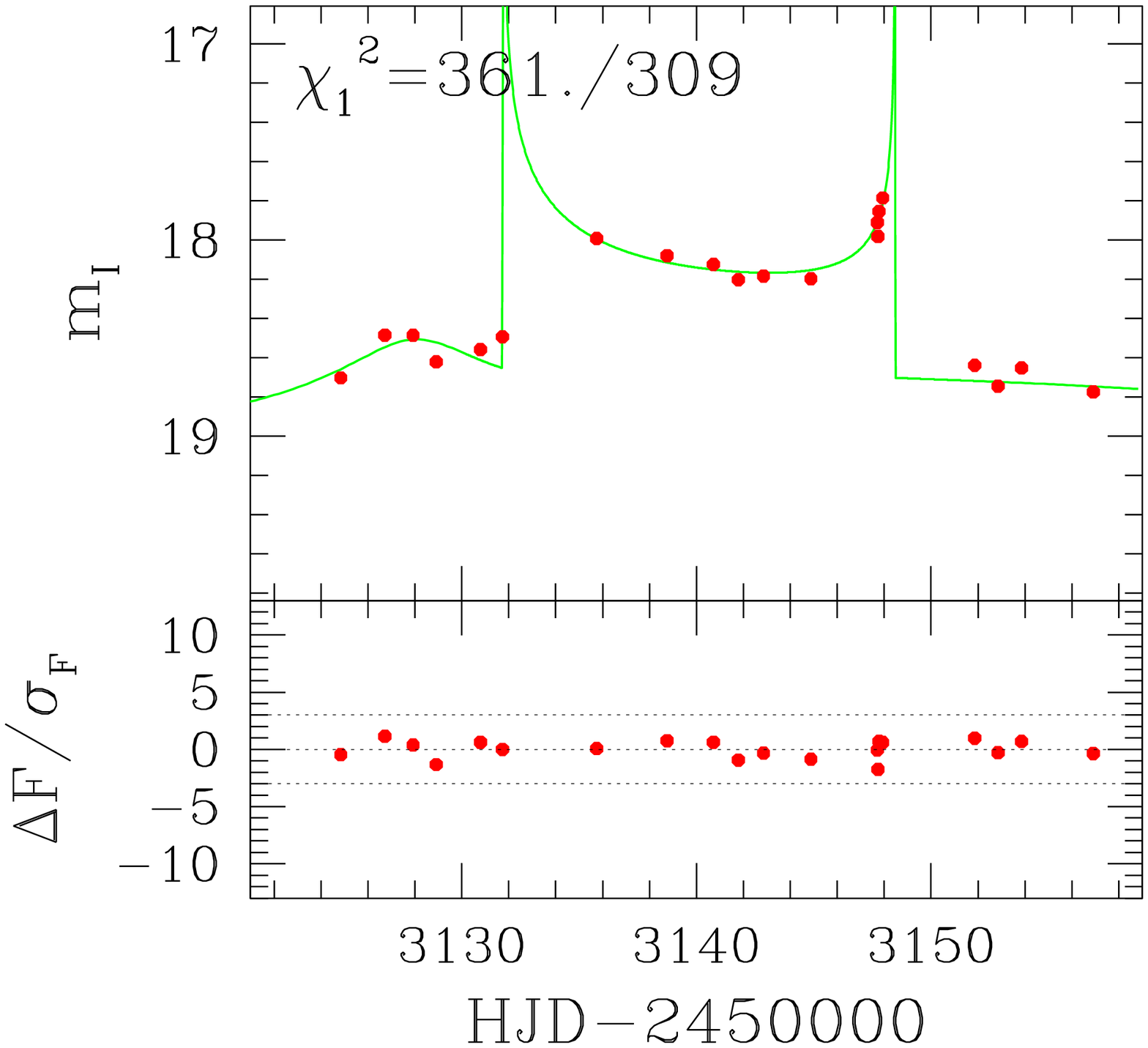}%

}

\noindent\parbox{12.75cm}{
\leftline {\bf OGLE 2004-BLG-226: II} 

 \includegraphics[height=63mm,width=62mm]{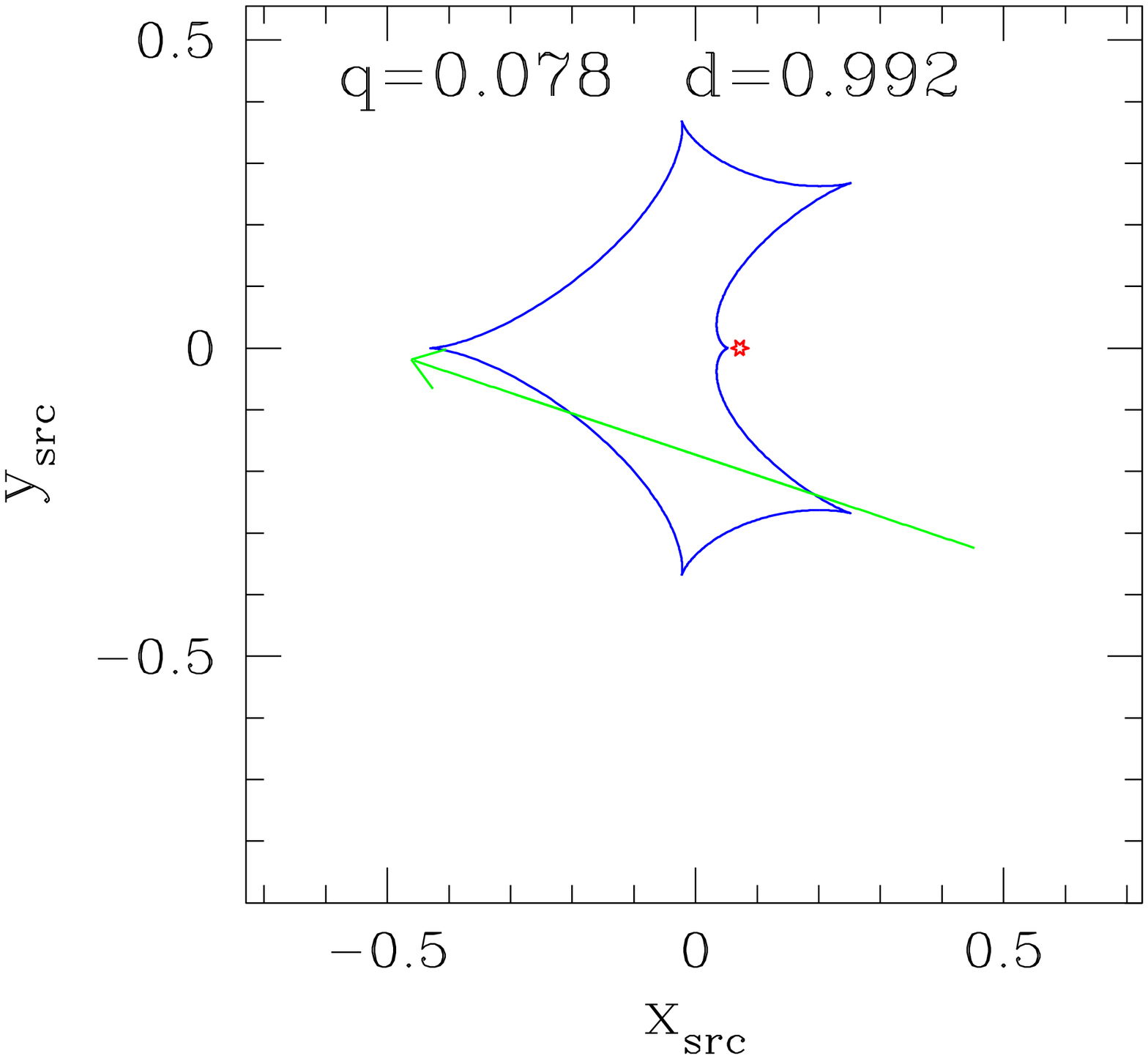}%
 \includegraphics[height=63mm,width=62mm]{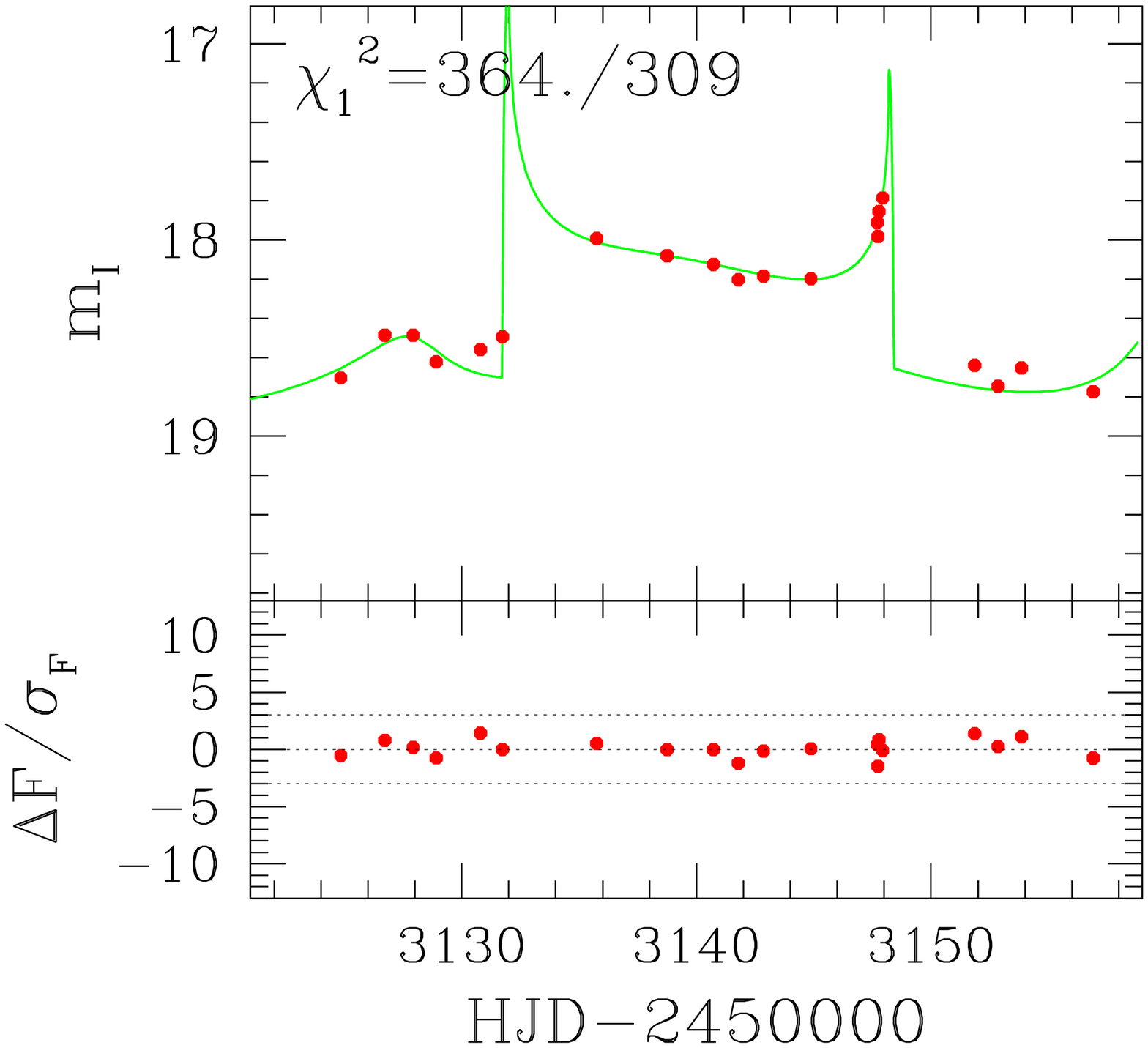}%

}

\noindent\parbox{12.75cm}{
\leftline {\bf OGLE 2004-BLG-250} 

 \includegraphics[height=63mm,width=62mm]{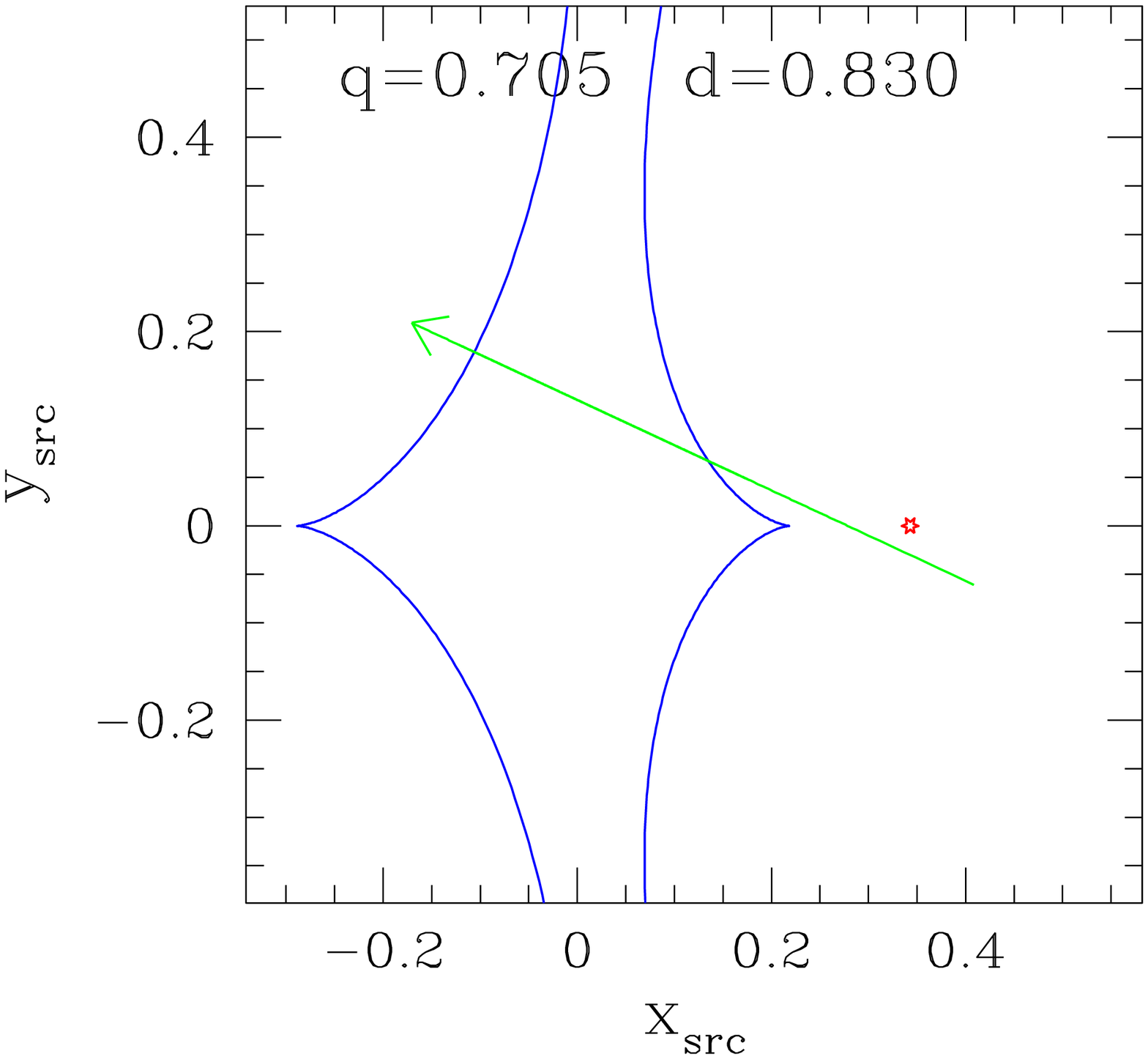}%
 \includegraphics[height=63mm,width=62mm]{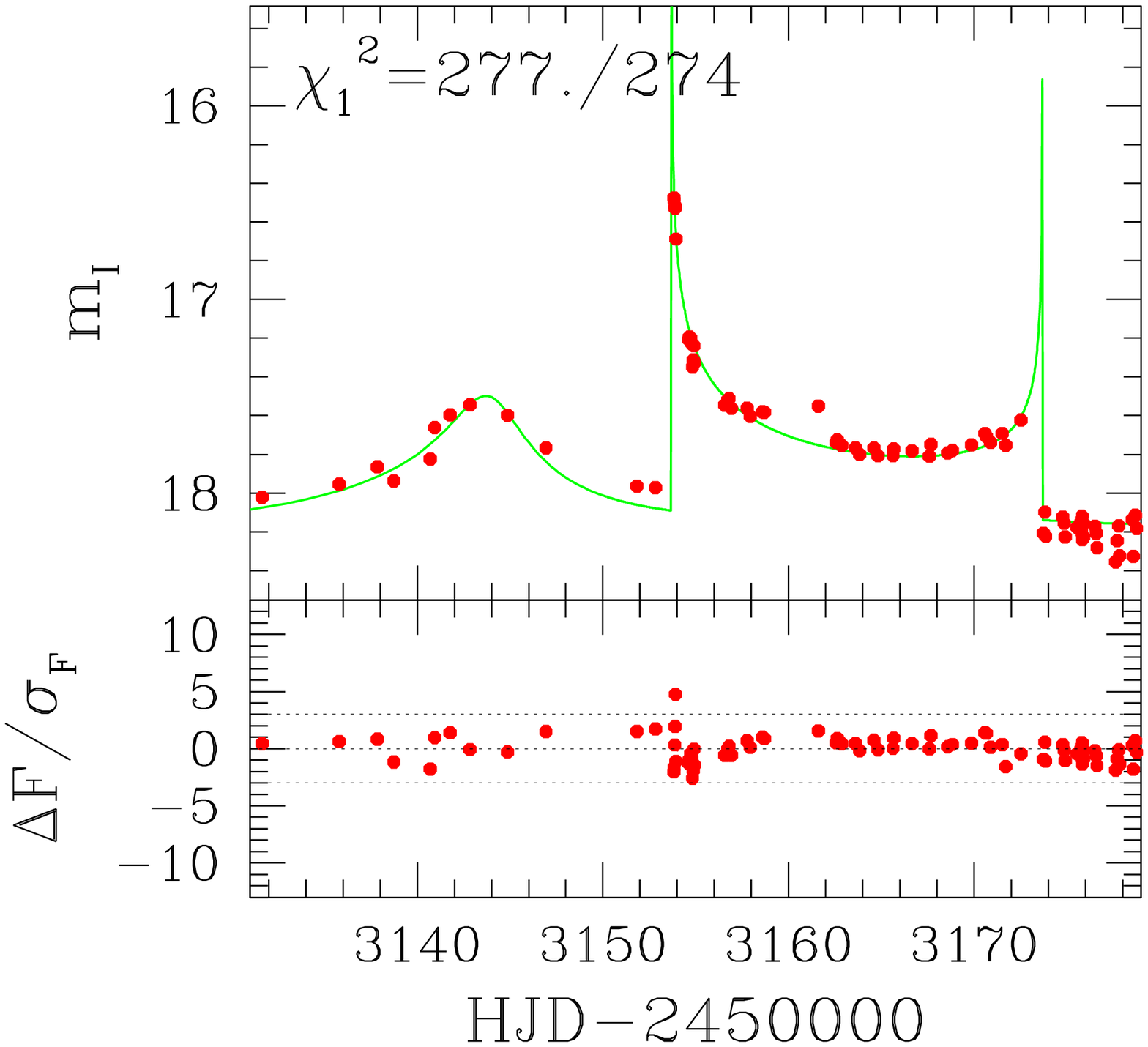}%

}

\noindent\parbox{12.75cm}{
\leftline {\bf OGLE 2004-BLG-273} 

 \includegraphics[height=63mm,width=62mm]{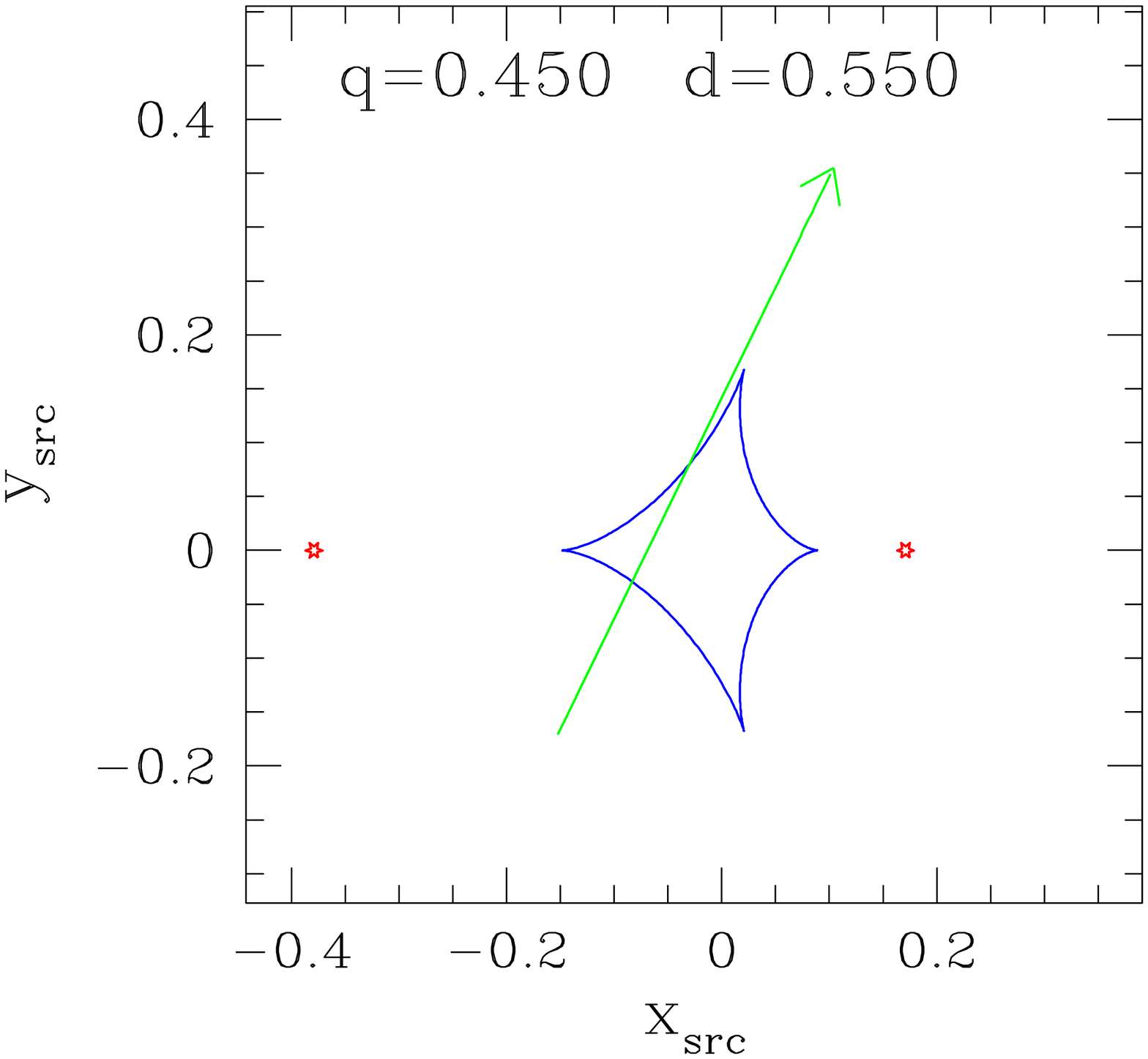}%
 \includegraphics[height=63mm,width=62mm]{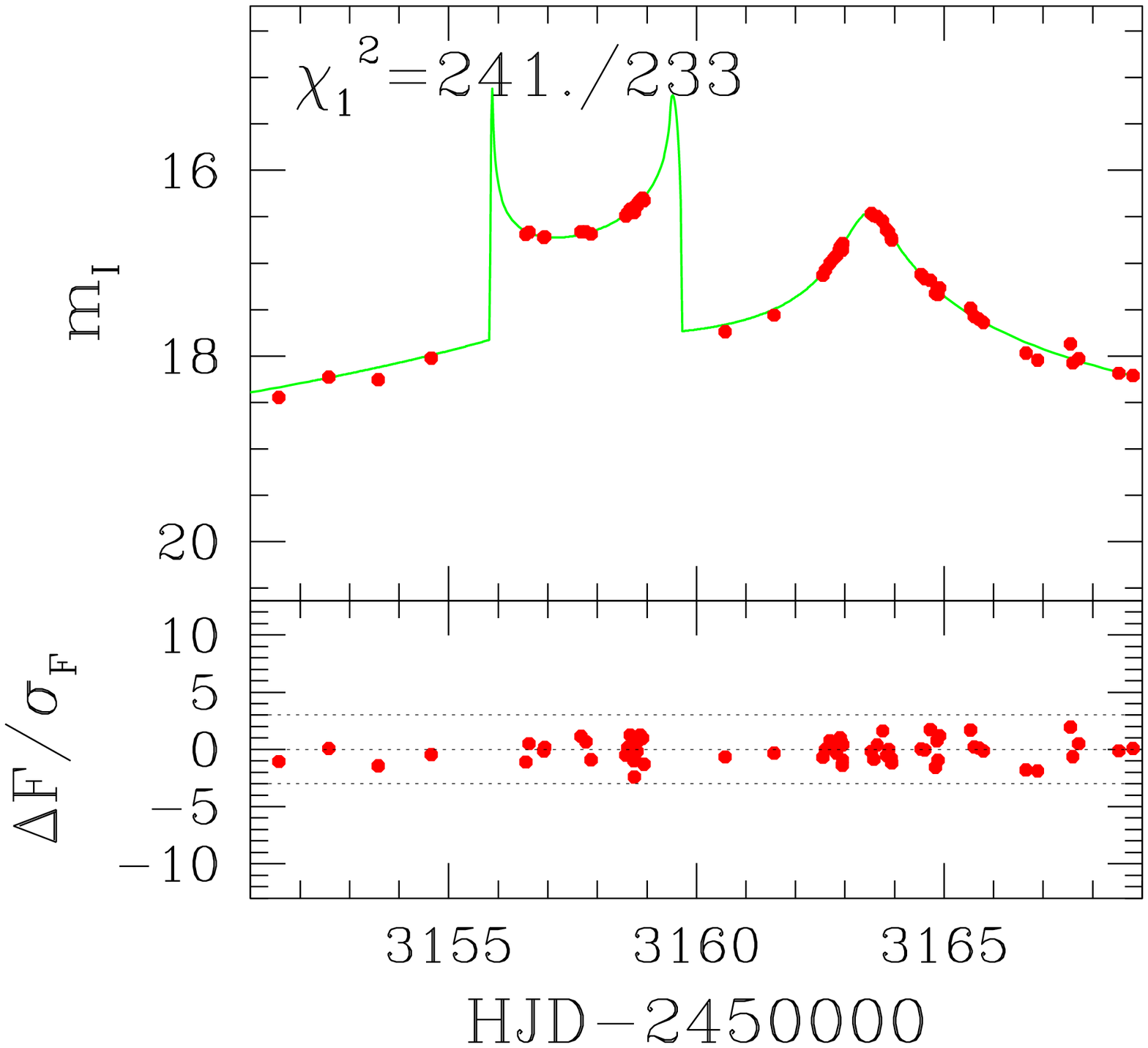}%

}

\noindent\parbox{12.75cm}{
\leftline {\bf OGLE 2004-BLG-280} 

 \includegraphics[height=63mm,width=62mm]{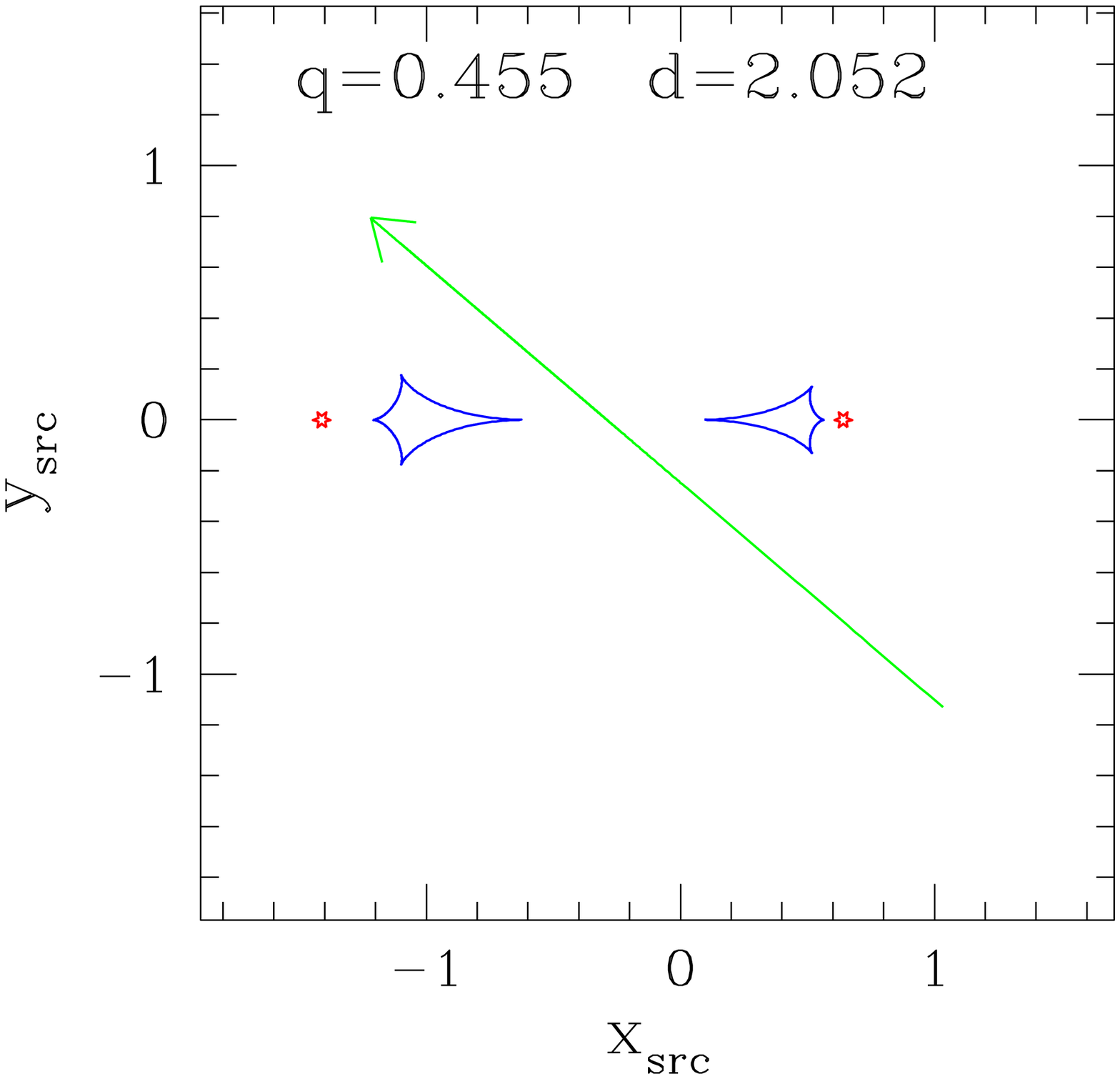}%
 \includegraphics[height=63mm,width=62mm]{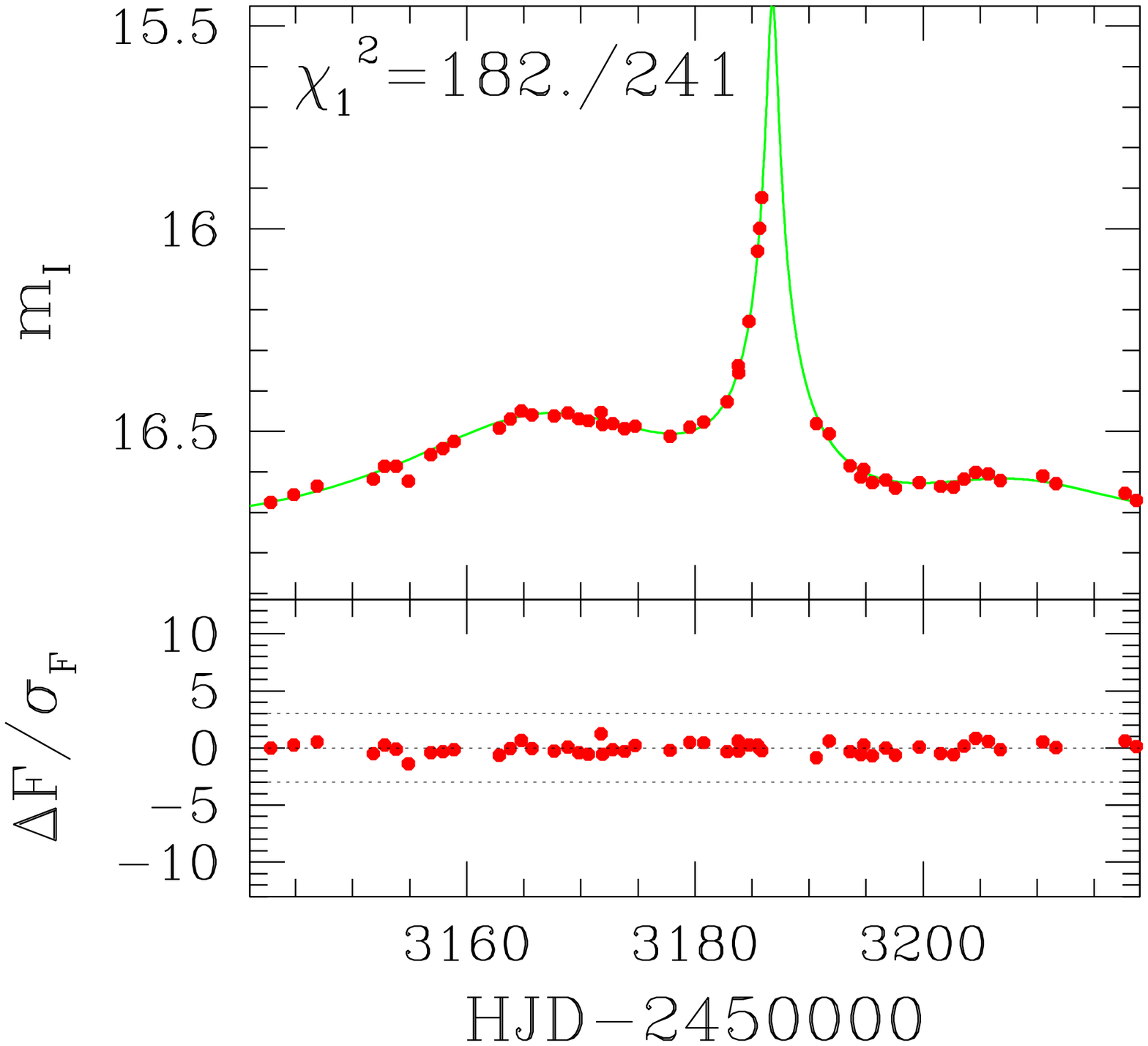}%

}

\noindent\parbox{12.75cm}{
\leftline {\bf OGLE 2004-BLG-309} 

 \includegraphics[height=63mm,width=62mm]{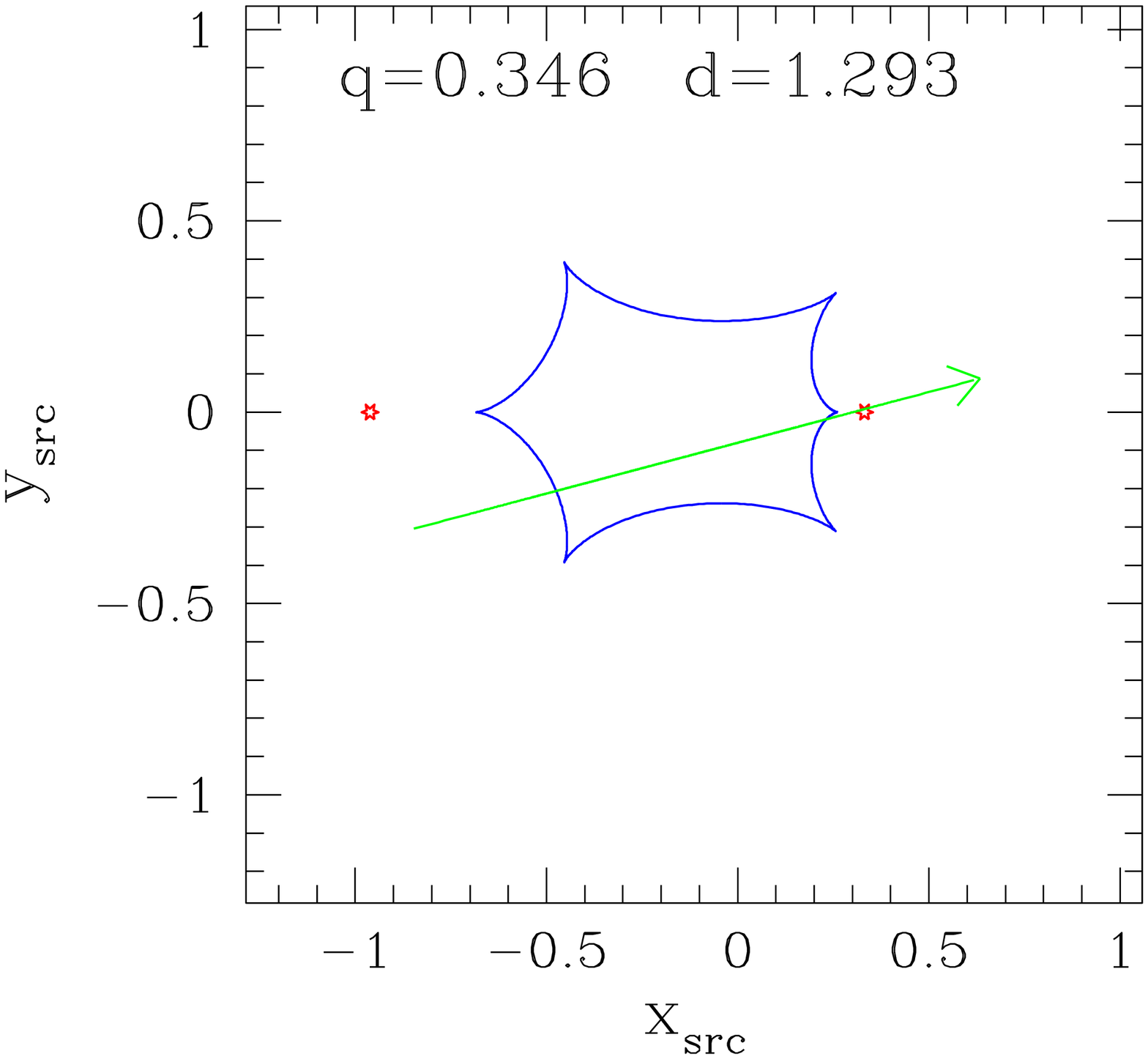}%
 \includegraphics[height=63mm,width=62mm]{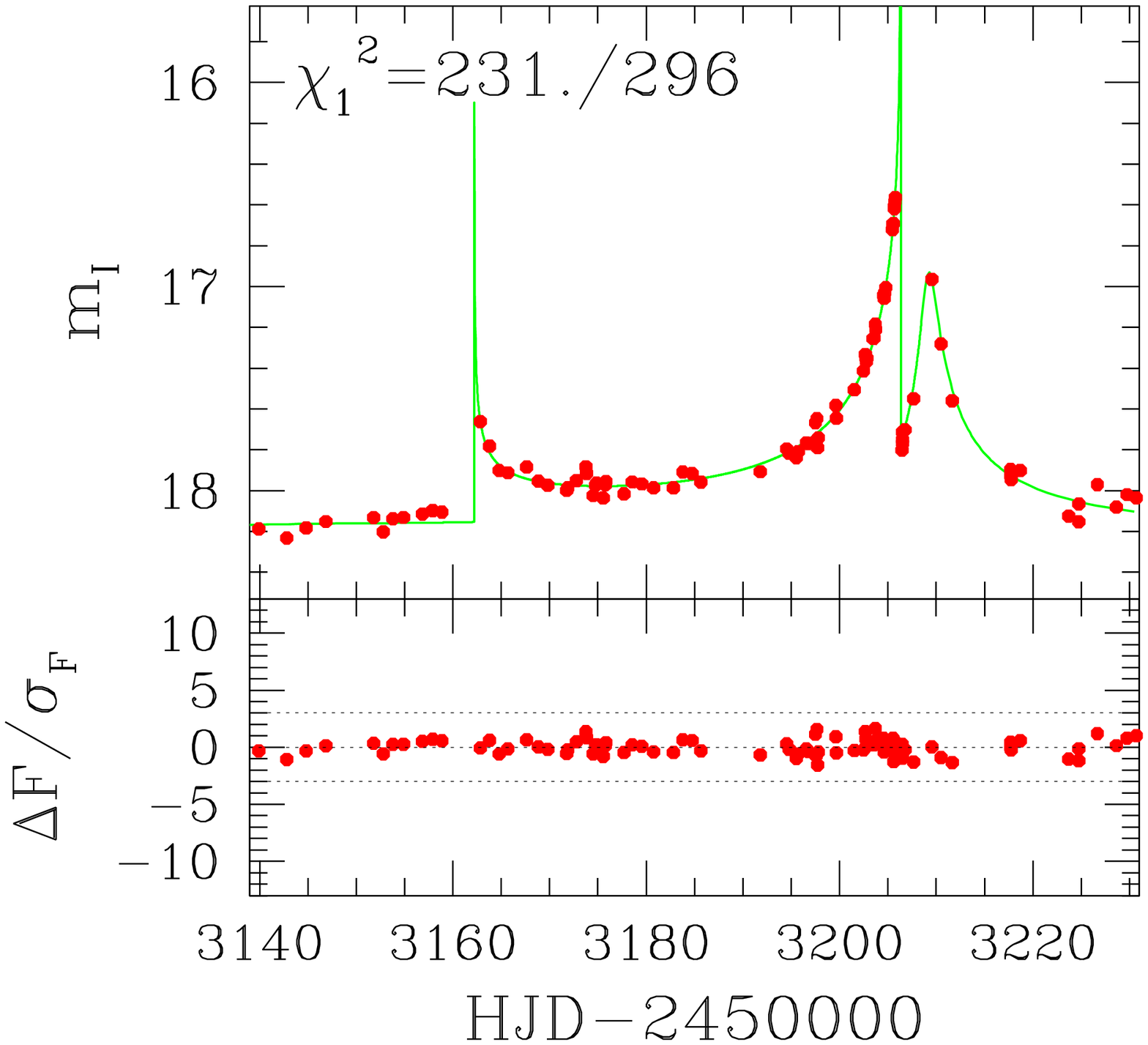}%

}

\noindent\parbox{12.75cm}{
\leftline {\bf OGLE 2004-BLG-325} 

 \includegraphics[height=63mm,width=62mm]{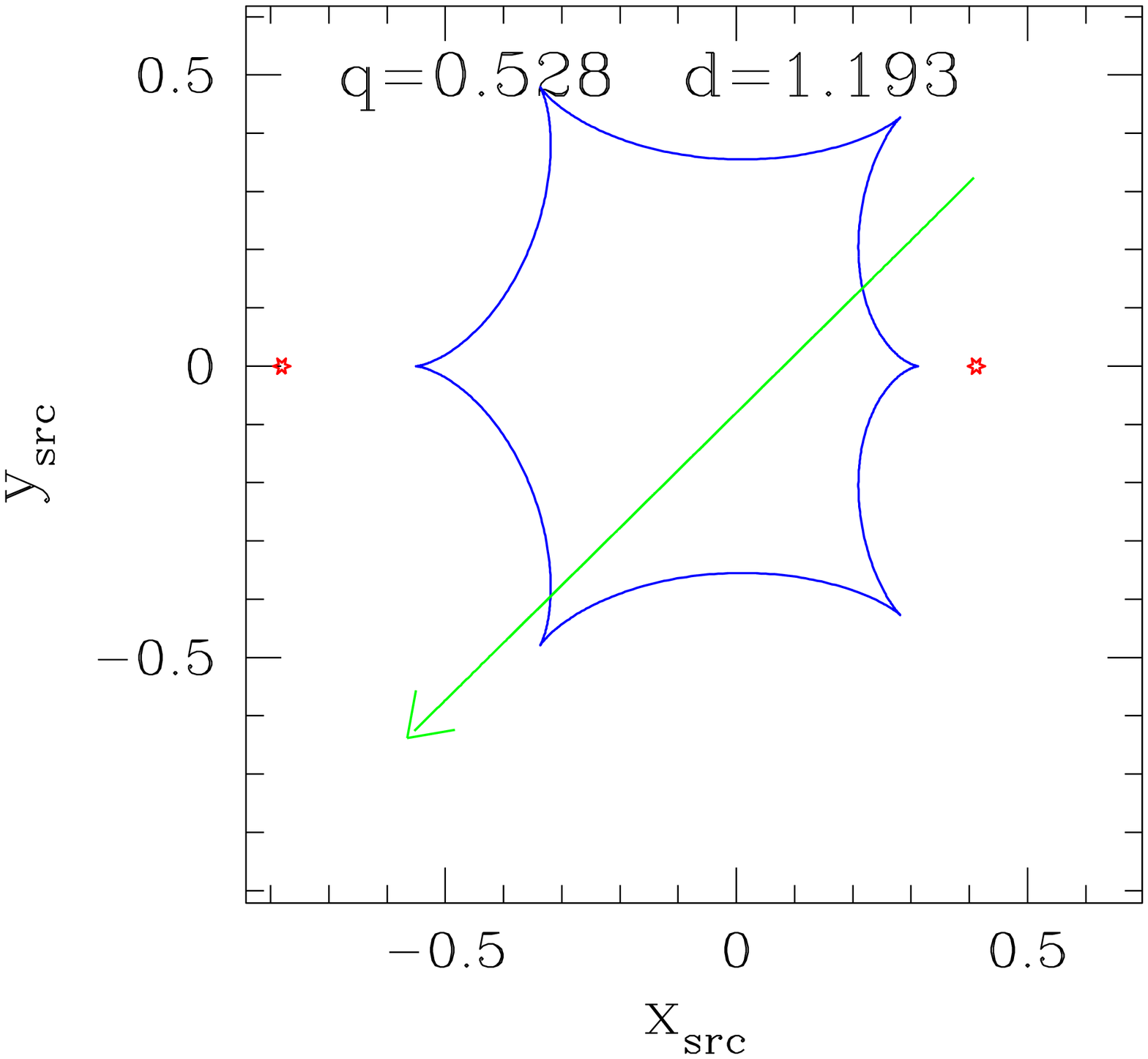}%
 \includegraphics[height=63mm,width=62mm]{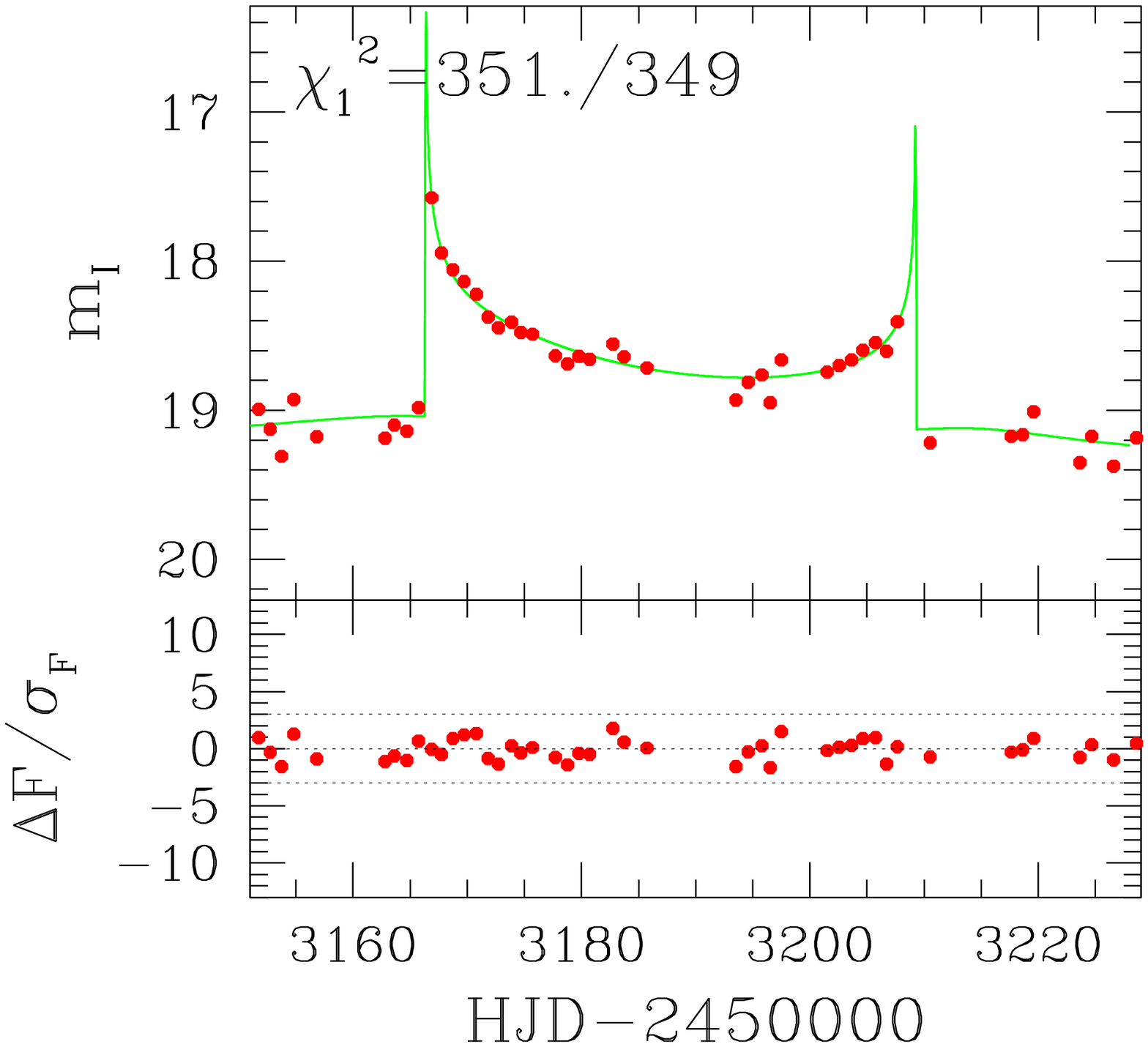}%

}

\noindent\parbox{12.75cm}{
\leftline {\bf OGLE 2004-BLG-347: I} 

 \includegraphics[height=63mm,width=62mm]{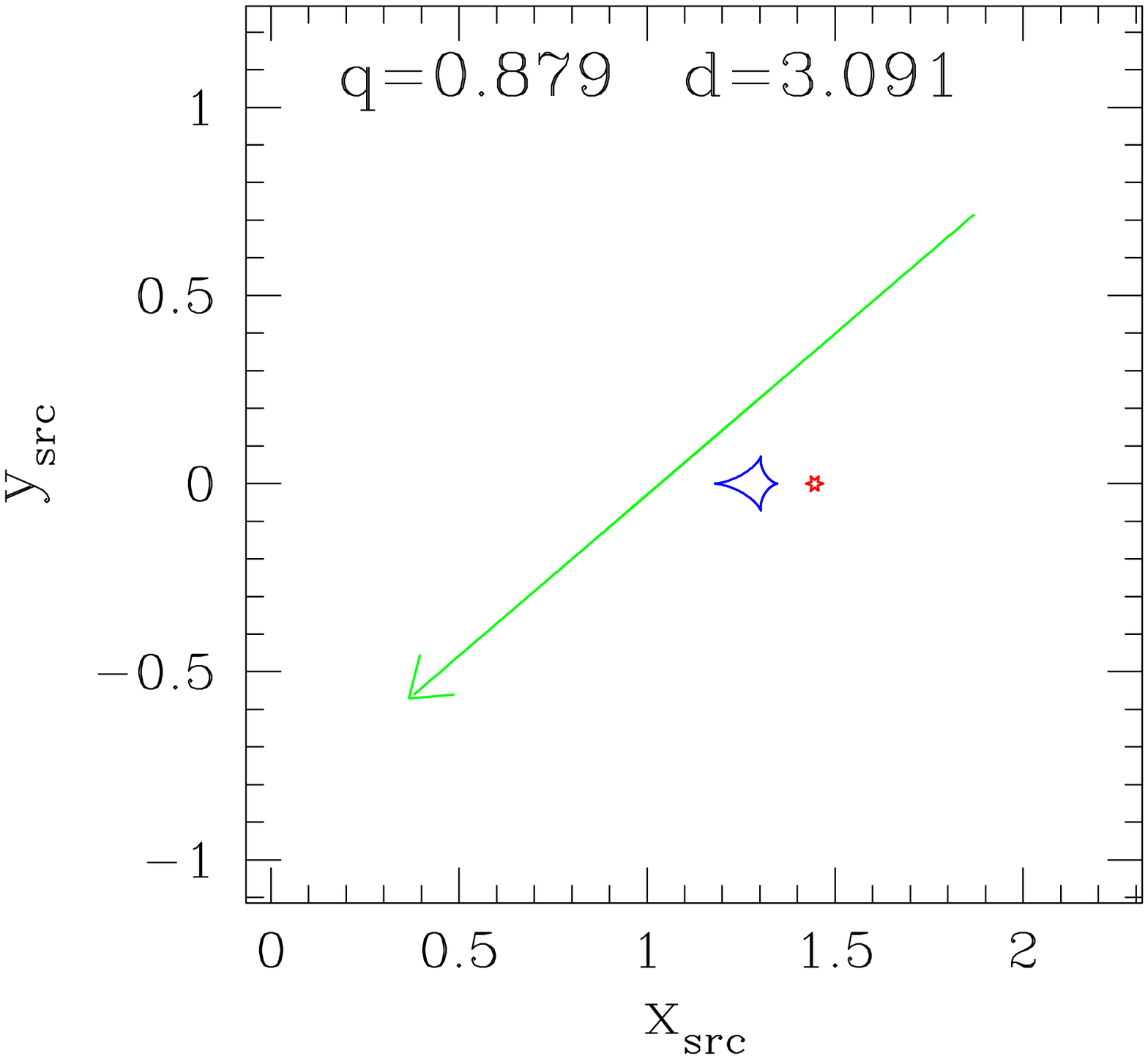}%
 \includegraphics[height=63mm,width=62mm]{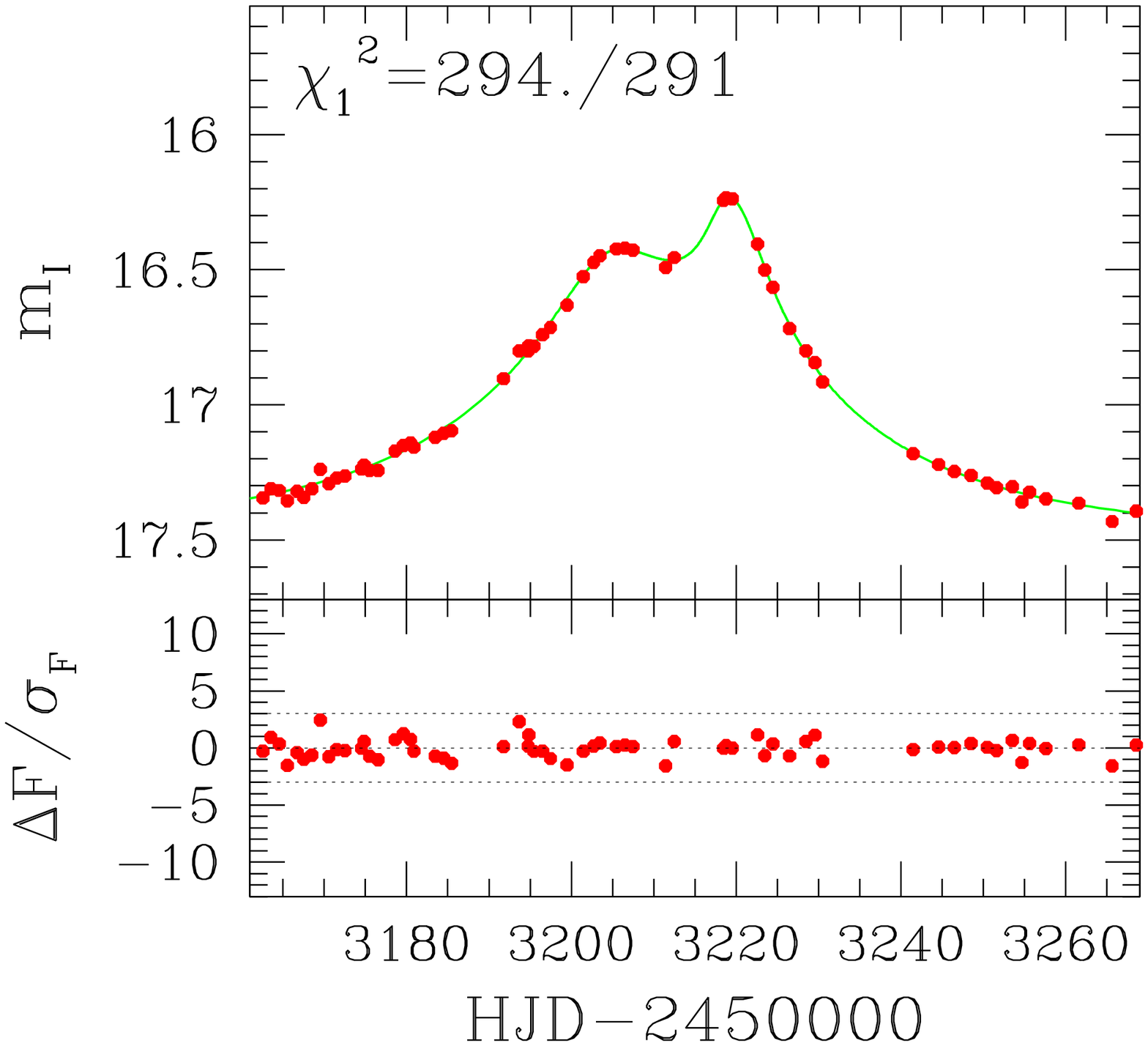}%

}

\noindent\parbox{12.75cm}{
\leftline {\bf OGLE 2004-BLG-347: II} 

 \includegraphics[height=63mm,width=62mm]{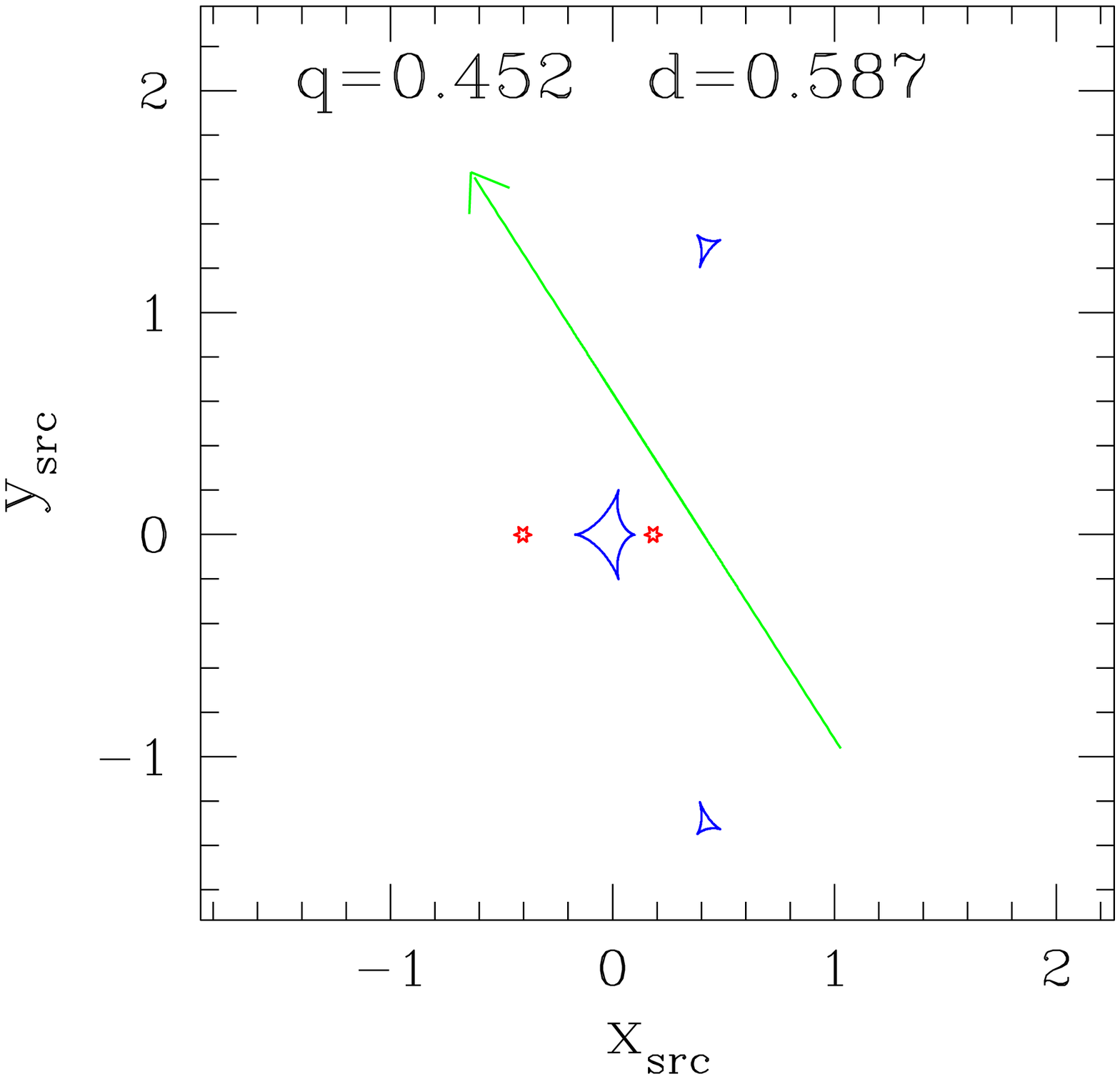}%
 \includegraphics[height=63mm,width=62mm]{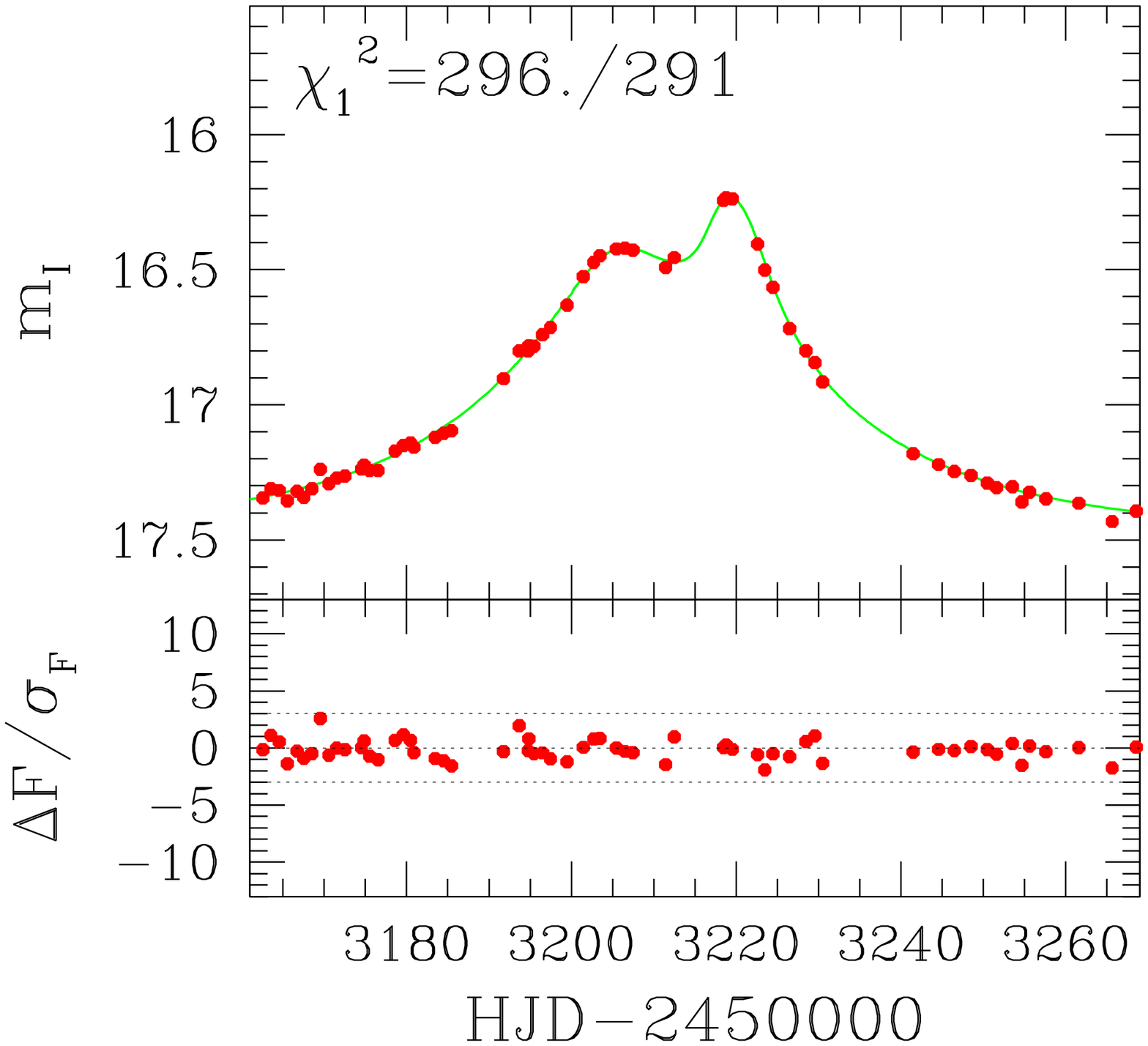}%

}

\noindent\parbox{12.75cm}{
\leftline {\bf OGLE 2004-BLG-354} 

 \includegraphics[height=63mm,width=62mm]{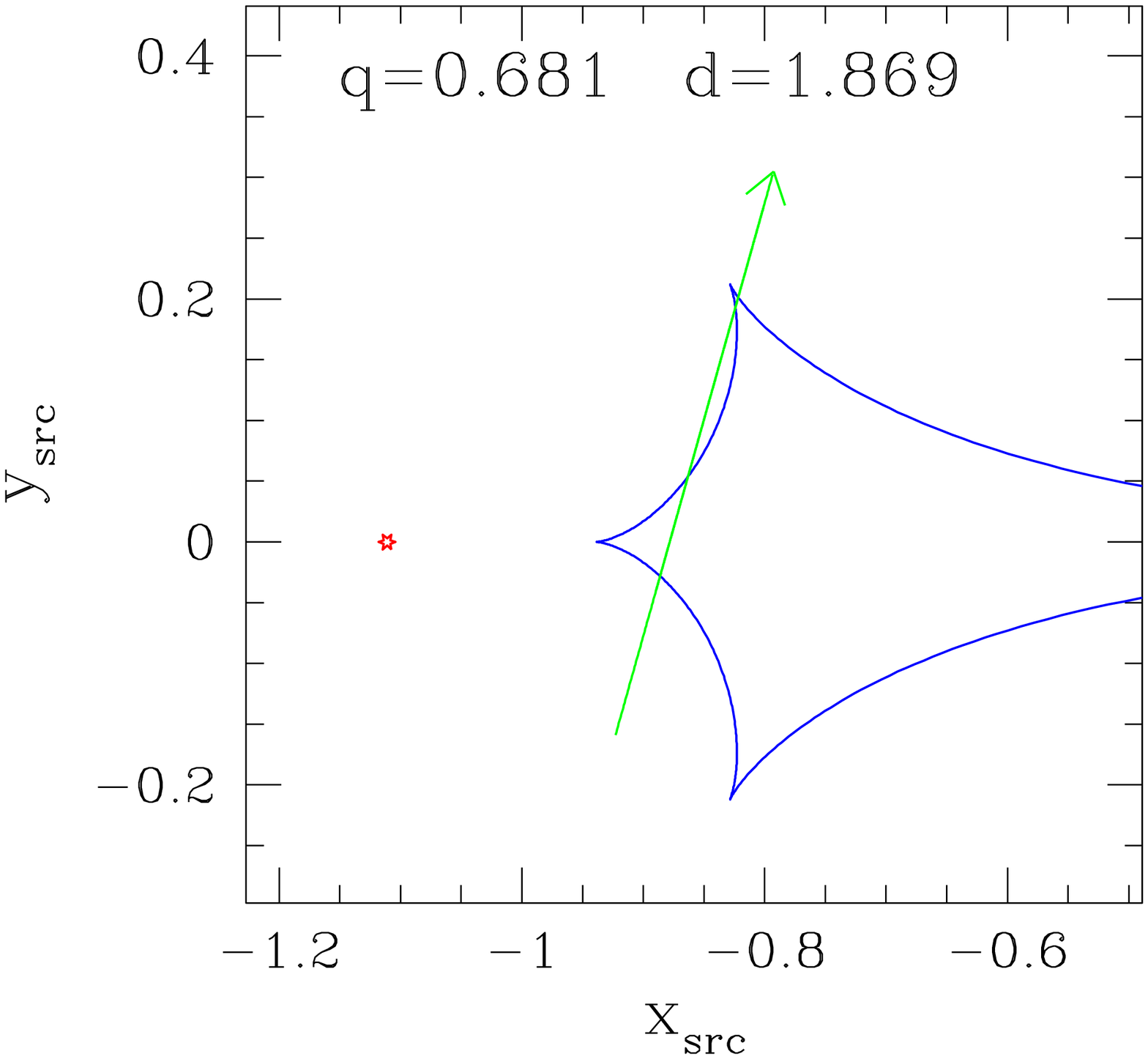}%
 \includegraphics[height=63mm,width=62mm]{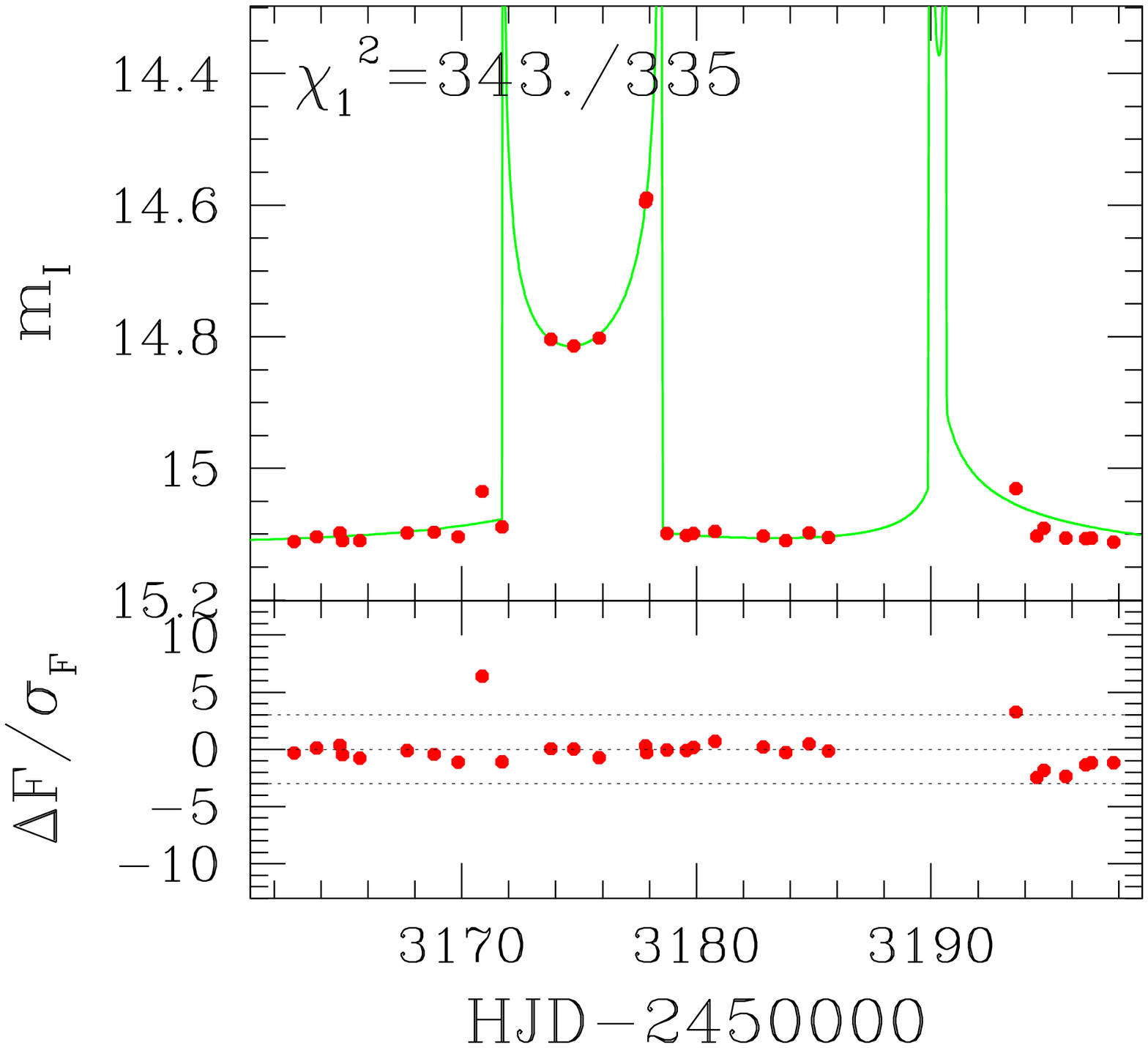}%

}

\noindent\parbox{12.75cm}{
\leftline {\bf OGLE 2004-BLG-362} 

 \includegraphics[height=63mm,width=62mm]{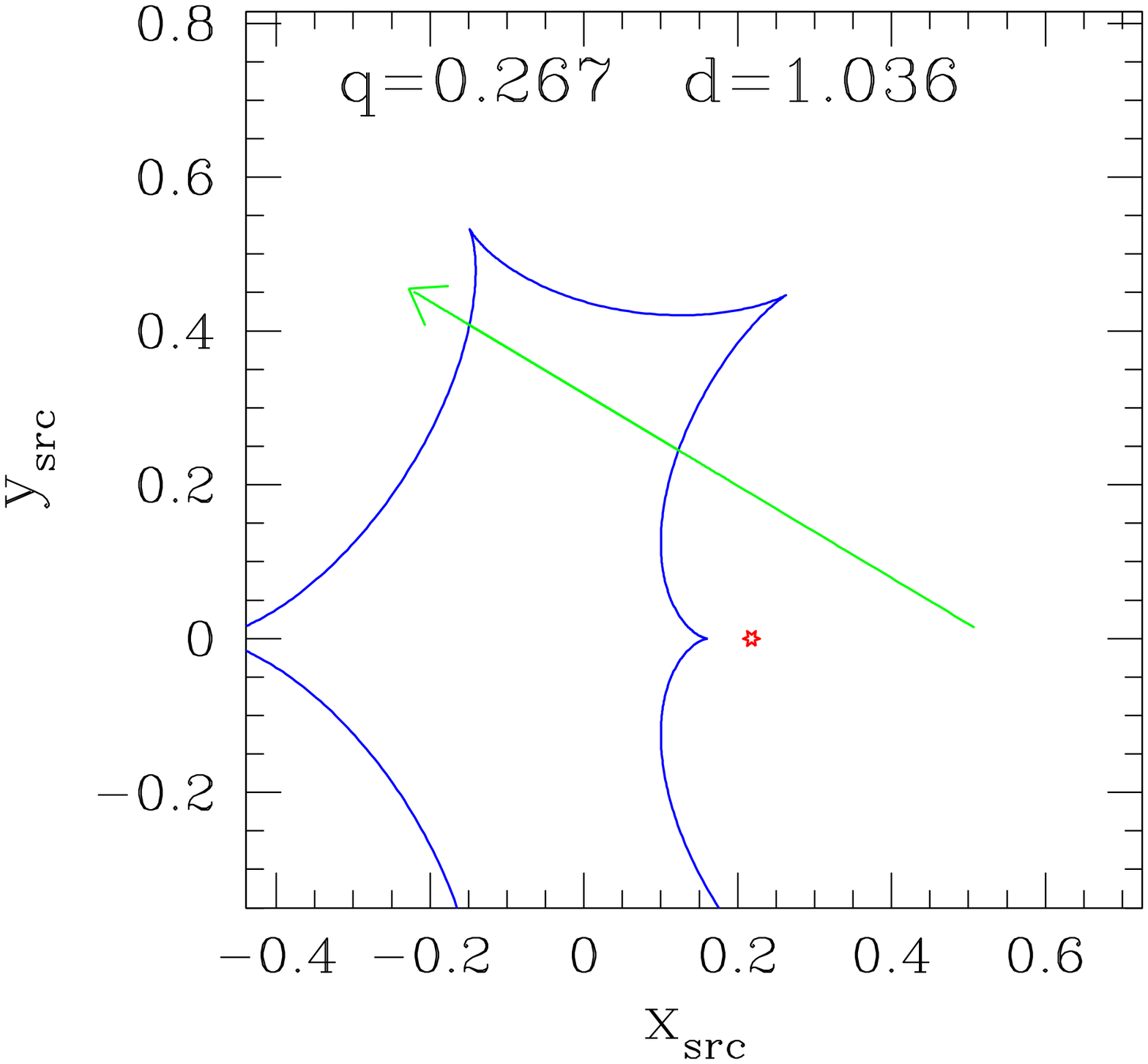}%
 \includegraphics[height=63mm,width=62mm]{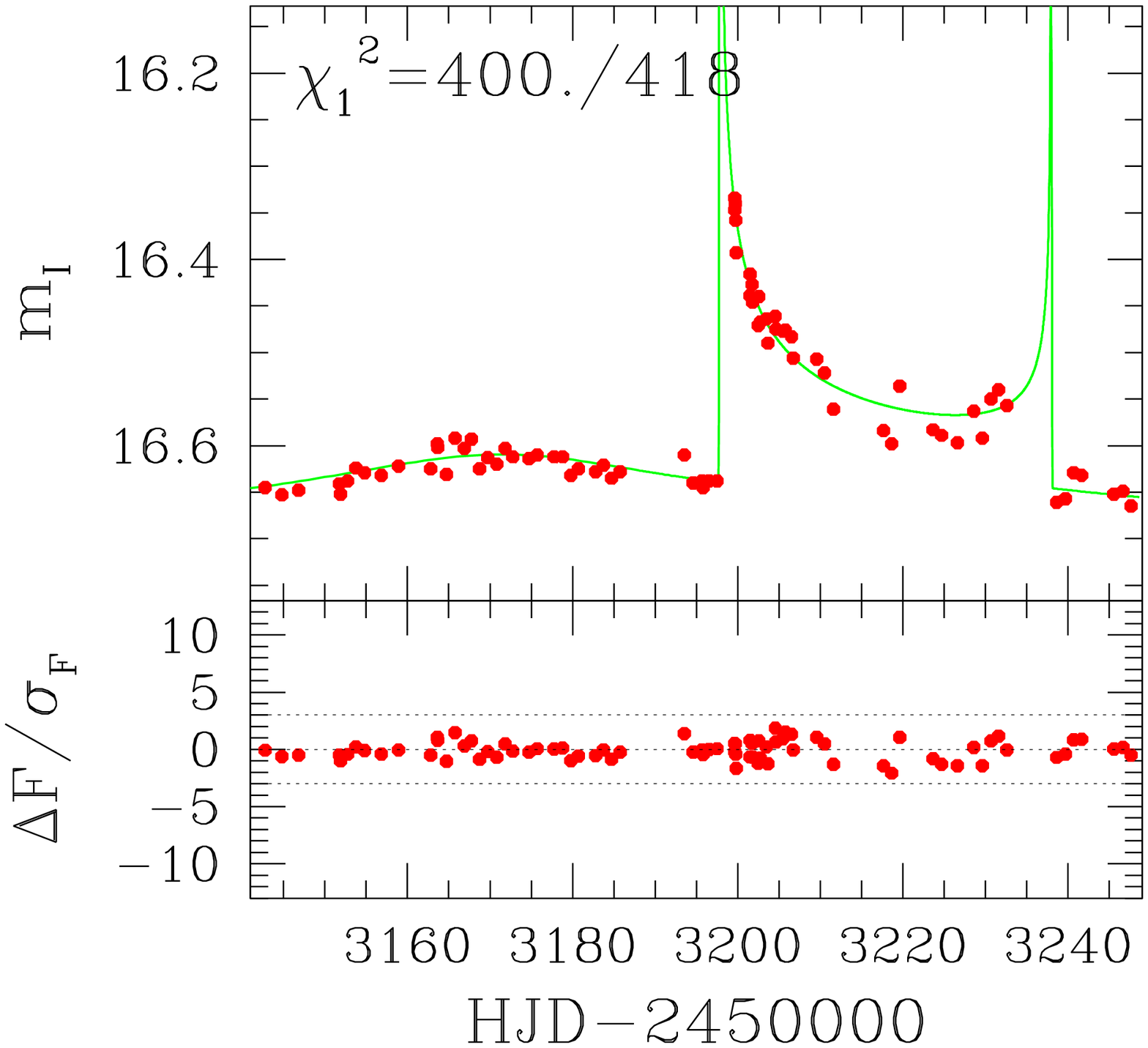}%

}

\noindent\parbox{12.75cm}{
\leftline {\bf OGLE 2004-BLG-366} 

 \includegraphics[height=63mm,width=62mm]{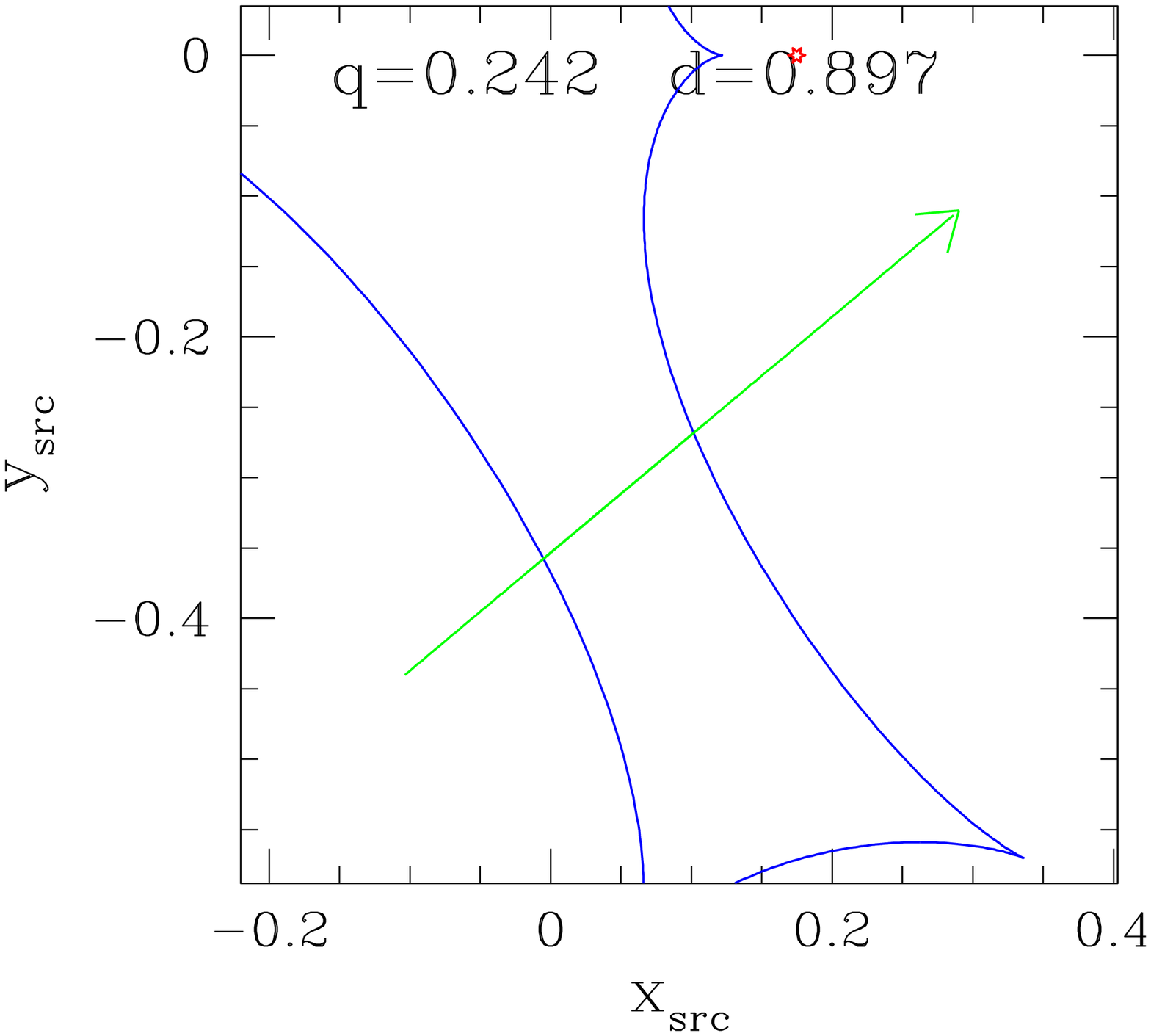}%
 \includegraphics[height=63mm,width=62mm]{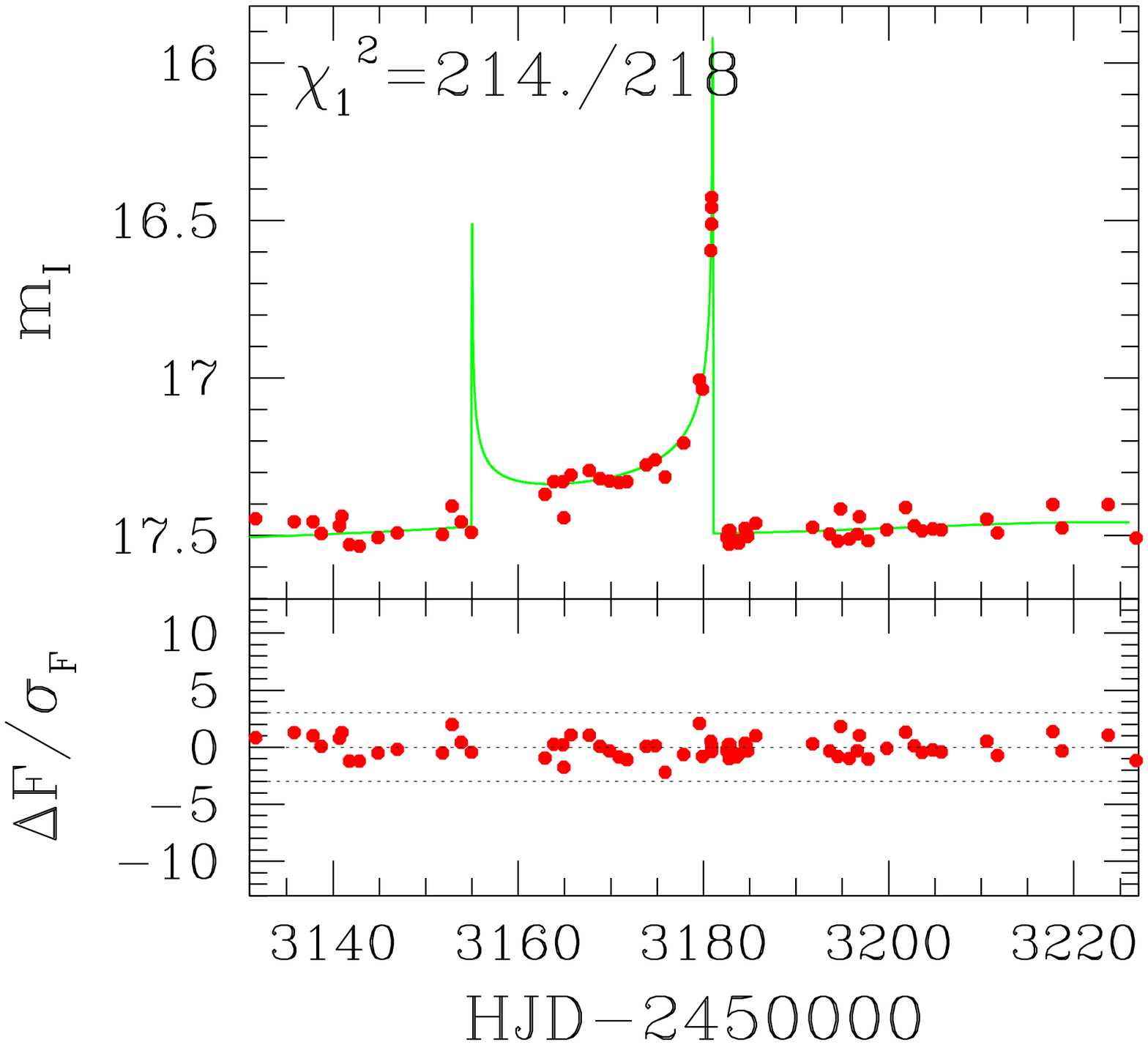}%

}

\noindent\parbox{12.75cm}{
\leftline {\bf OGLE 2004-BLG-367} 

 \includegraphics[height=63mm,width=62mm]{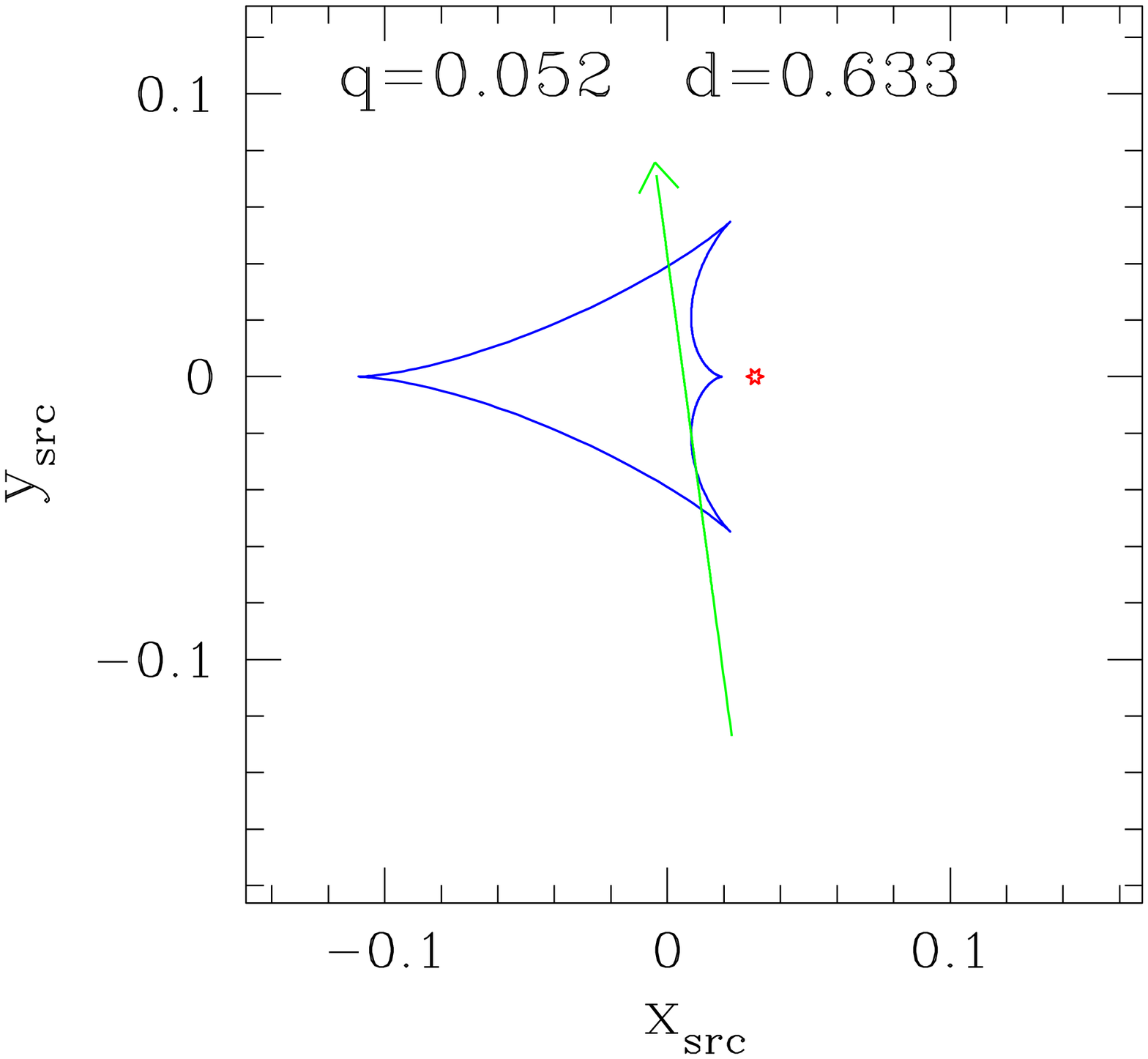}%
 \includegraphics[height=63mm,width=62mm]{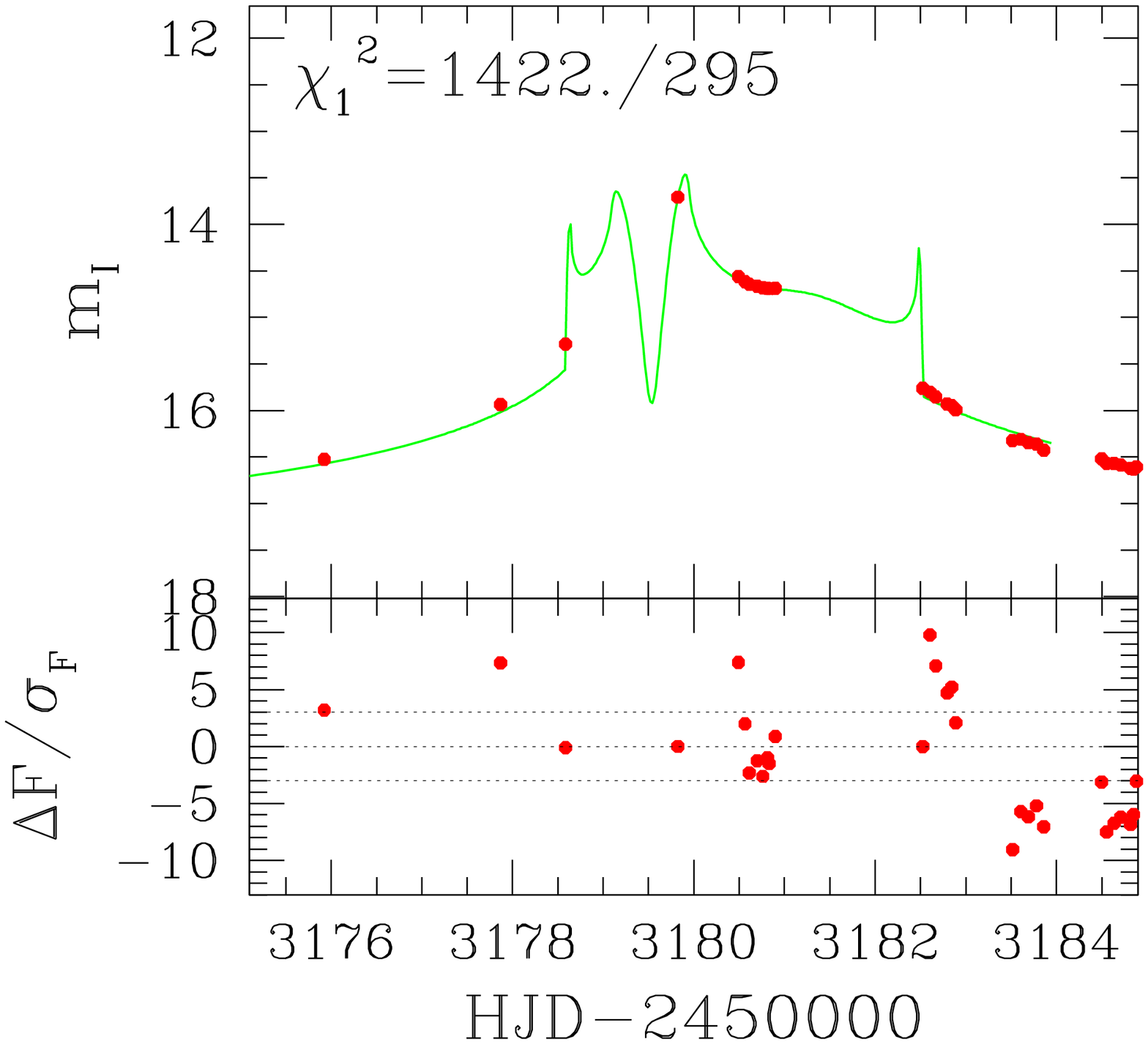}%

}

\noindent\parbox{12.75cm}{
\leftline {\bf OGLE 2004-BLG-373} 

 \includegraphics[height=63mm,width=62mm]{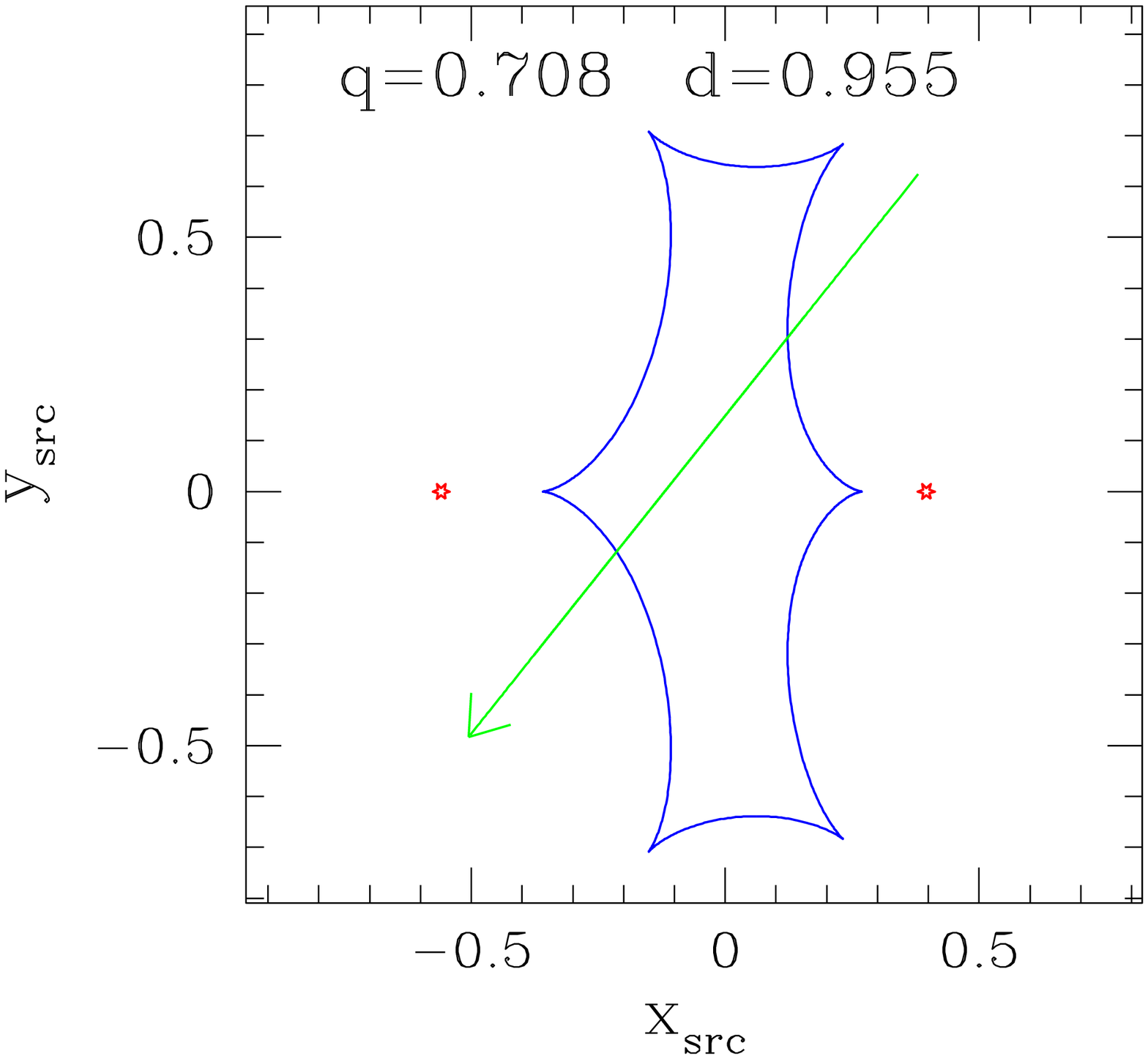}%
 \includegraphics[height=63mm,width=62mm]{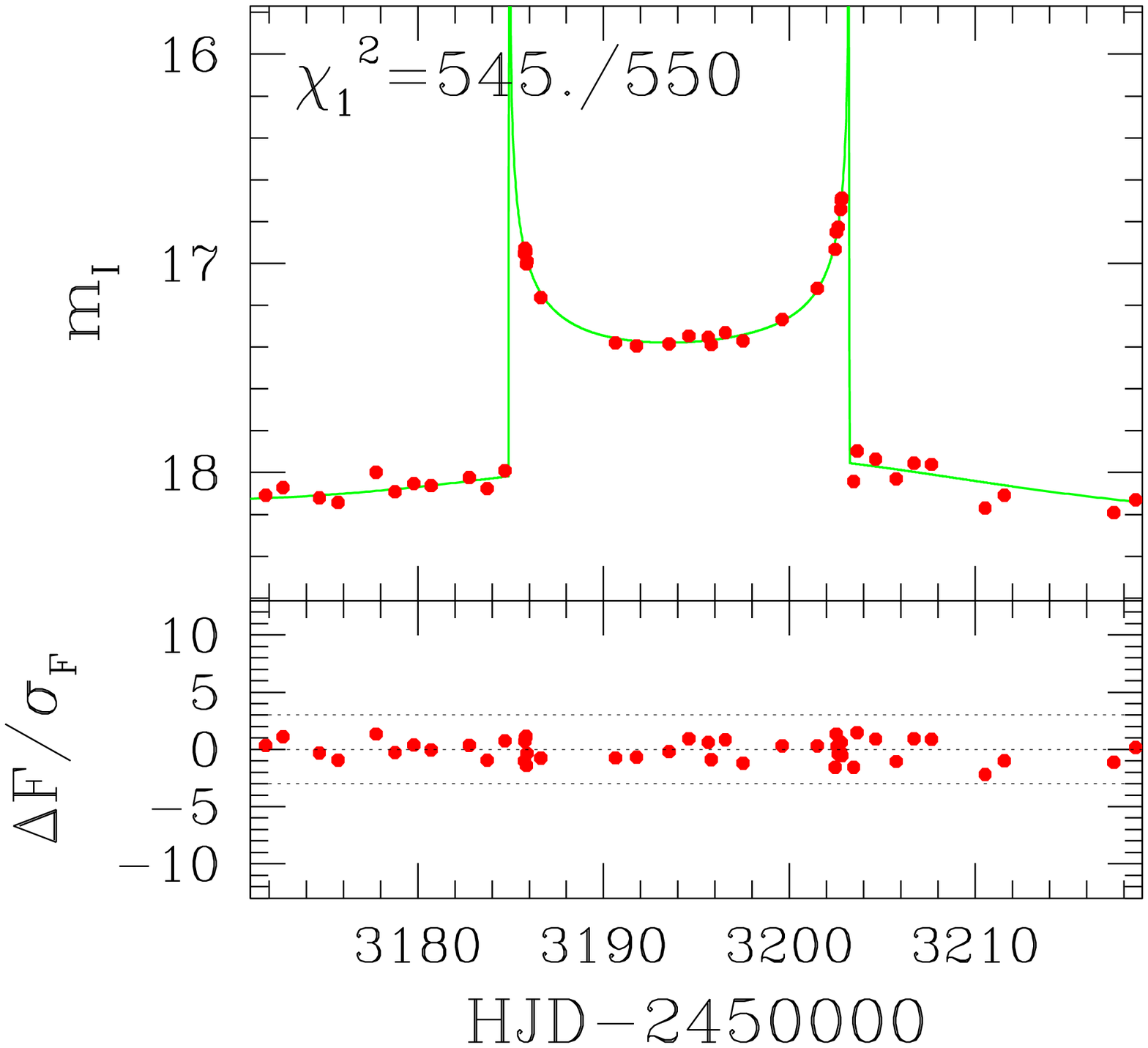}%

}

\noindent\parbox{12.75cm}{
\leftline {\bf OGLE 2004-BLG-379} 

 \includegraphics[height=63mm,width=62mm]{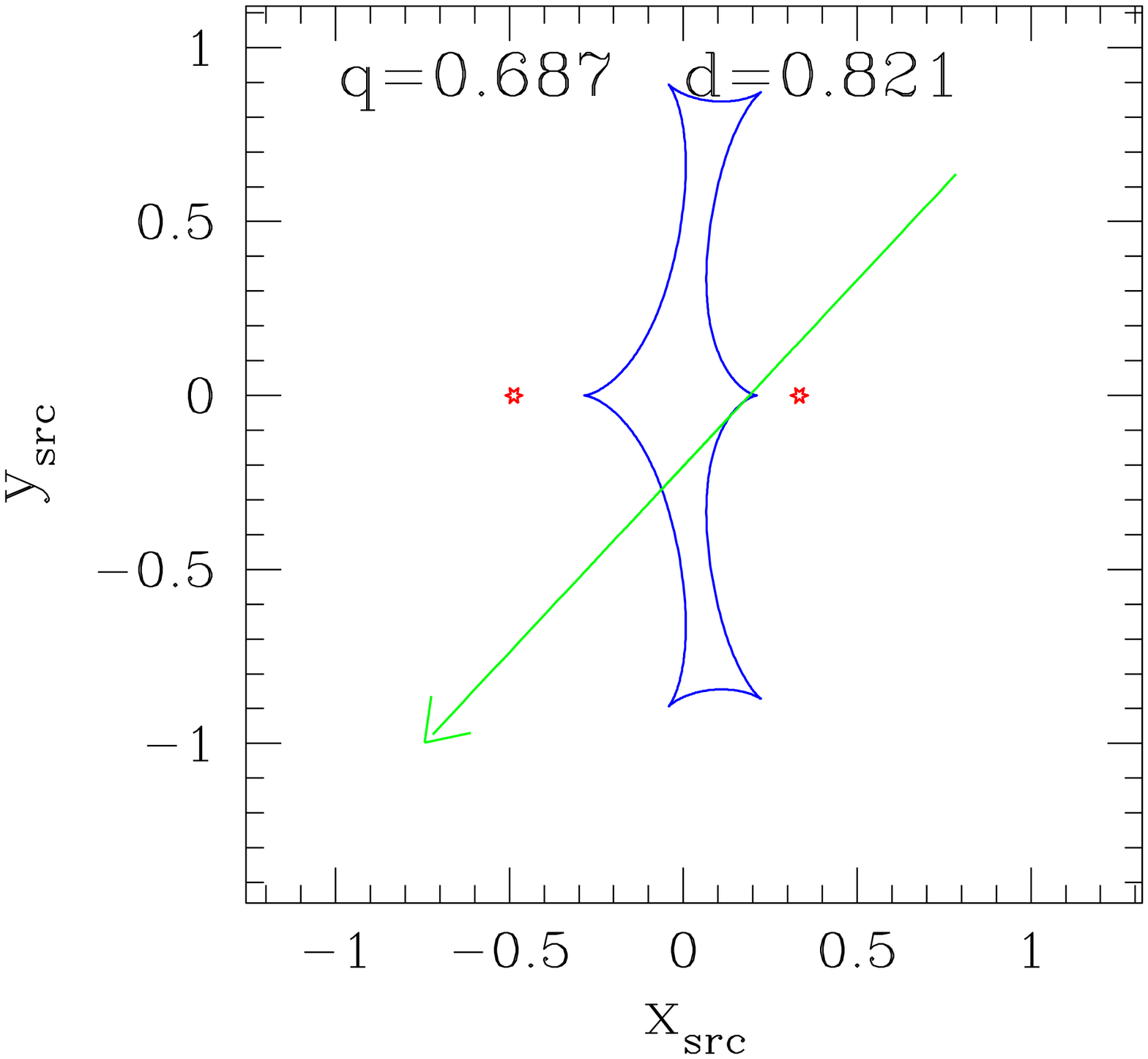}%
 \includegraphics[height=63mm,width=62mm]{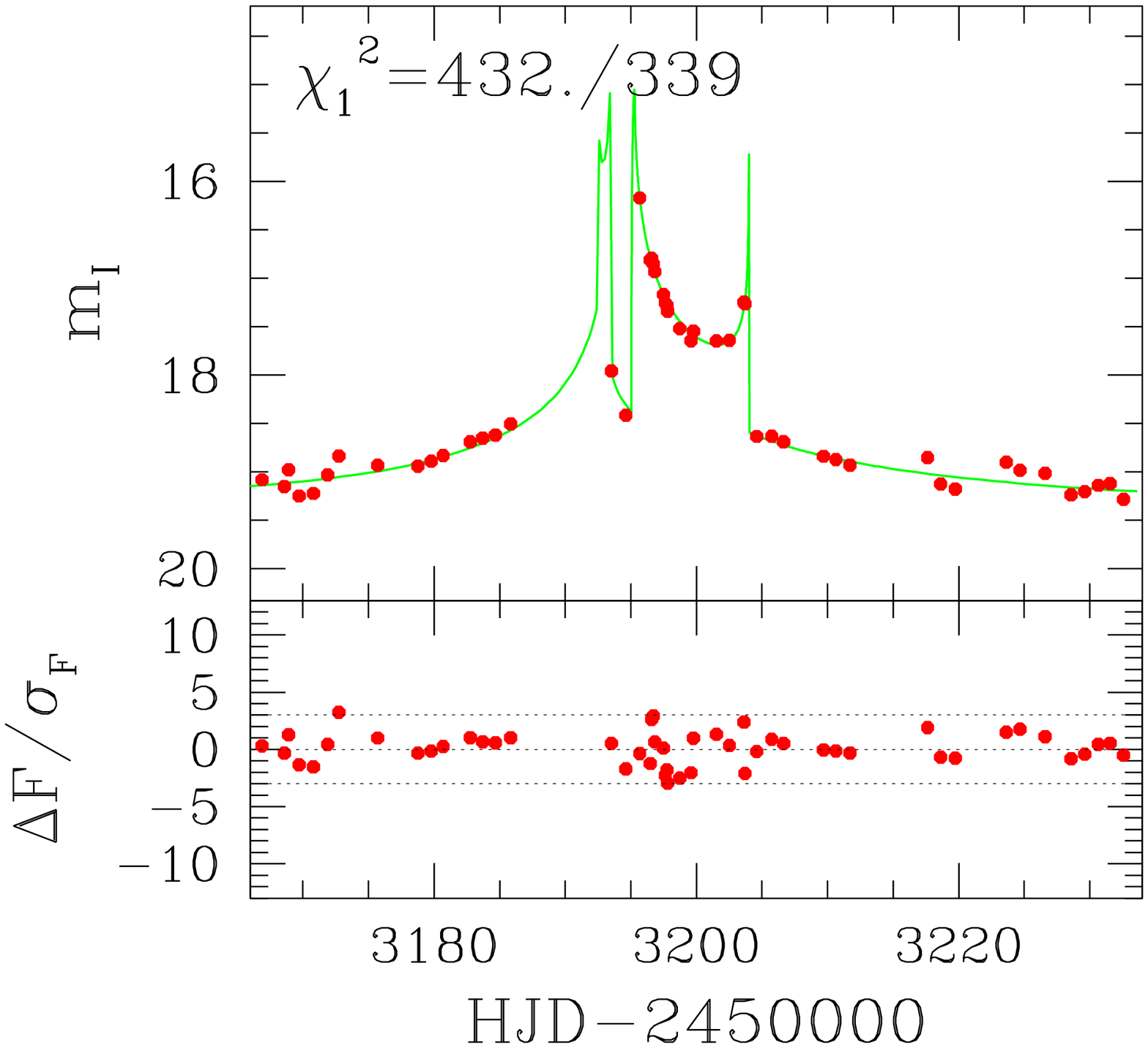}%

}

\noindent\parbox{12.75cm}{
\leftline {\bf OGLE 2004-BLG-406: I} 

 \includegraphics[height=63mm,width=62mm]{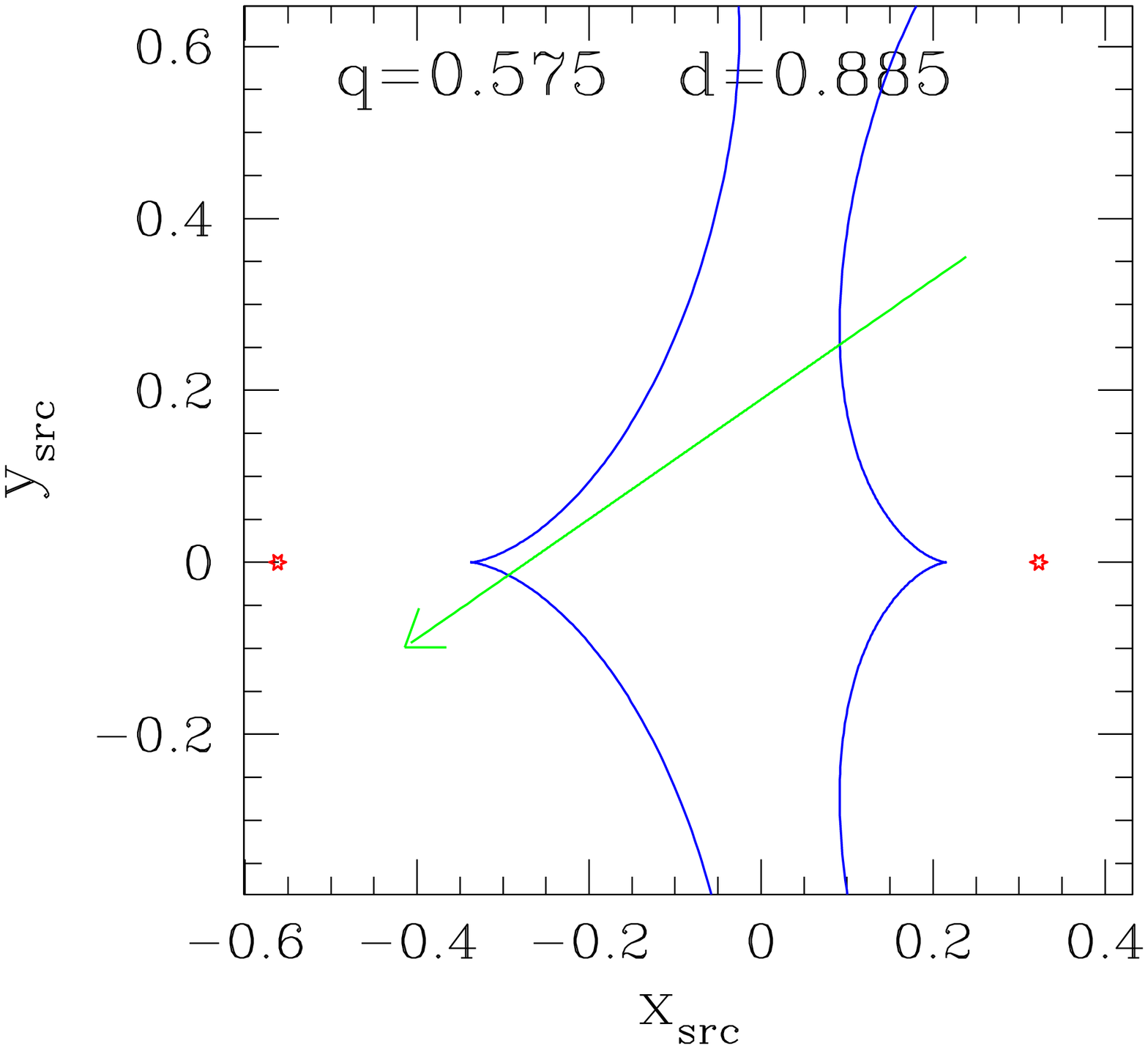}%
 \includegraphics[height=63mm,width=62mm]{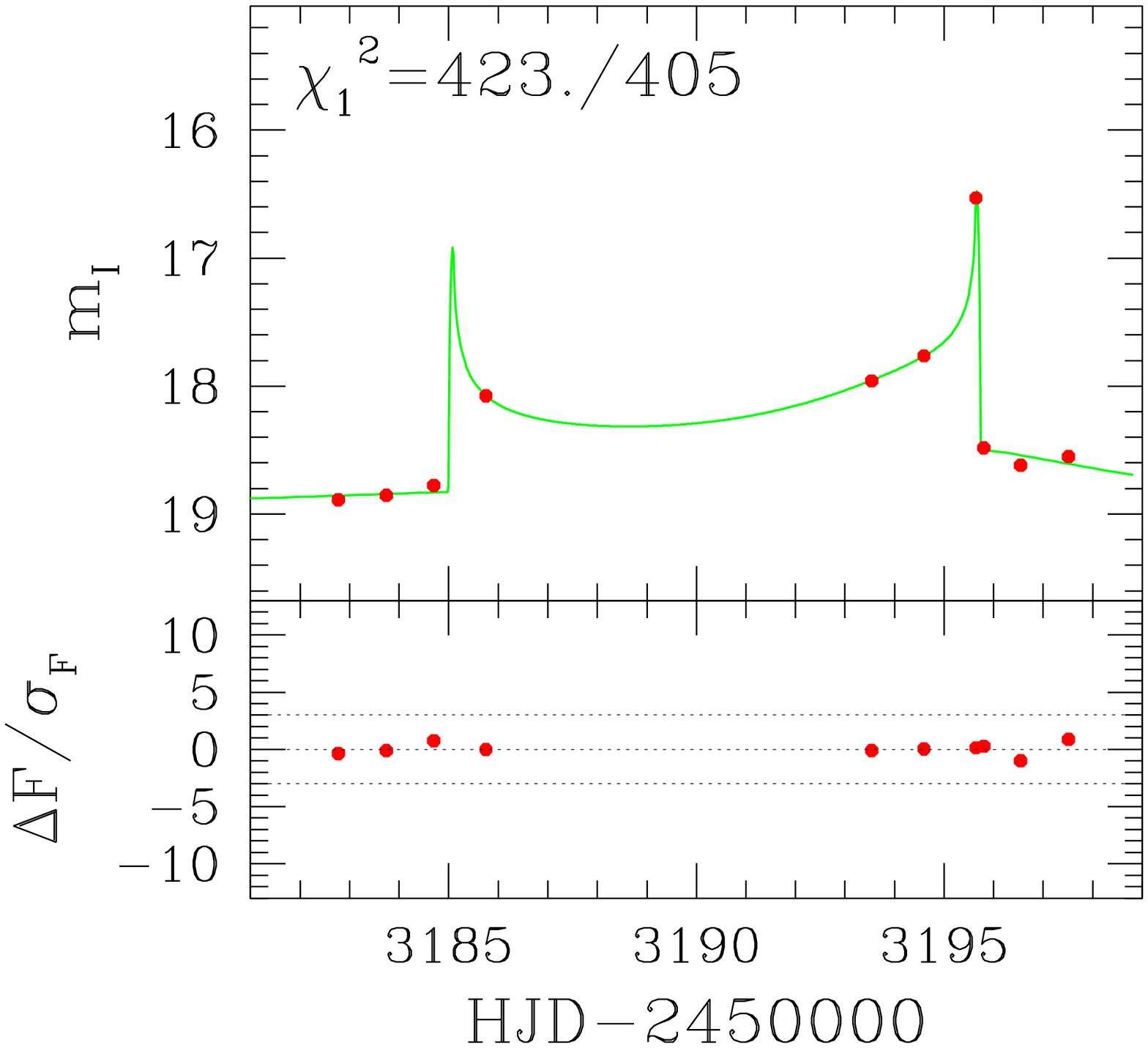}%

}

\noindent\parbox{12.75cm}{
\leftline {\bf OGLE 2004-BLG-406: II} 

 \includegraphics[height=63mm,width=62mm]{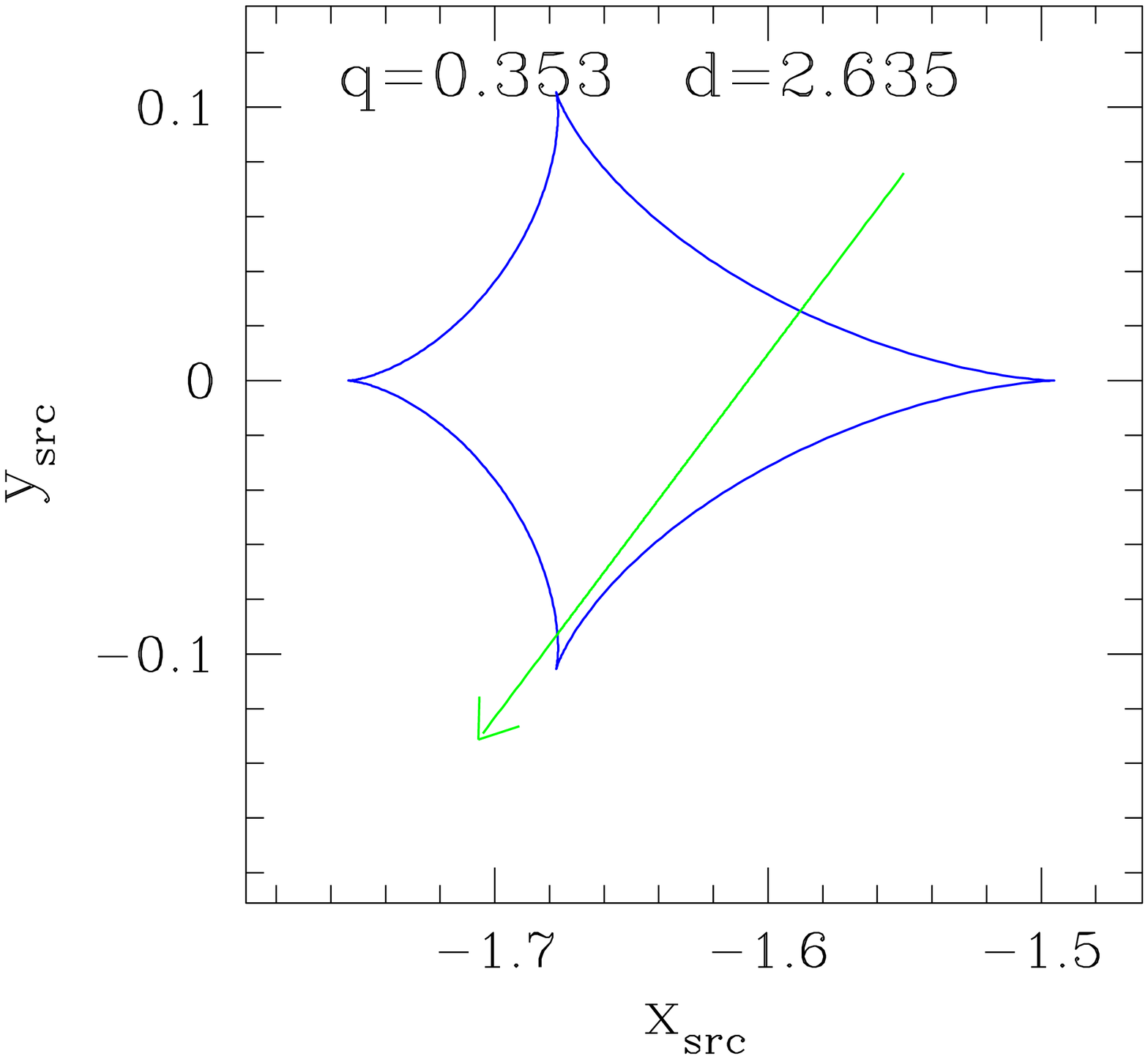}%
 \includegraphics[height=63mm,width=62mm]{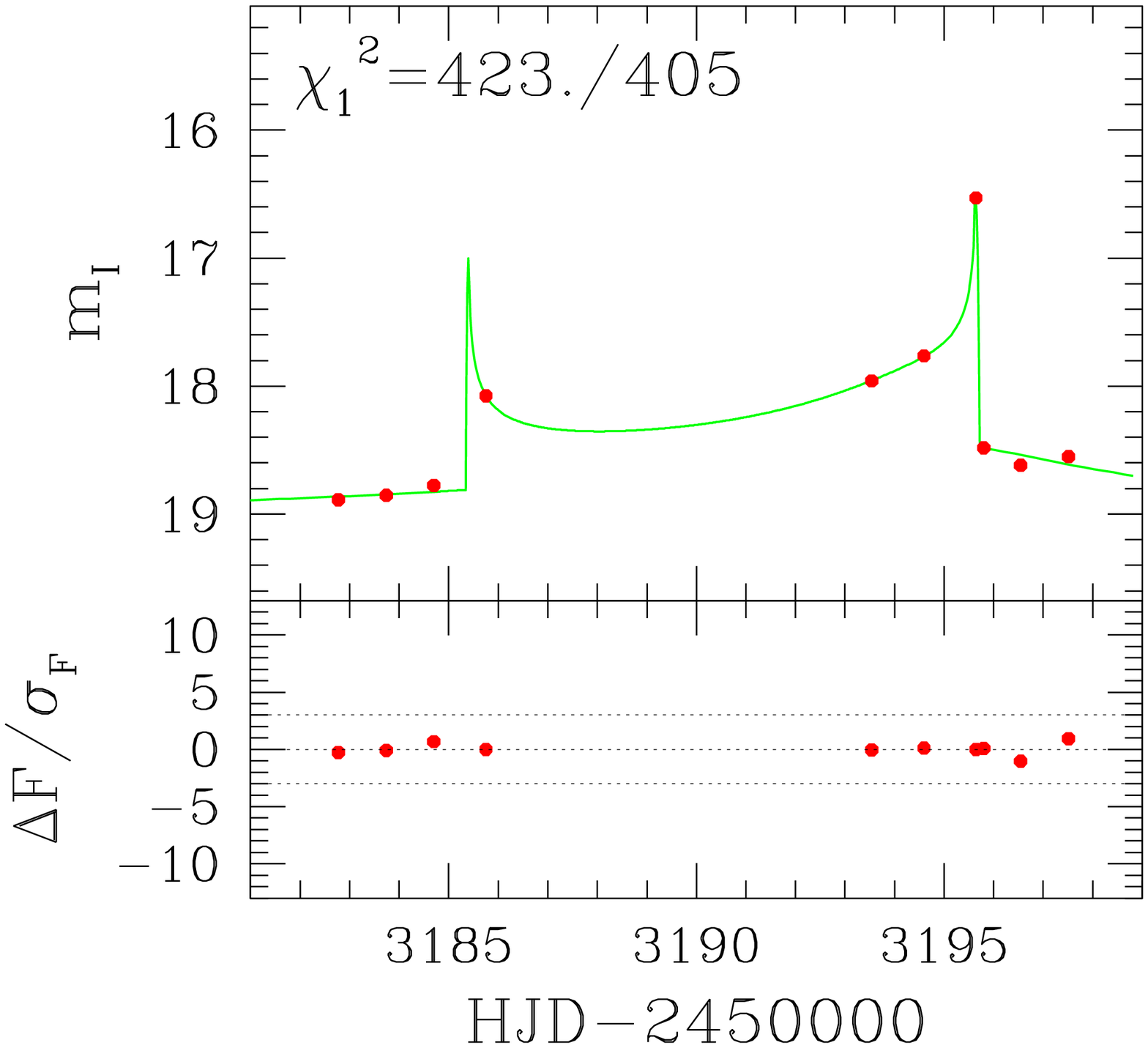}%

}

\noindent\parbox{12.75cm}{
\leftline {\bf OGLE 2004-BLG-444} 

 \includegraphics[height=63mm,width=62mm]{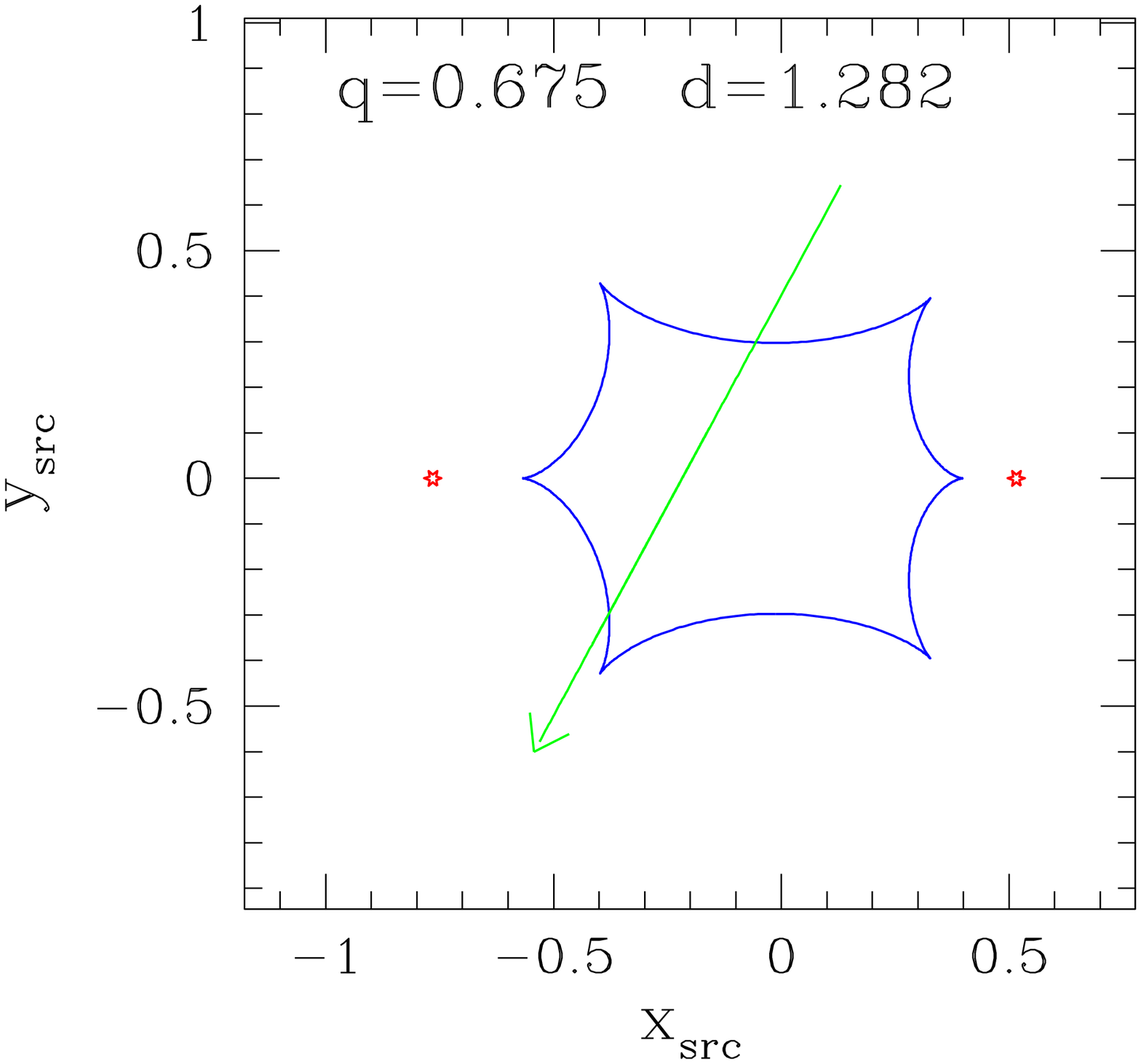}%
 \includegraphics[height=63mm,width=62mm]{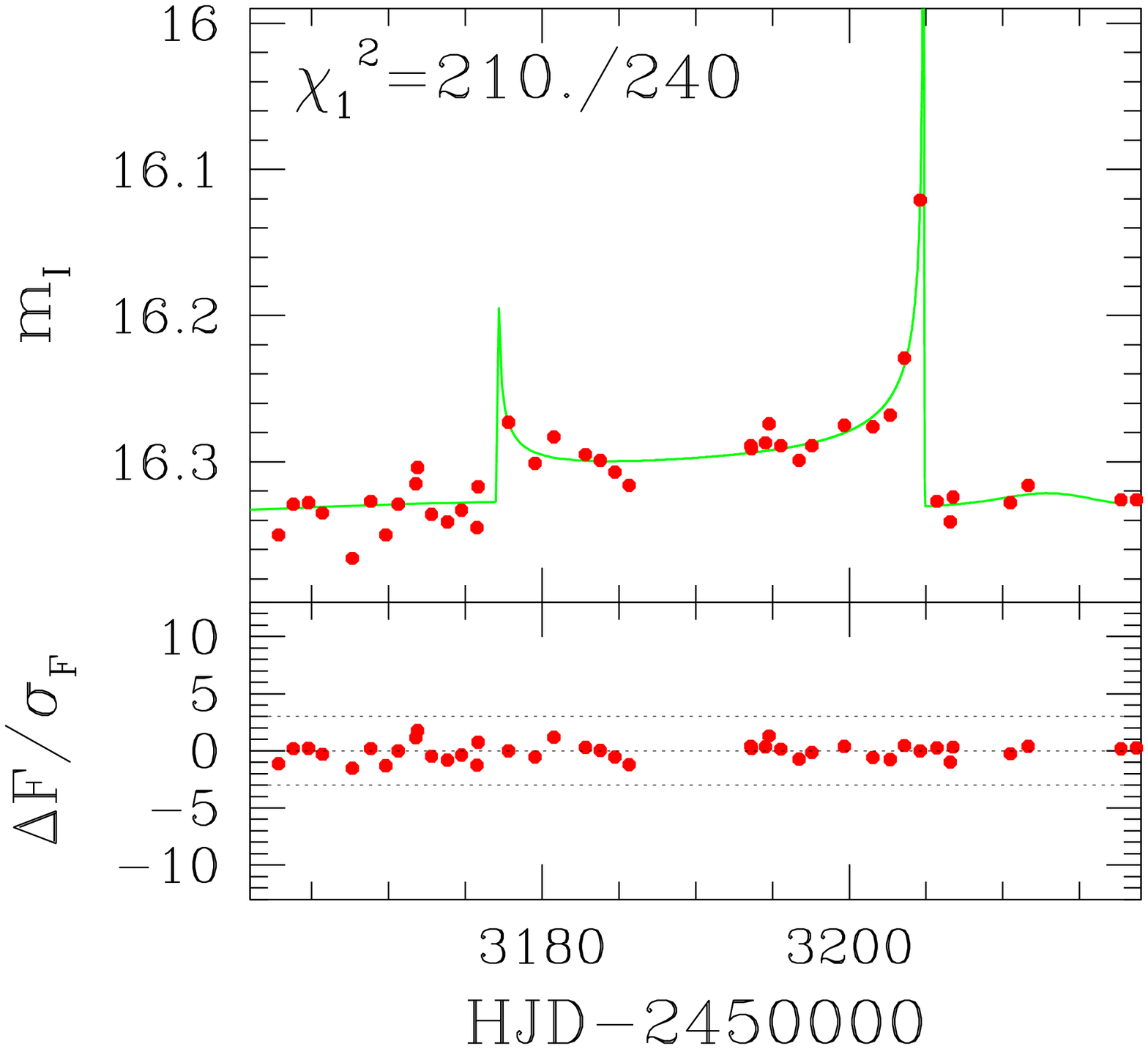}%

}

\noindent\parbox{12.75cm}{
\leftline {\bf OGLE 2004-BLG-451} 

 \includegraphics[height=63mm,width=62mm]{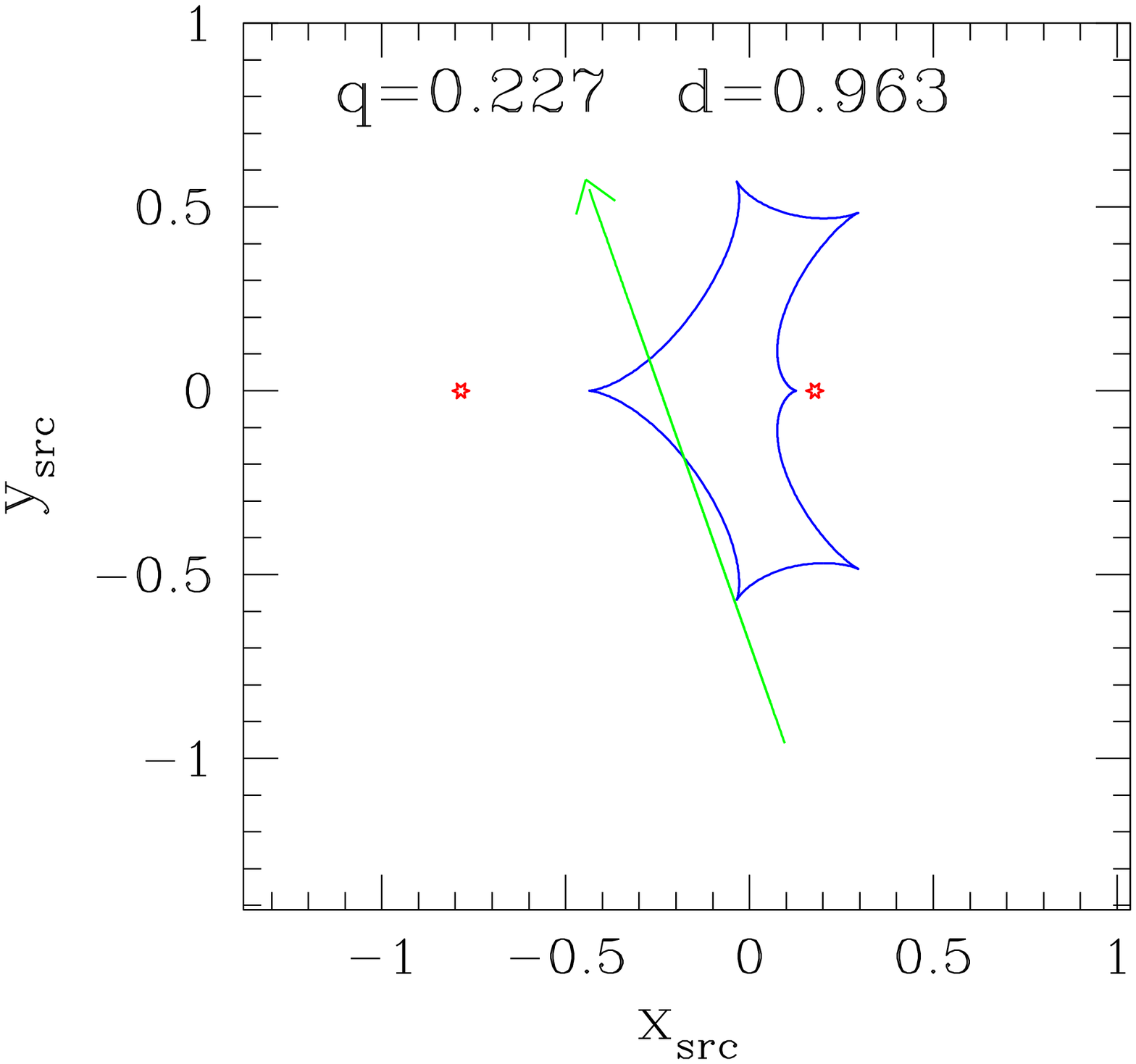}%
 \includegraphics[height=63mm,width=62mm]{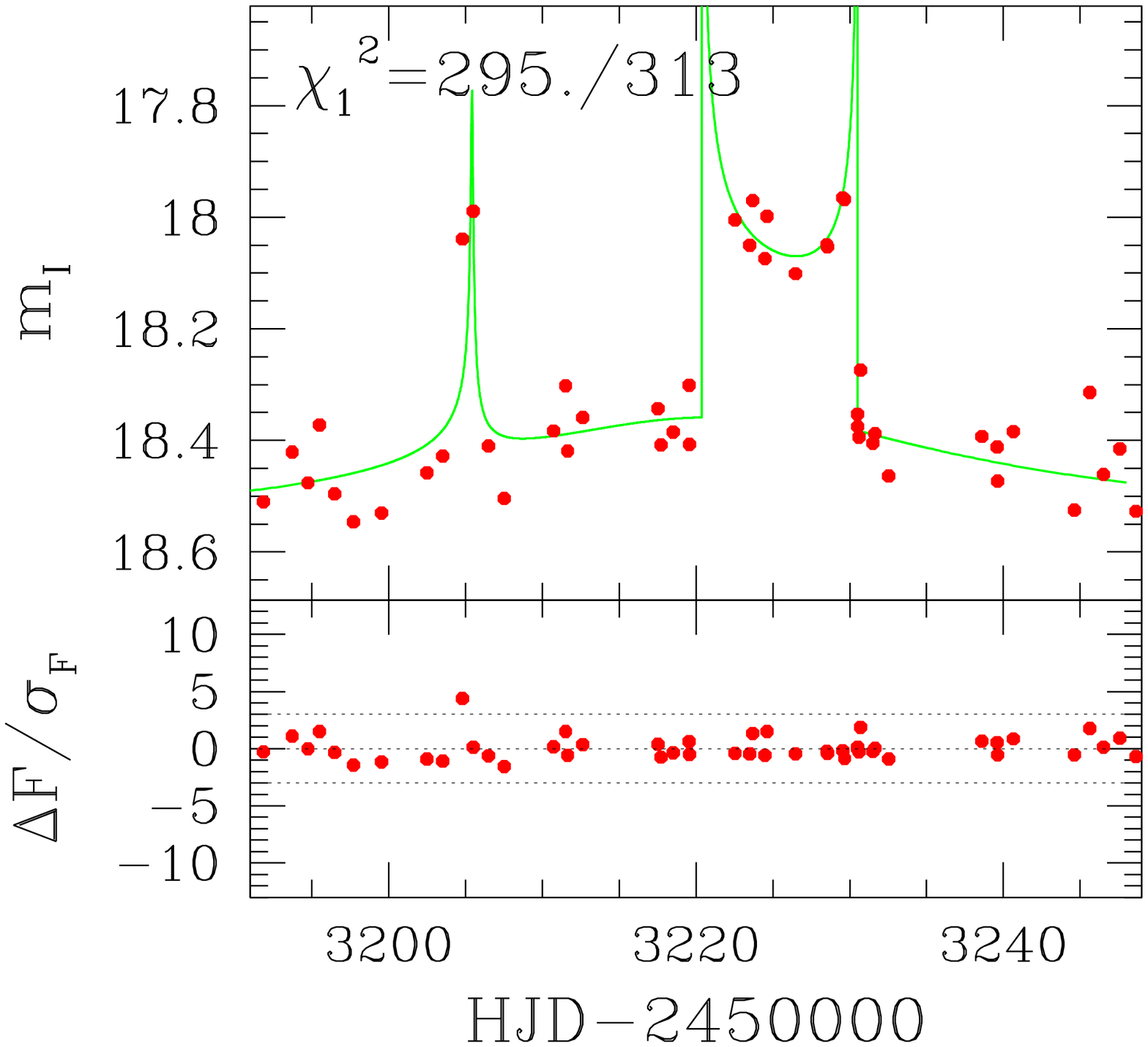}%

}

\noindent\parbox{12.75cm}{
\leftline {\bf OGLE 2004-BLG-460: I} 

 \includegraphics[height=63mm,width=62mm]{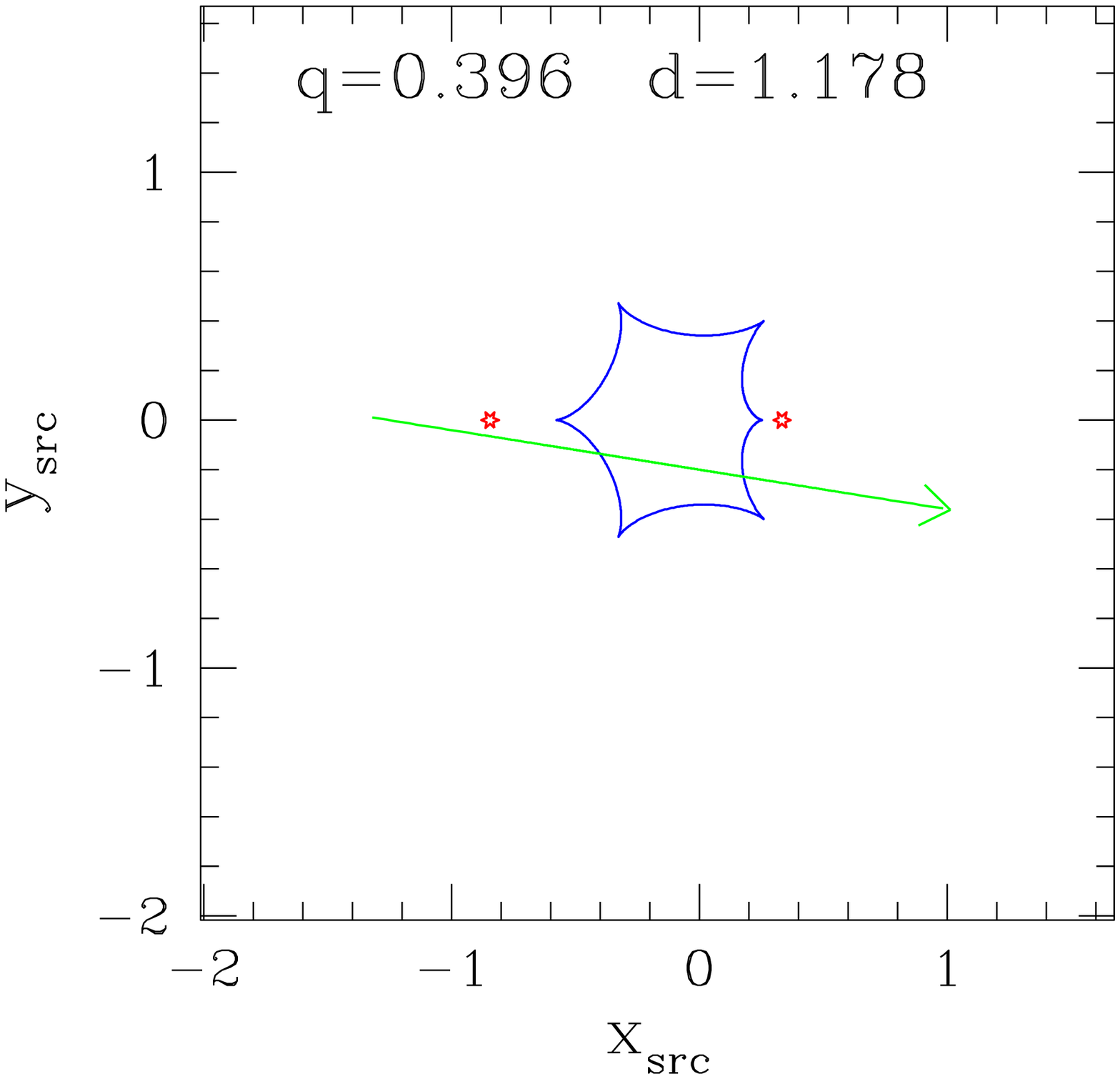}%
 \includegraphics[height=63mm,width=62mm]{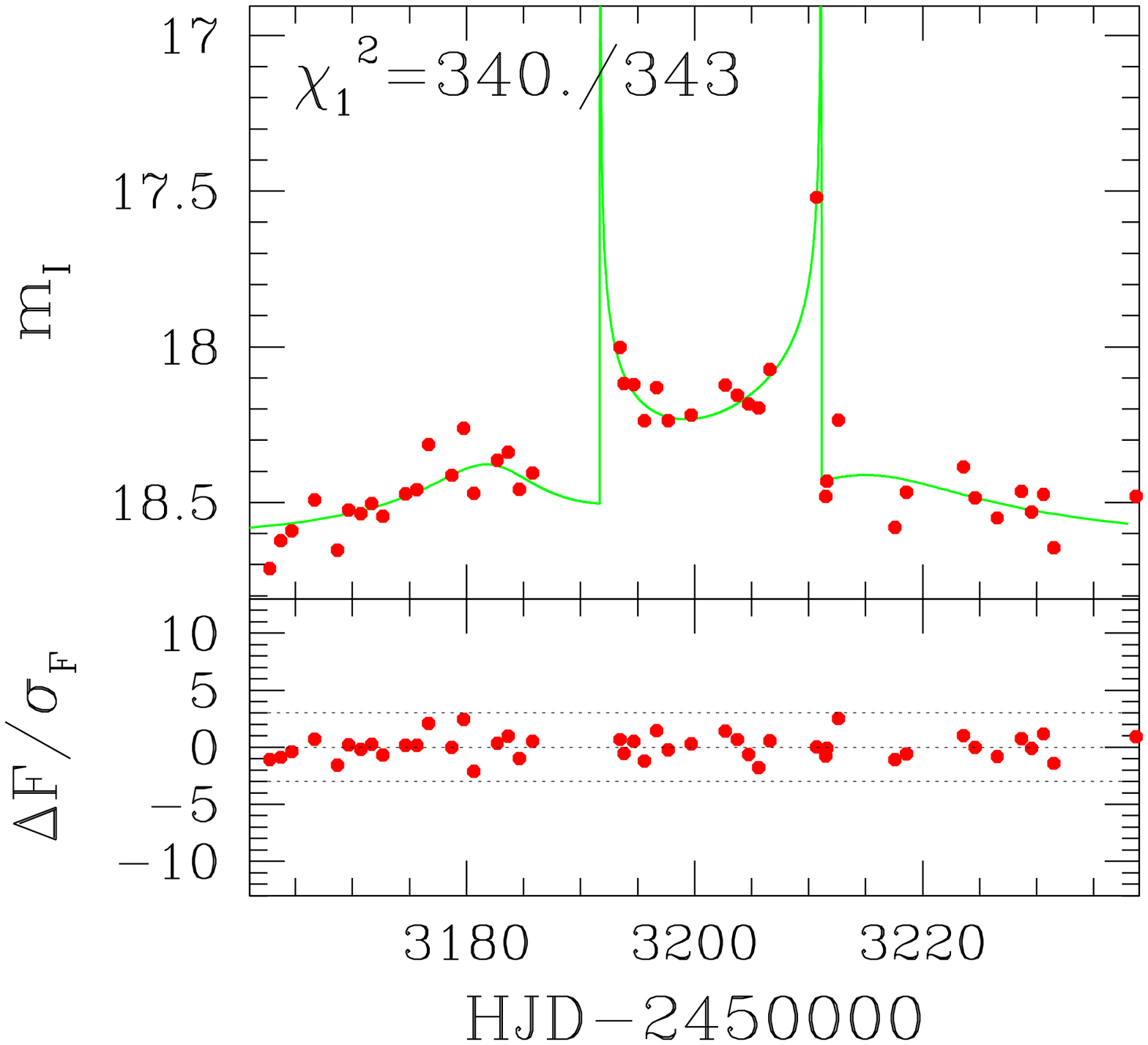}%

}

\noindent\parbox{12.75cm}{
\leftline {\bf OGLE 2004-BLG-460: II} 

 \includegraphics[height=63mm,width=62mm]{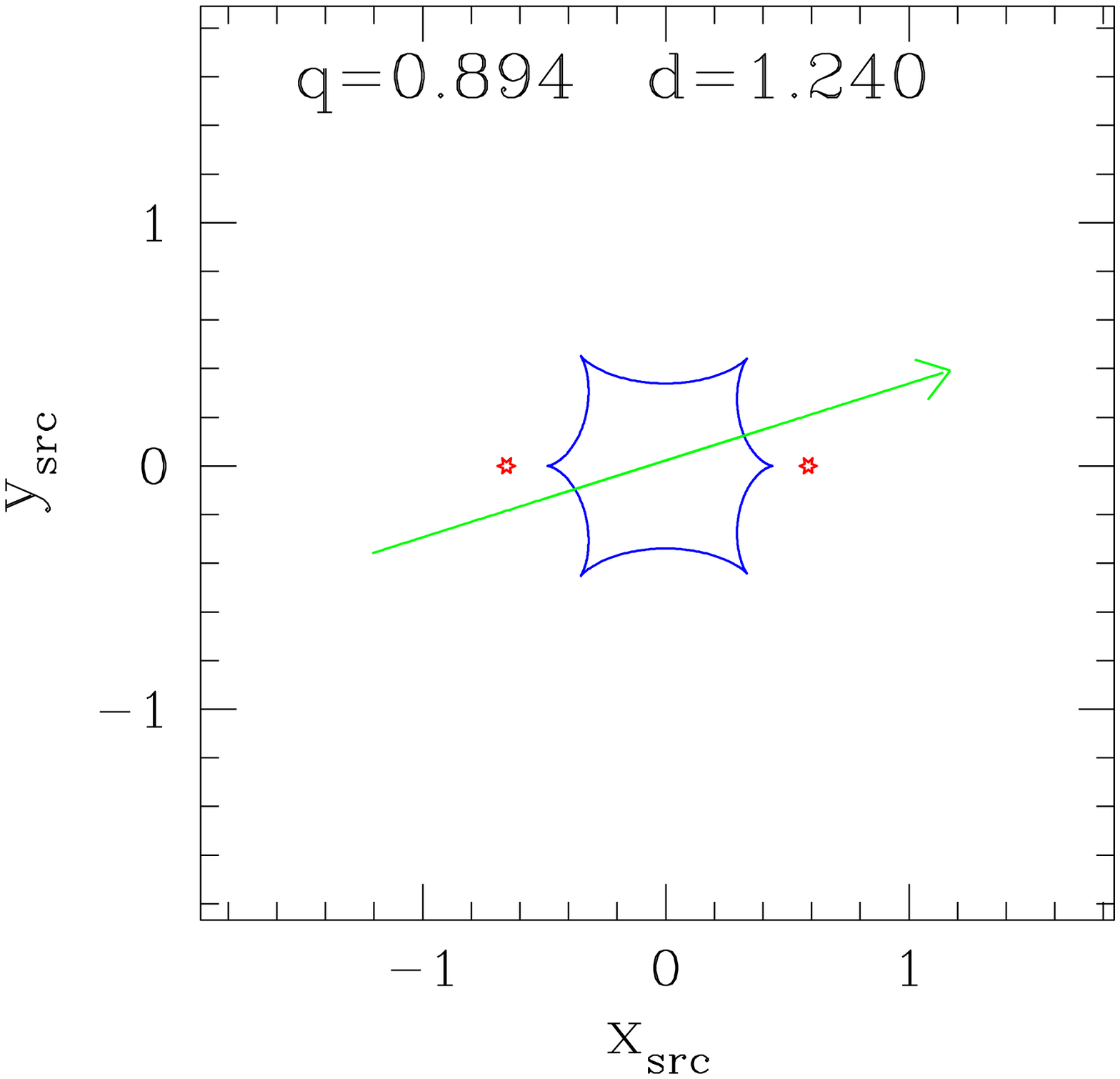}%
 \includegraphics[height=63mm,width=62mm]{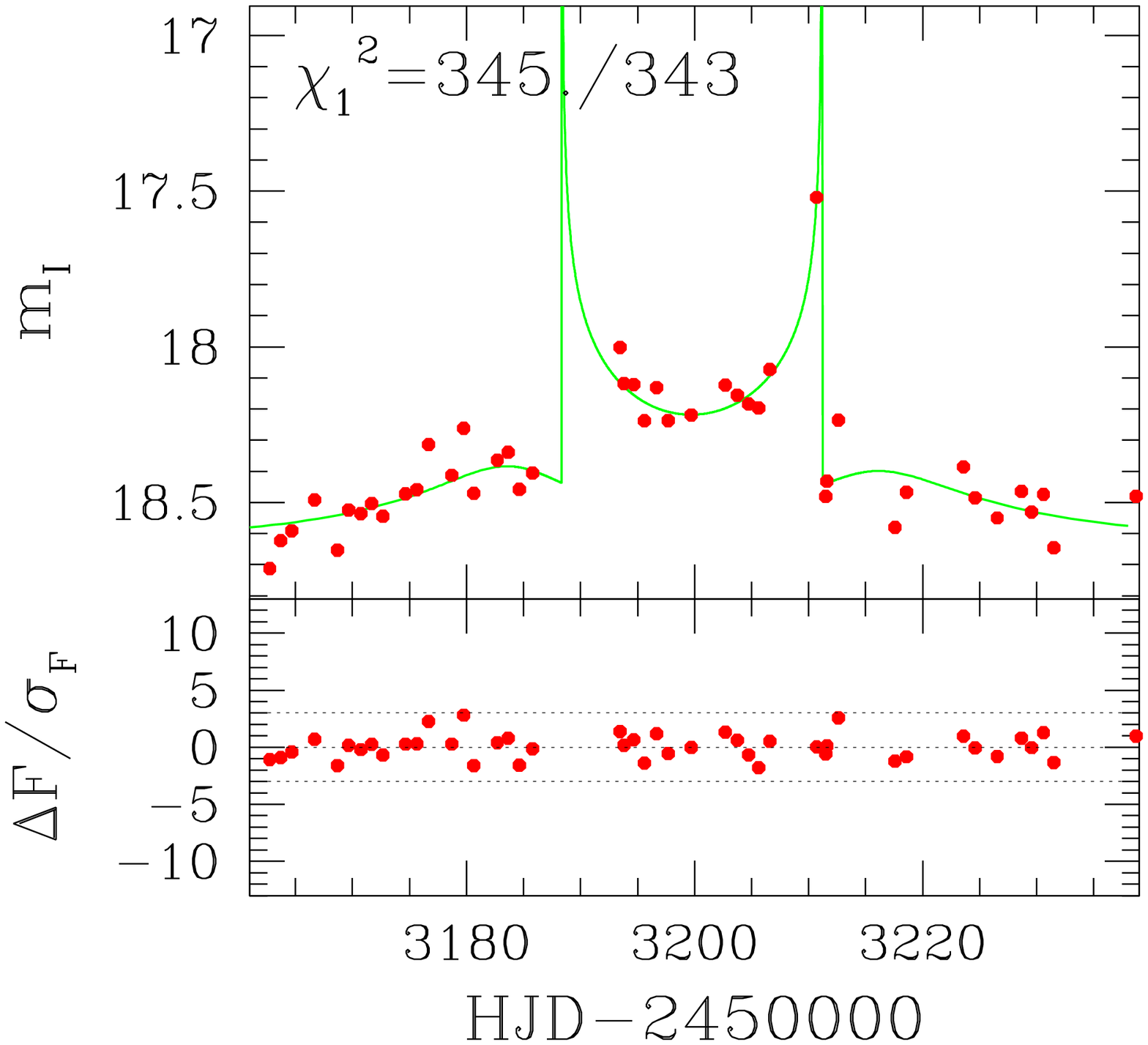}%

}

\noindent\parbox{12.75cm}{
\leftline {\bf OGLE 2004-BLG-480: I} 

 \includegraphics[height=63mm,width=62mm]{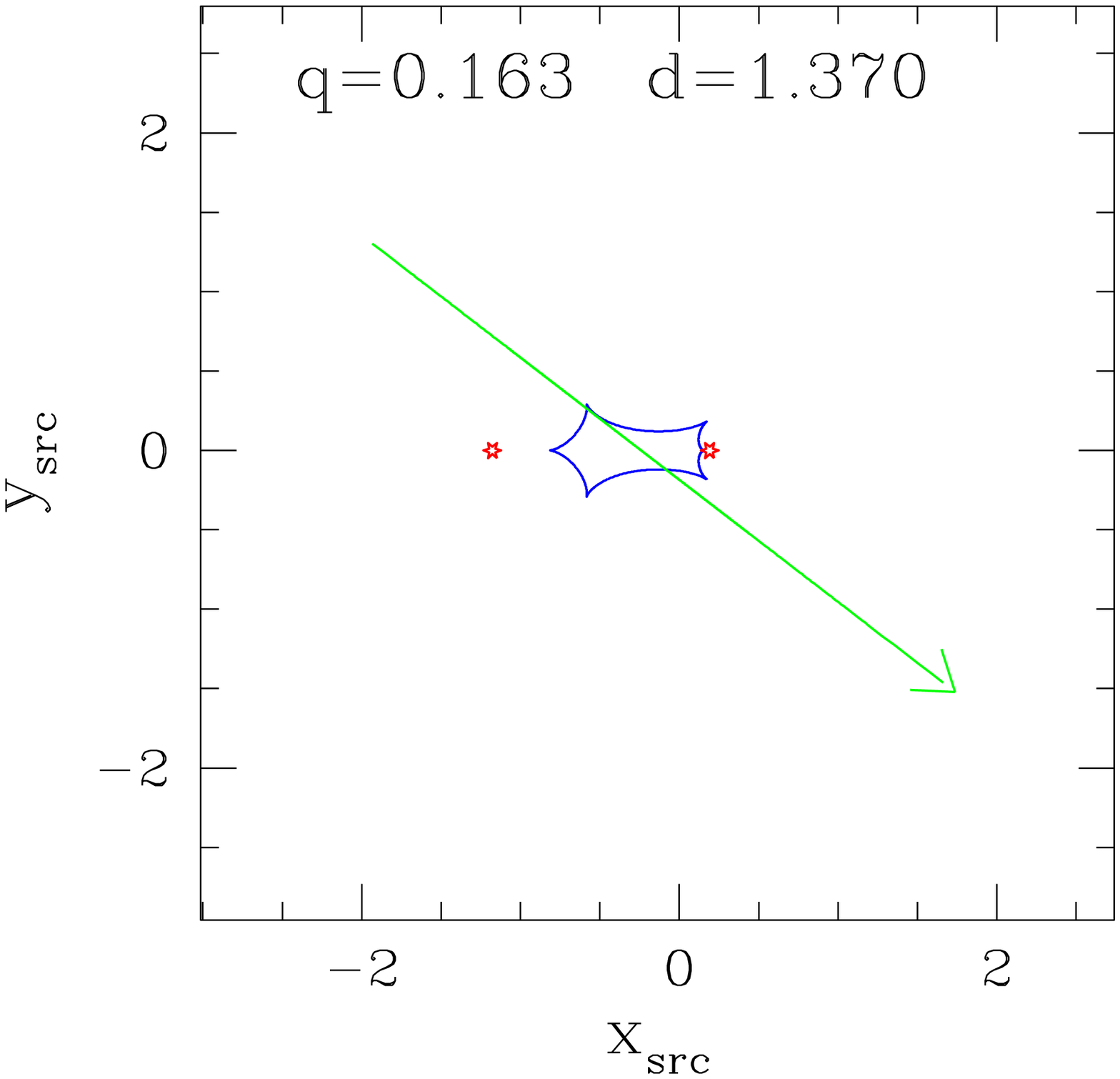}%
 \includegraphics[height=63mm,width=62mm]{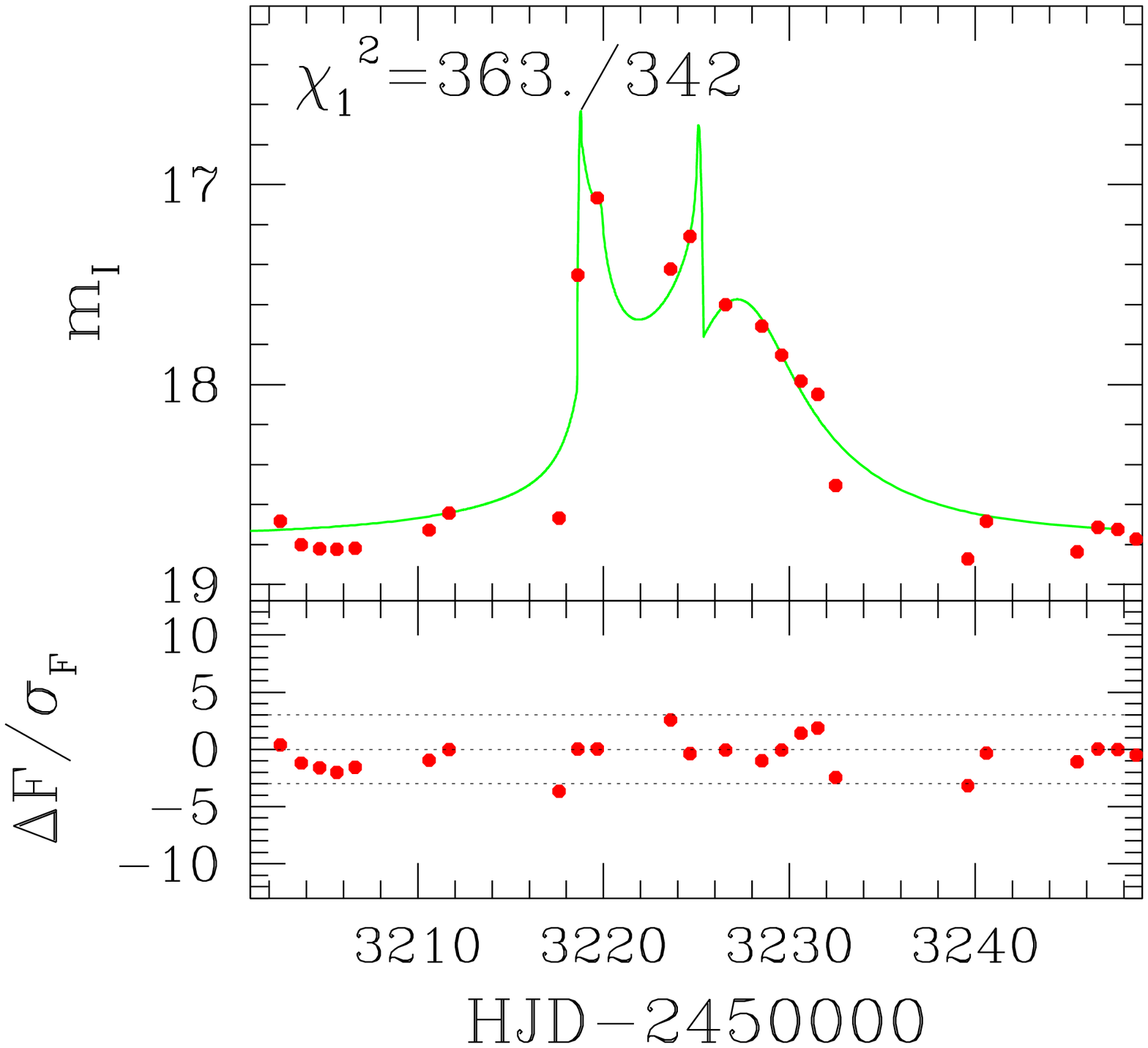}%

}

\noindent\parbox{12.75cm}{
\leftline {\bf OGLE 2004-BLG-480: II} 

 \includegraphics[height=63mm,width=62mm]{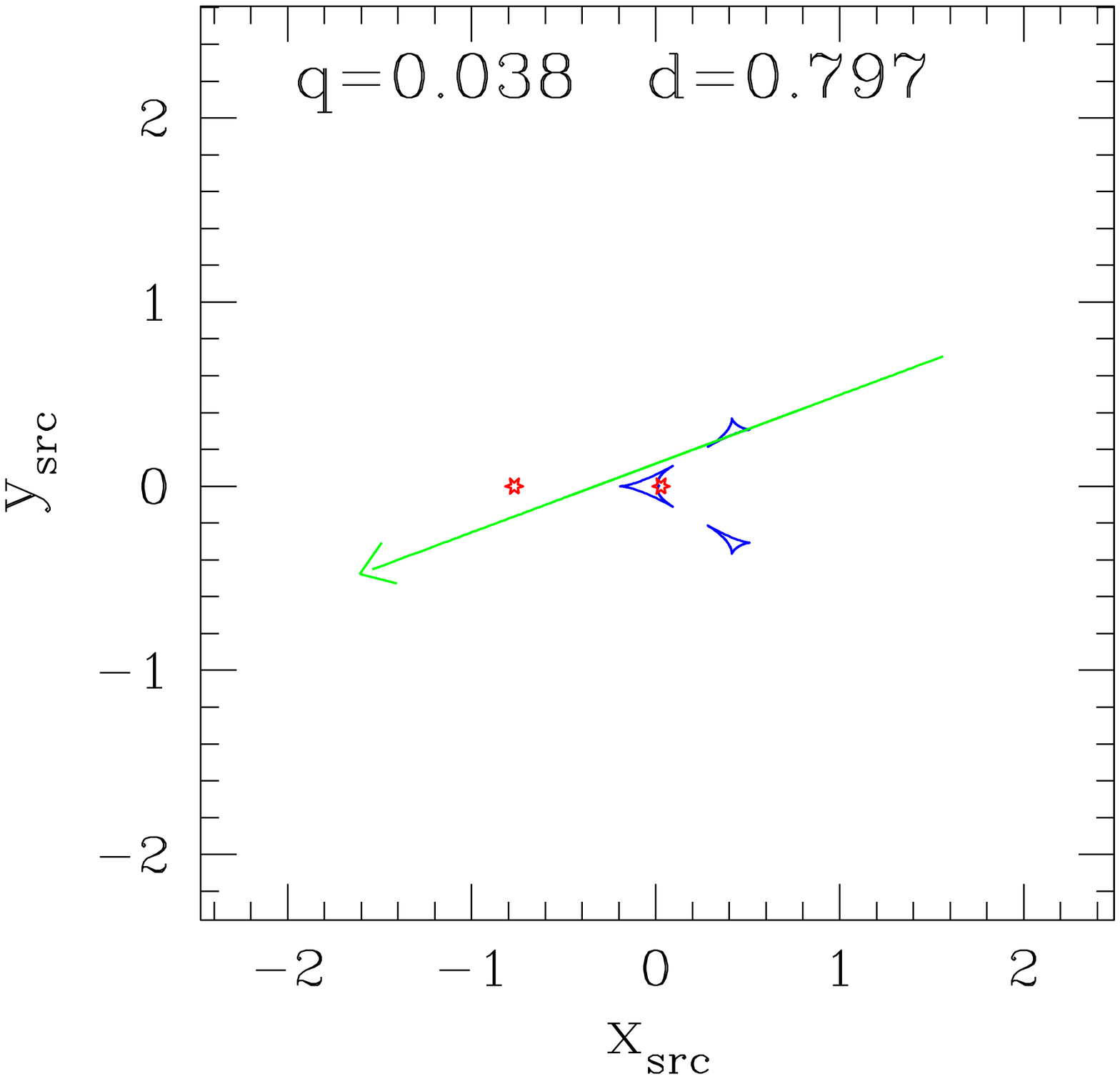}%
 \includegraphics[height=63mm,width=62mm]{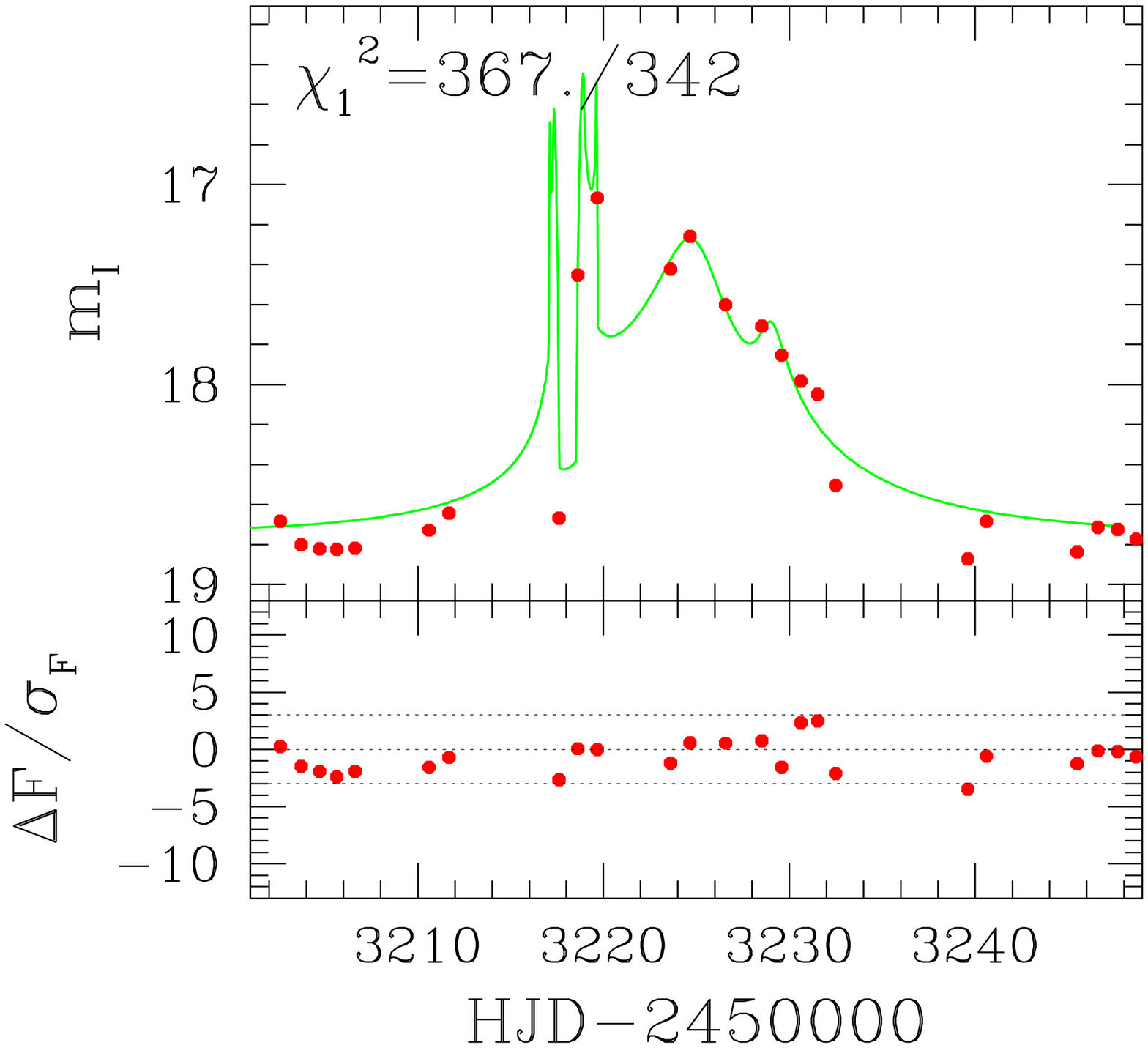}%
}

\noindent\parbox{12.75cm}{
\leftline {\bf OGLE 2004-BLG-490: I} 

 \includegraphics[height=63mm,width=62mm]{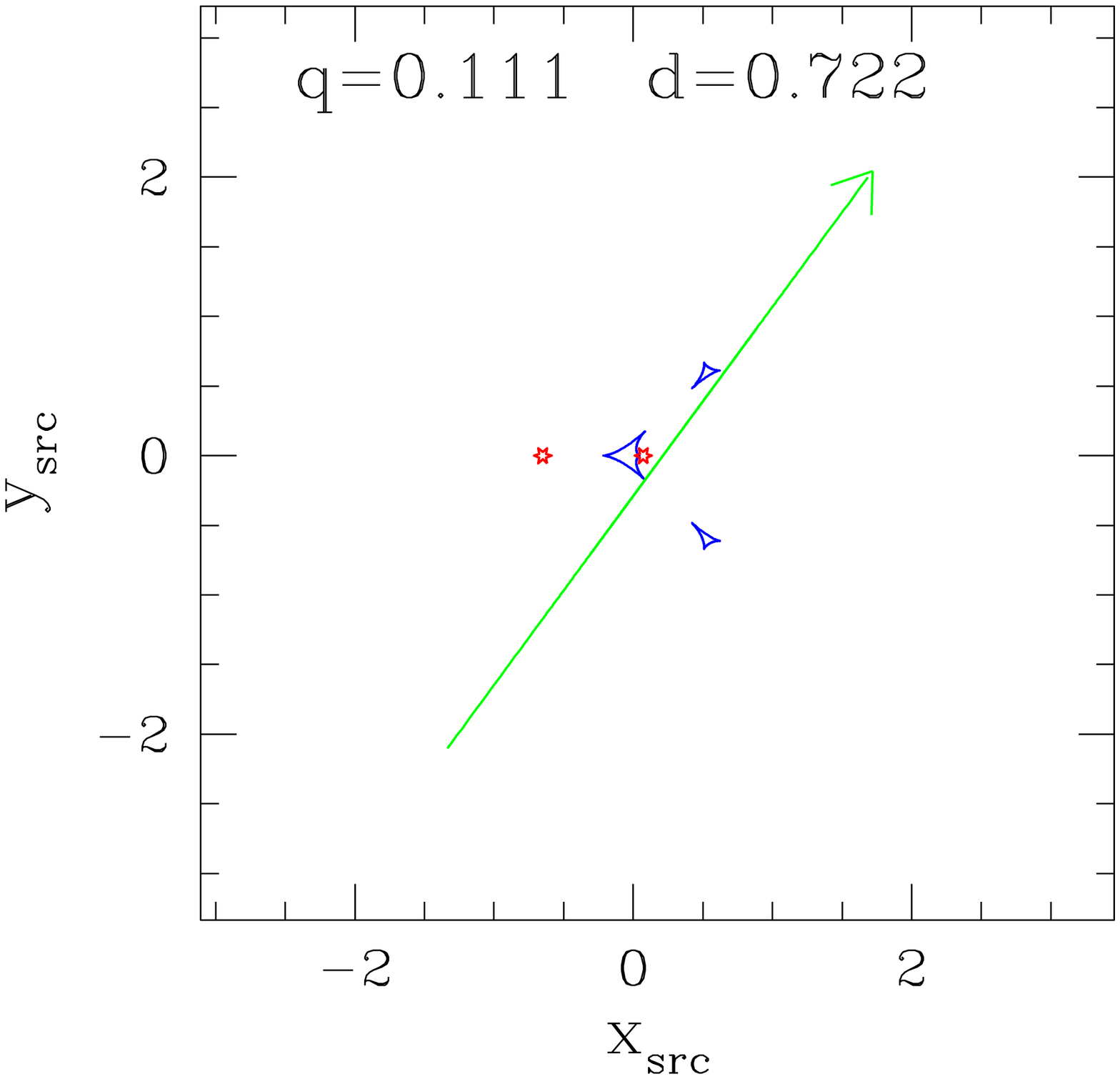}%
 \includegraphics[height=63mm,width=62mm]{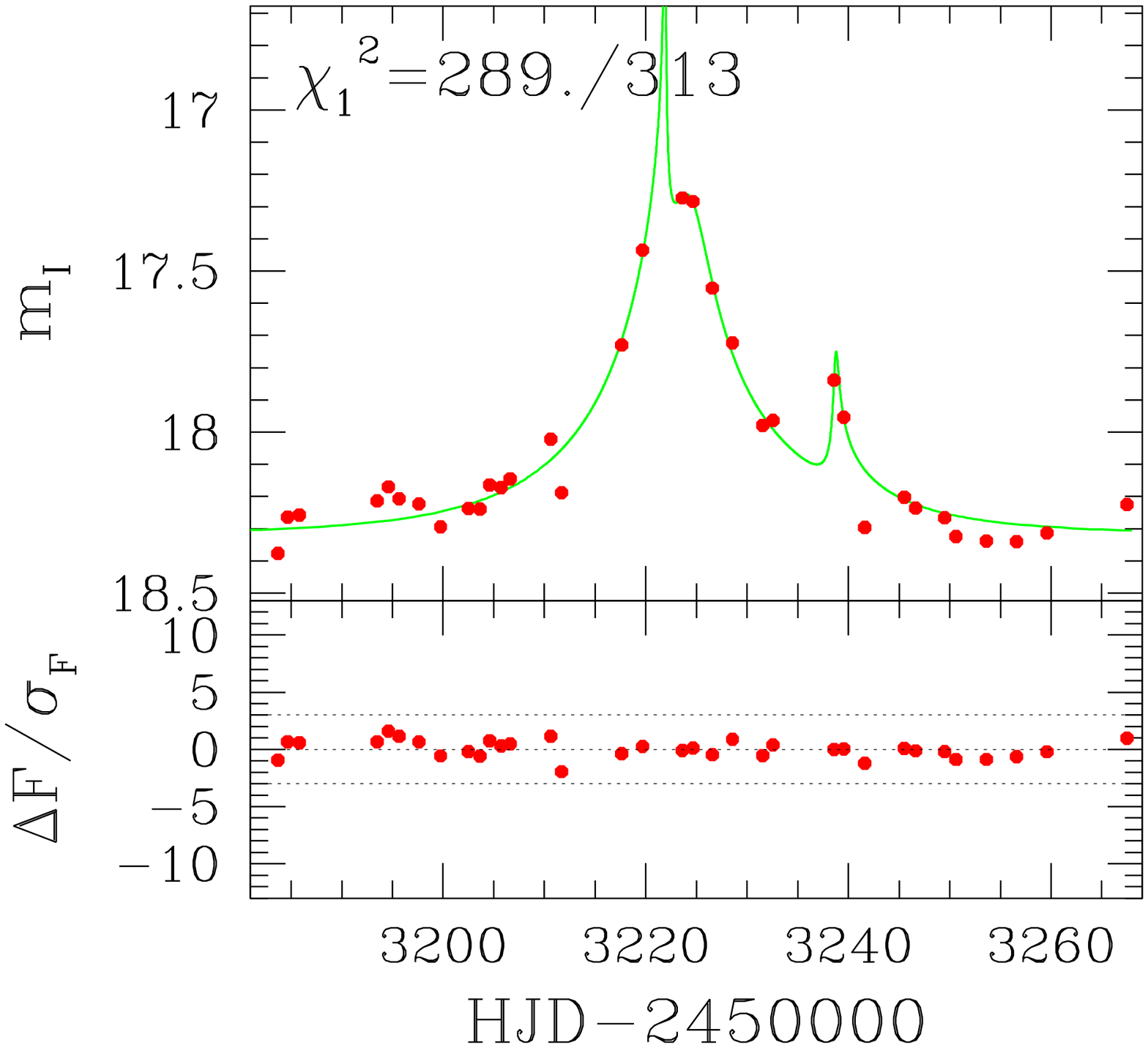}%

}

\noindent\parbox{12.75cm}{
\leftline {\bf OGLE 2004-BLG-490: II} 

 \includegraphics[height=63mm,width=62mm]{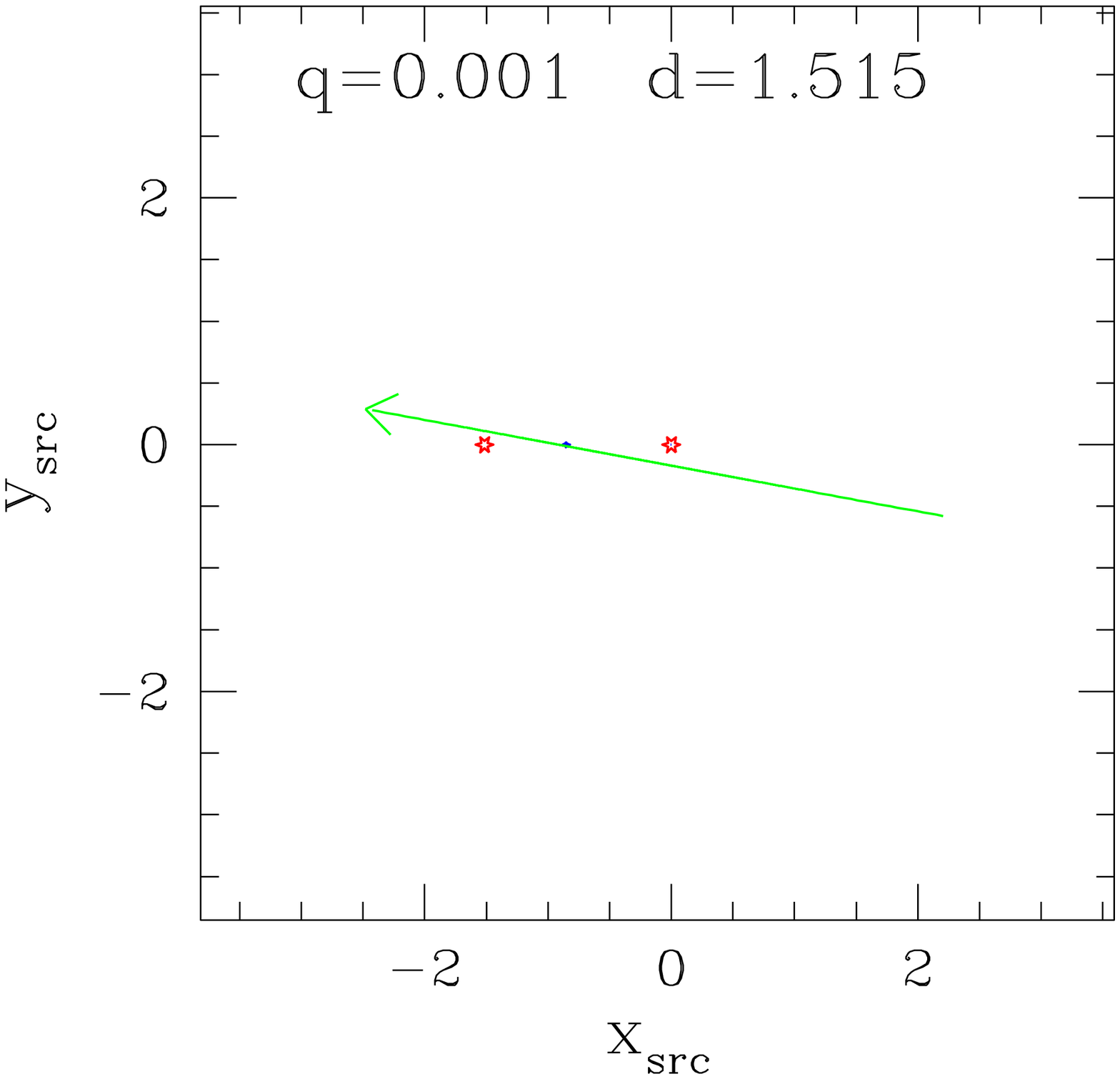}%
 \includegraphics[height=63mm,width=62mm]{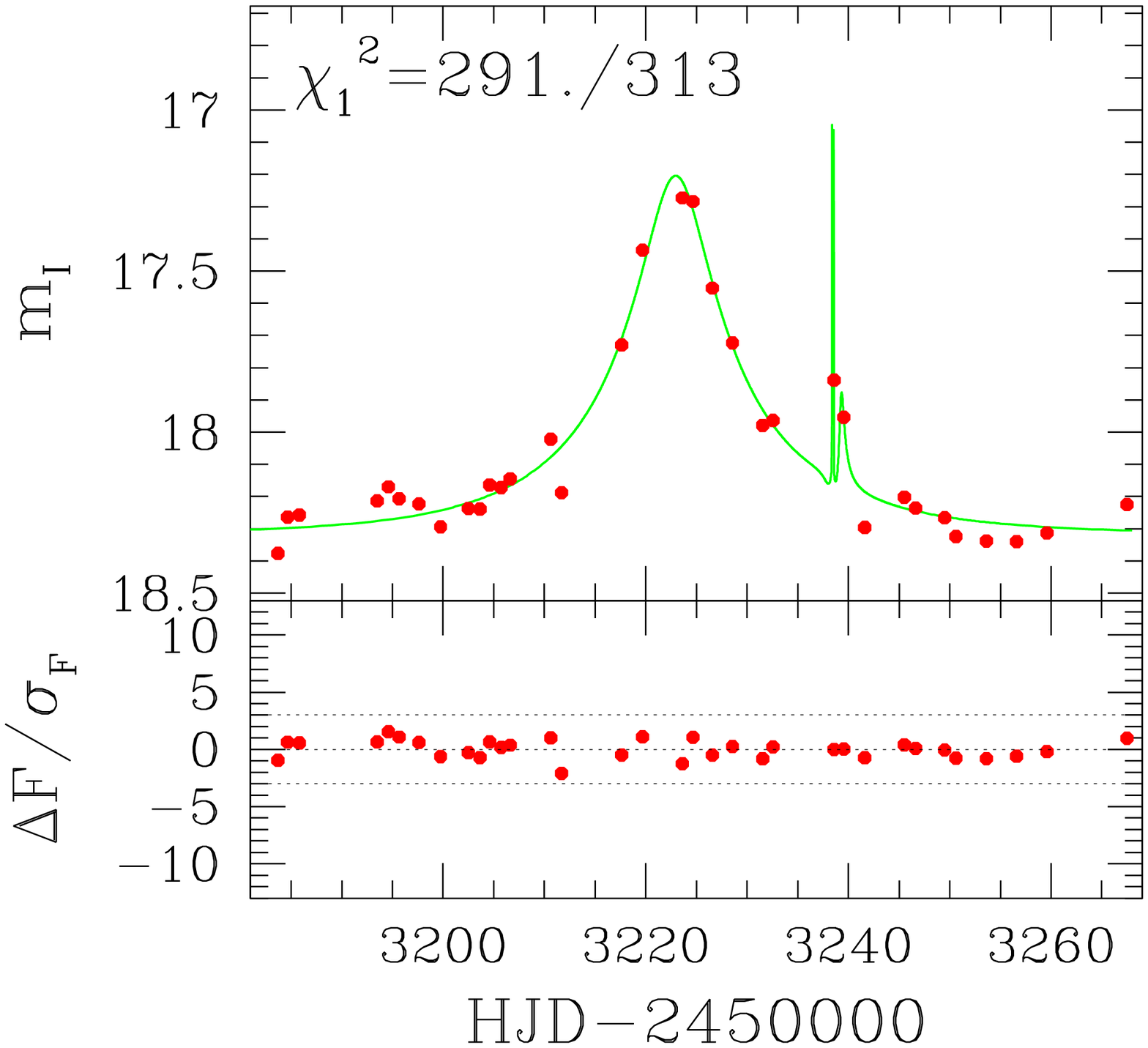}%

}

\noindent\parbox{12.75cm}{
\leftline {\bf OGLE 2004-BLG-559} 

 \includegraphics[height=63mm,width=62mm]{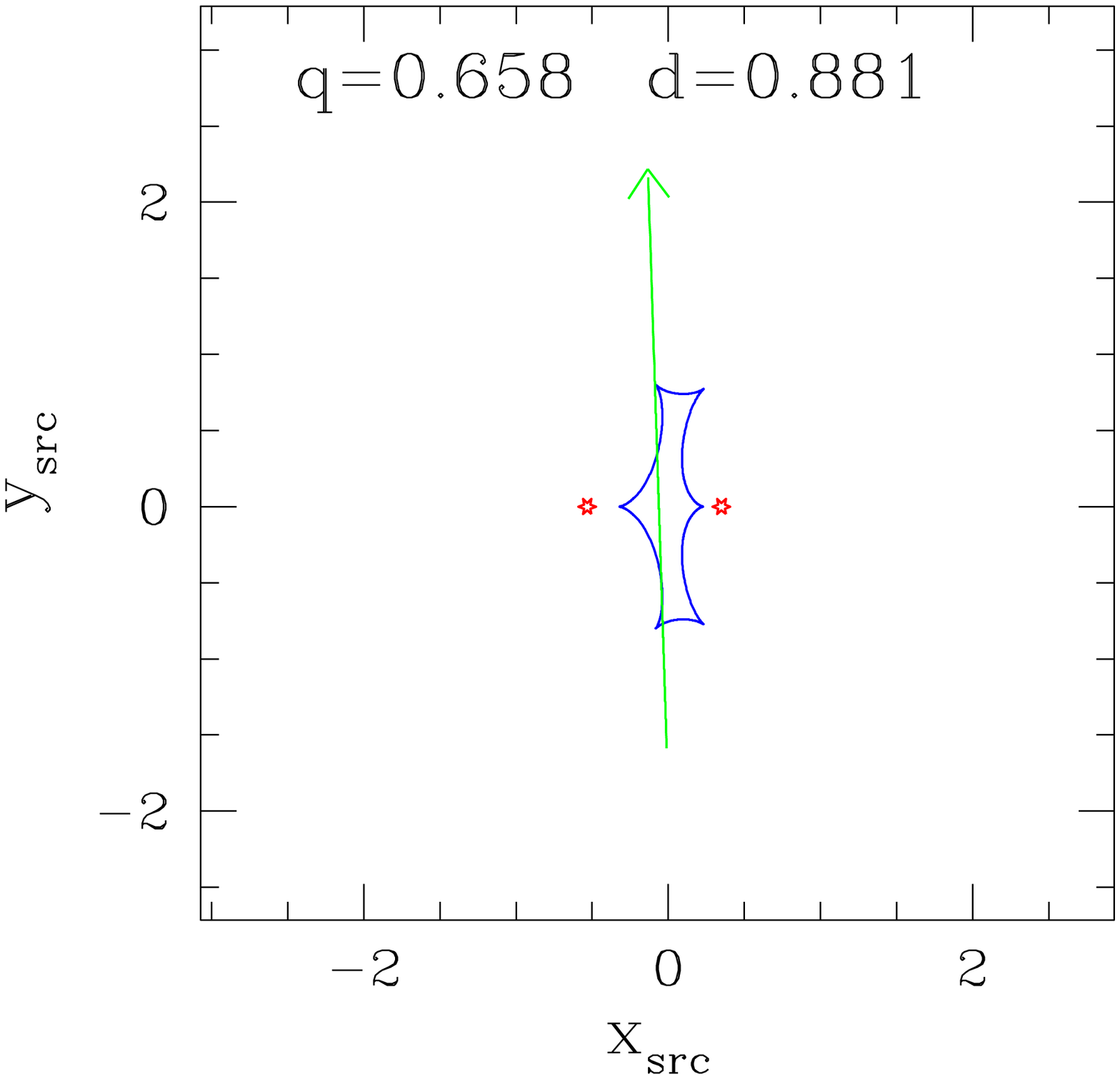}%
 \includegraphics[height=63mm,width=62mm]{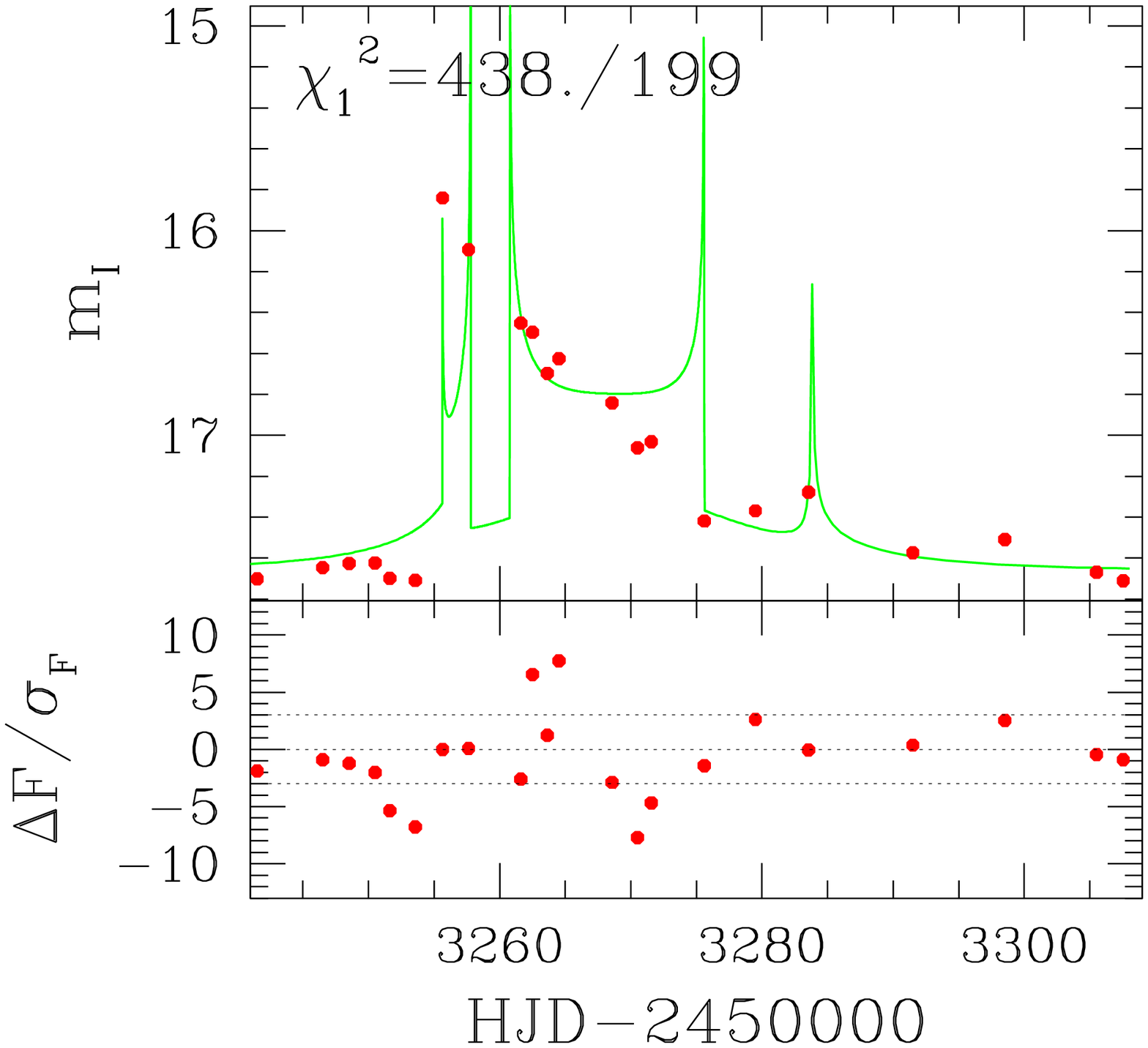}%

}

\noindent\parbox{12.75cm}{
\leftline {\bf OGLE 2004-BLG-572} 

 \includegraphics[height=63mm,width=62mm]{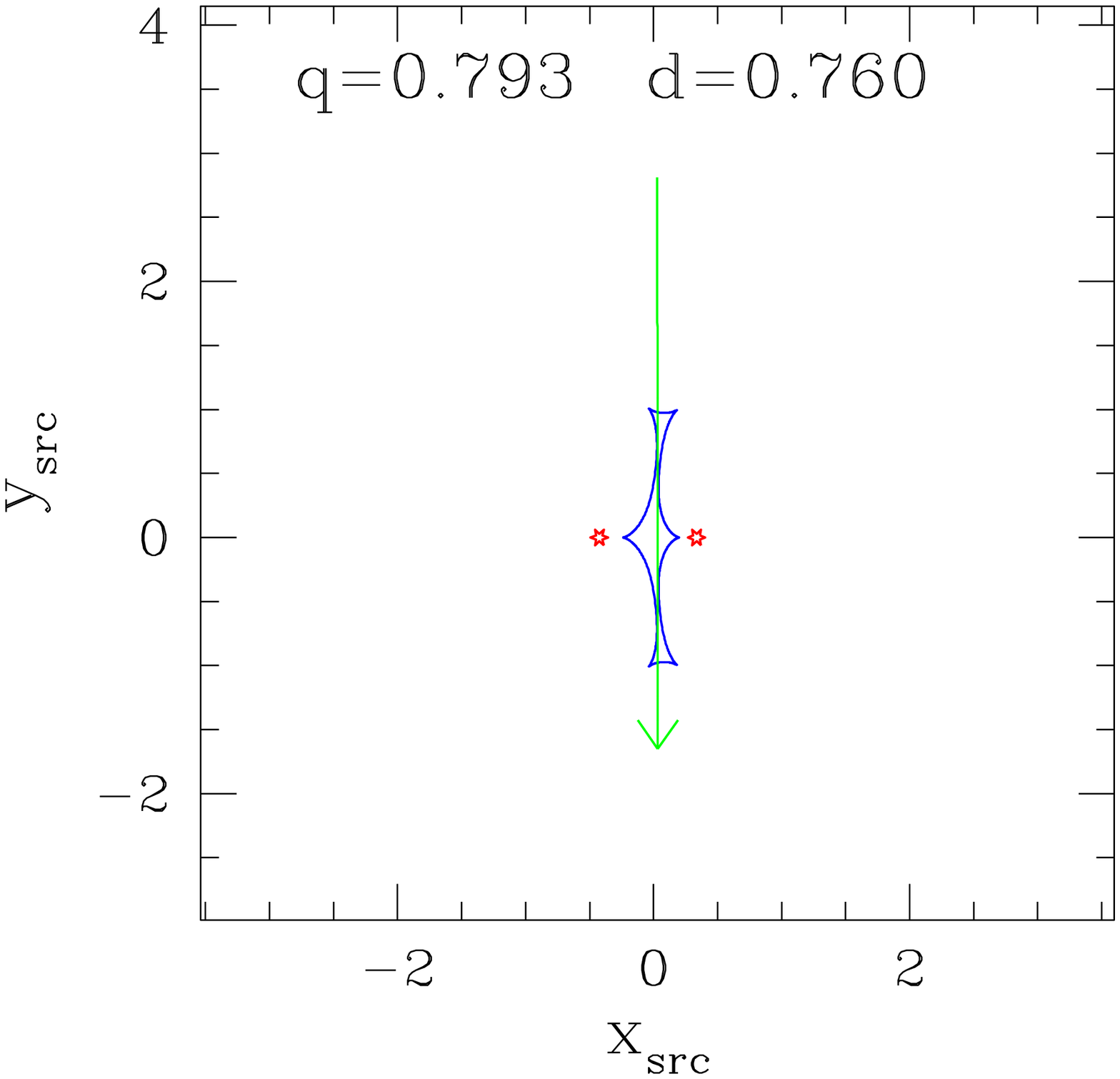}%
 \includegraphics[height=63mm,width=62mm]{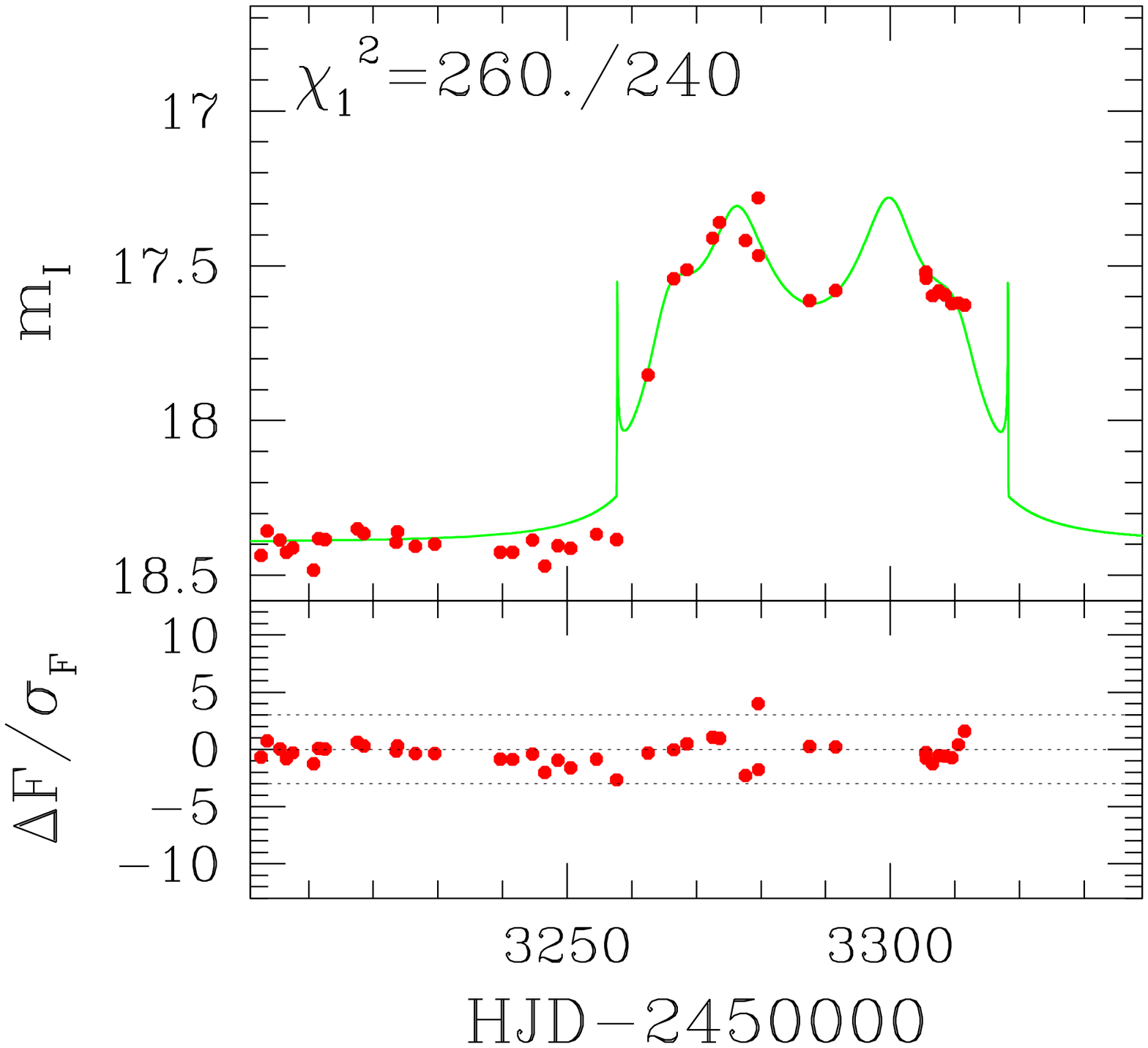}%

}

\noindent\parbox{12.75cm}{
\leftline {\bf OGLE 2004-BLG-605: I} 

 \includegraphics[height=63mm,width=62mm]{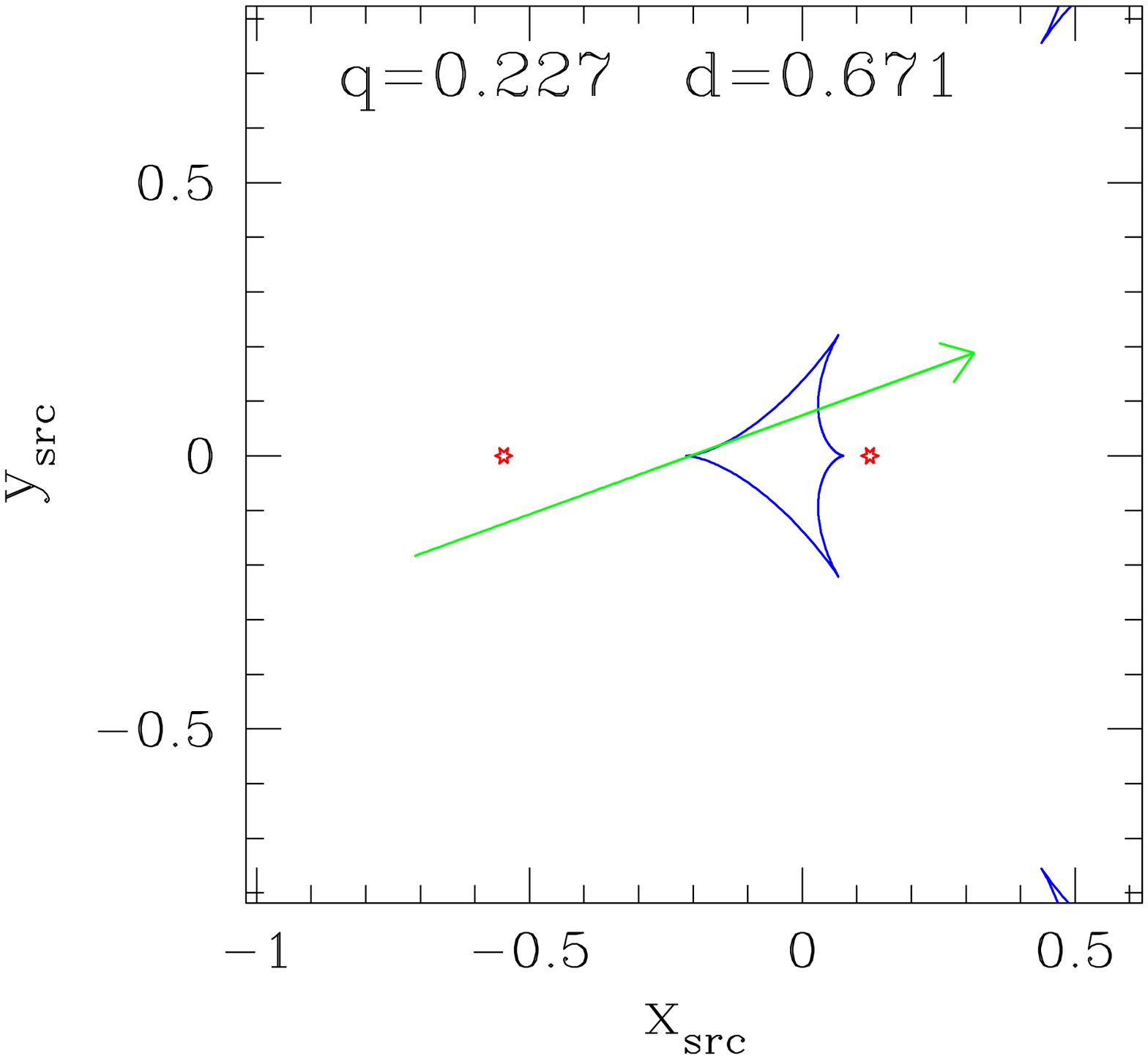}%
 \includegraphics[height=63mm,width=62mm]{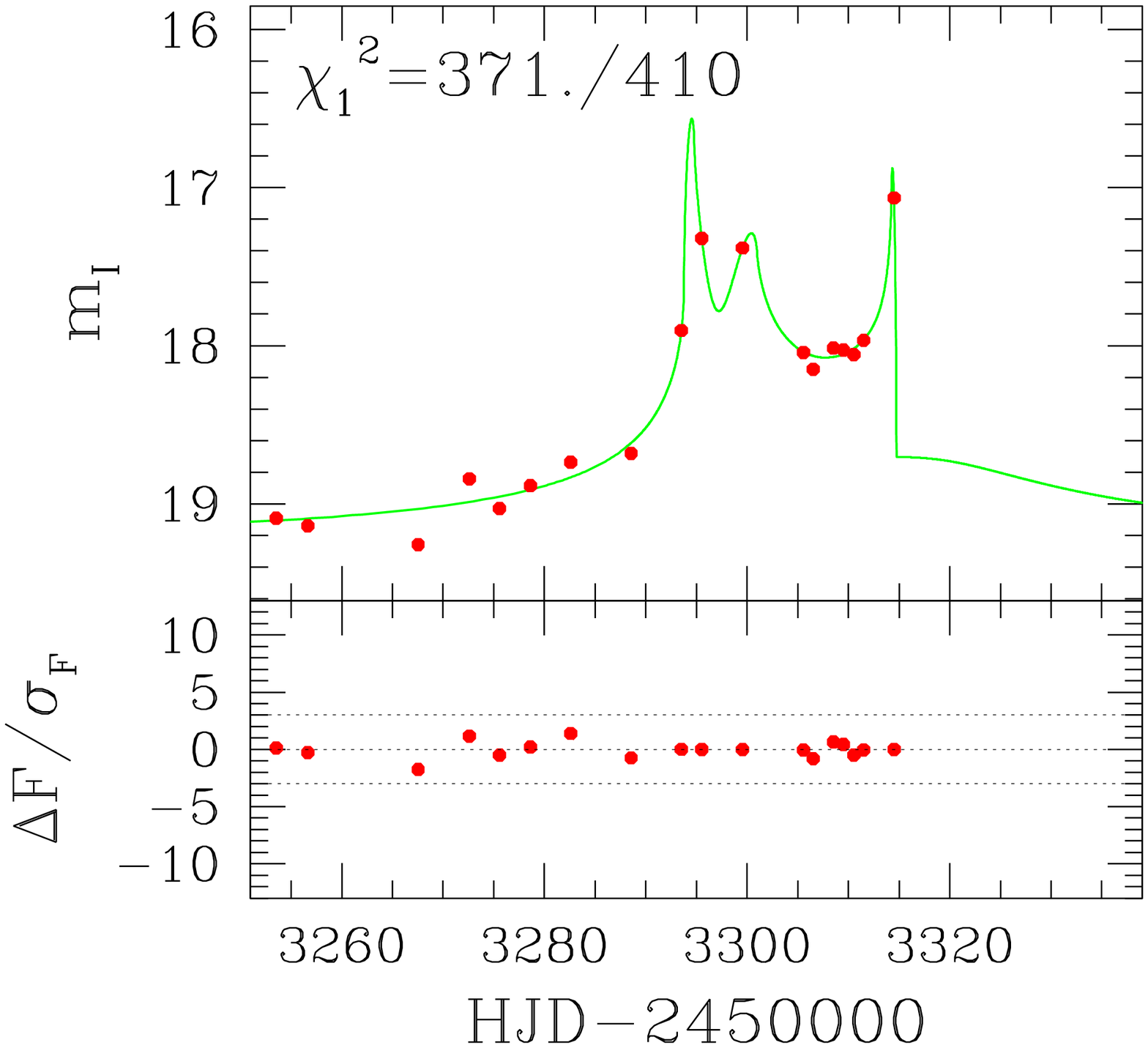}%

}

\noindent\parbox{12.75cm}{
\leftline {\bf OGLE 2004-BLG-605: II} 

 \includegraphics[height=63mm,width=62mm]{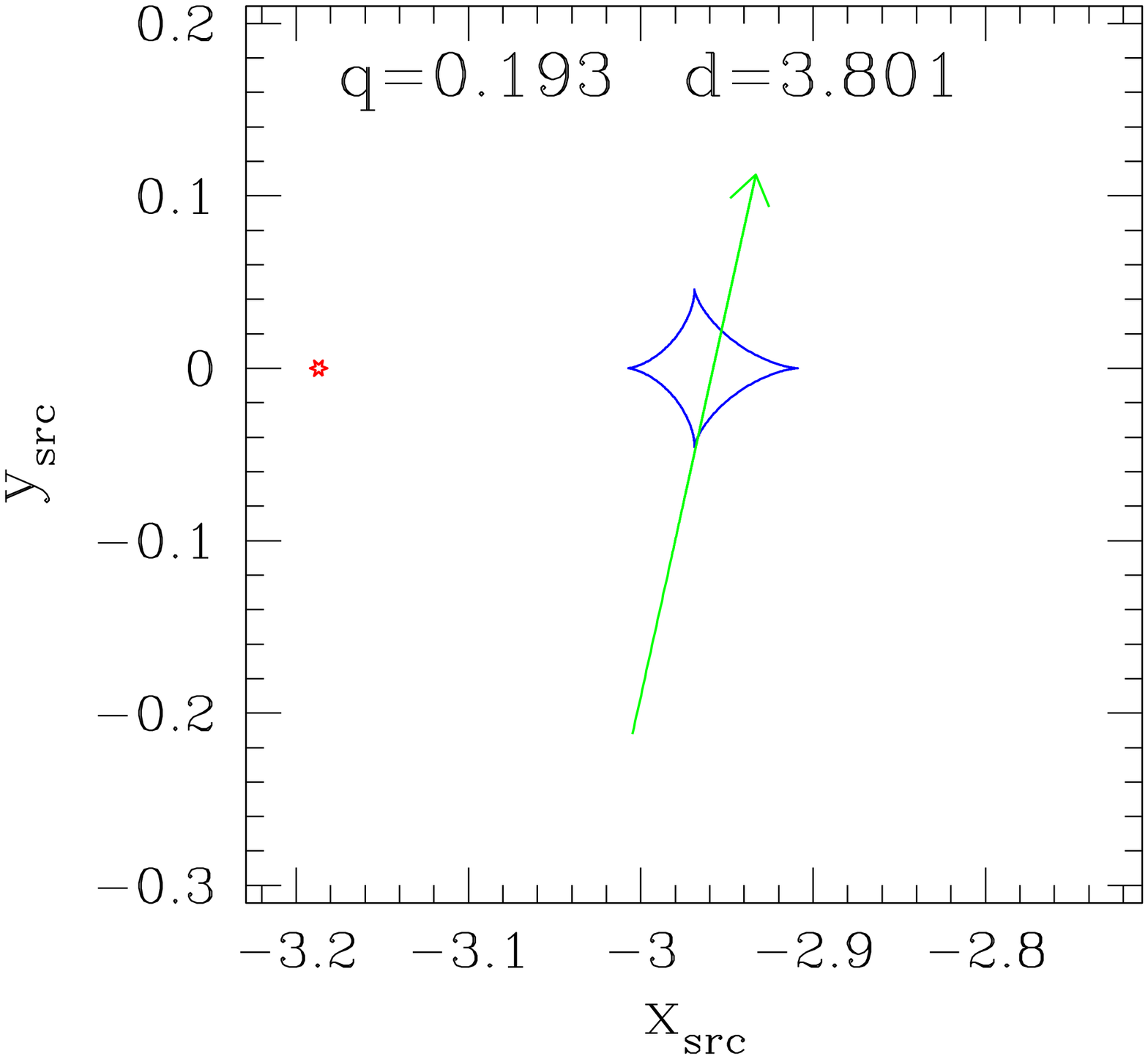}%
 \includegraphics[height=63mm,width=62mm]{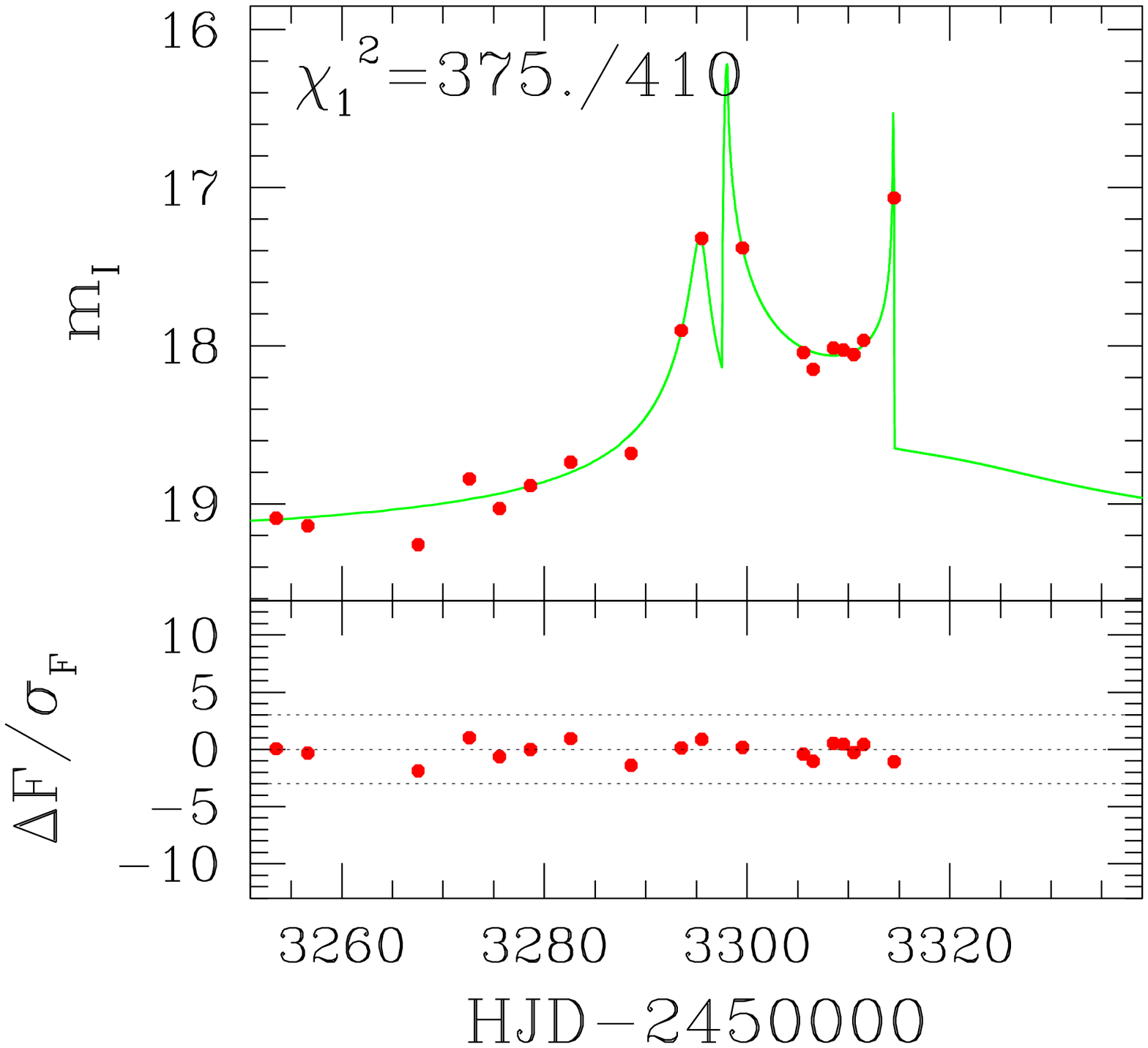}%

}
\vfill
\eject

\noindent 
{\bf Ambiguous Binary Lens/Double Source Events}

Below we show the binary lens (on the left) and double source (on the
right) models of the light curves for some of the considered events. 
The light curve in a double source model is a sum of the constant
blended flux plus the two single lens light curves for the source
components, each shown with dotted lines. We use fluxes for both kind of
models to ensure a better comparison. In majority of cases the binary
lens models give formally better fits as compared to the double source
models presented. On the other hand double source models, always
producing simpler light curves, look more natural in some cases.

\vskip 0.5cm
\noindent\parbox{12.75cm}{
\leftline {\bf OGLE 2004-BLG-226} 

\includegraphics[height=63mm,width=62mm]{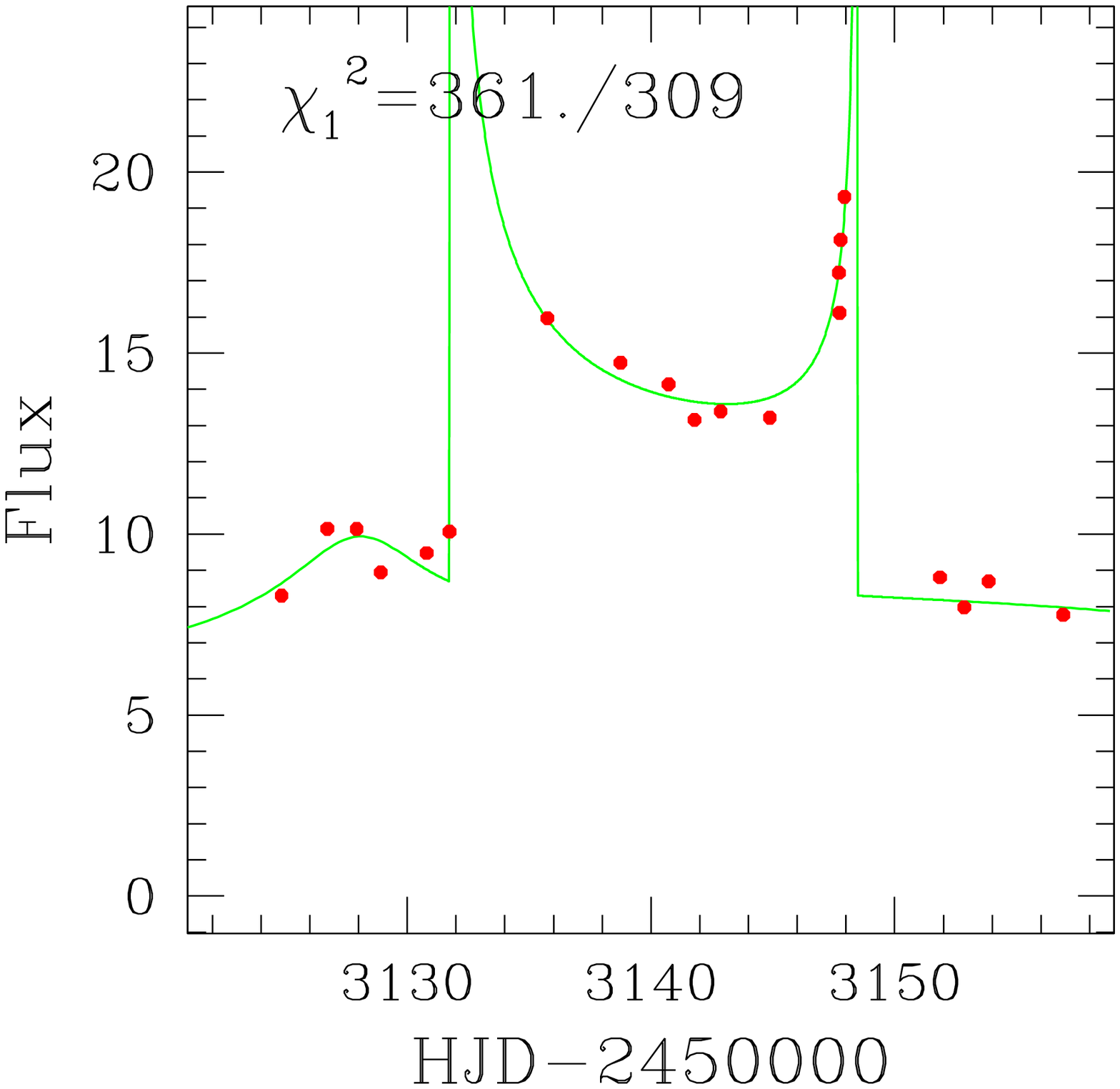}%
\includegraphics[height=63mm,width=62mm]{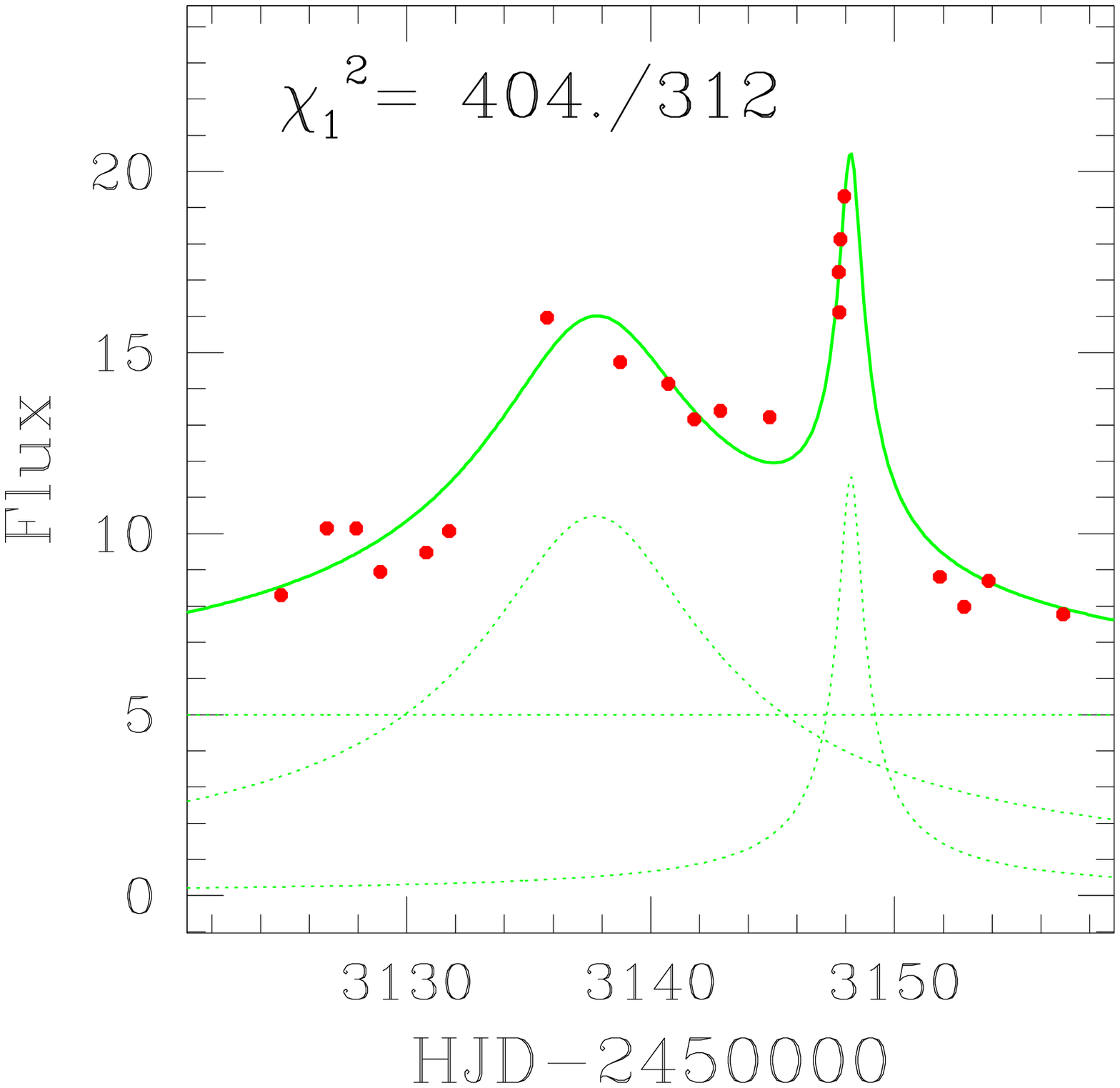}%

}

\noindent\parbox{12.75cm}{
\leftline {\bf OGLE 2004-BLG-280} 

 \includegraphics[height=63mm,width=62mm]{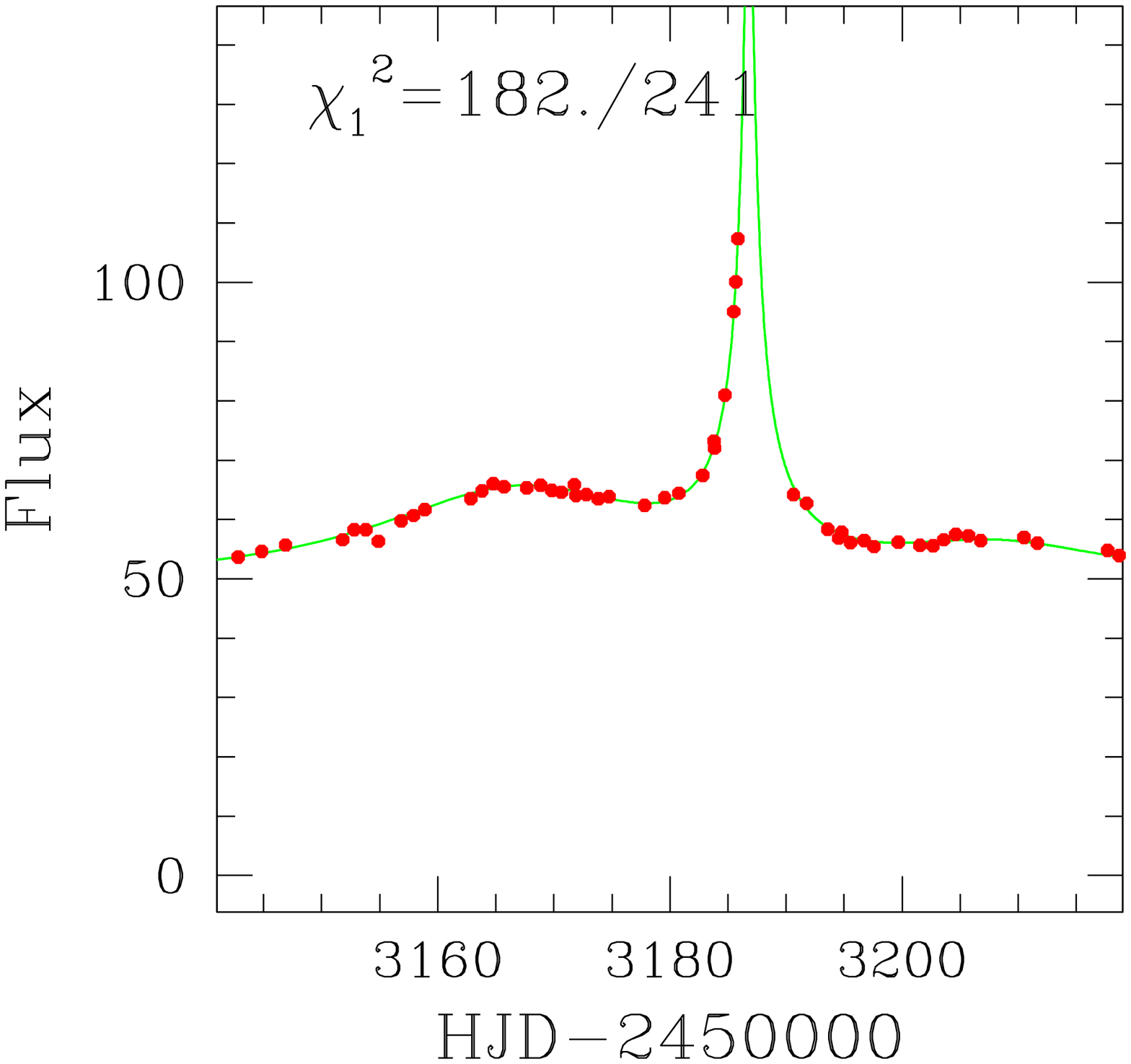}%
 \includegraphics[height=63mm,width=62mm]{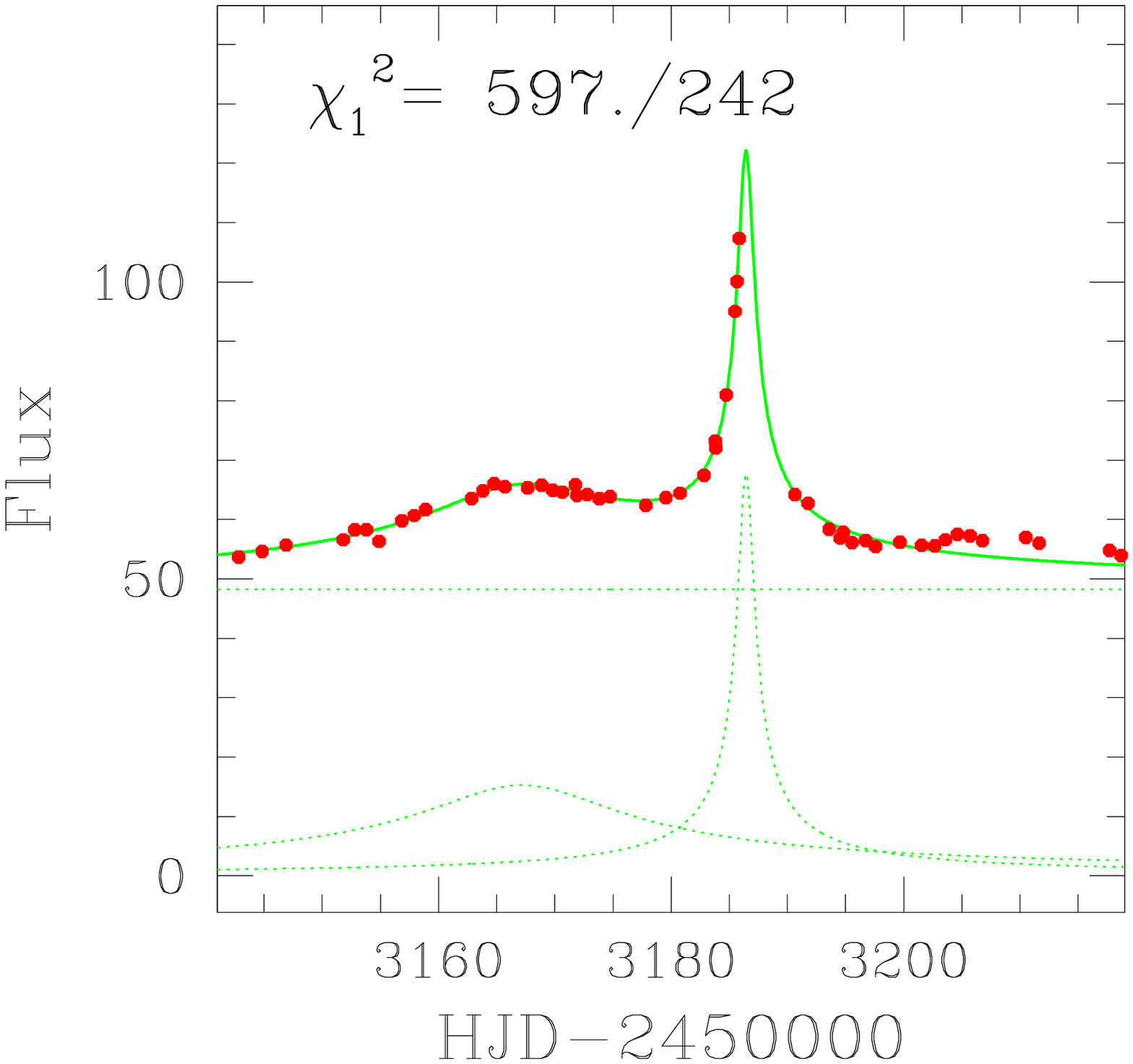}%

}

\noindent\parbox{12.75cm}{
\leftline {\bf OGLE 2004-BLG-347} 

 \includegraphics[height=63mm,width=62mm]{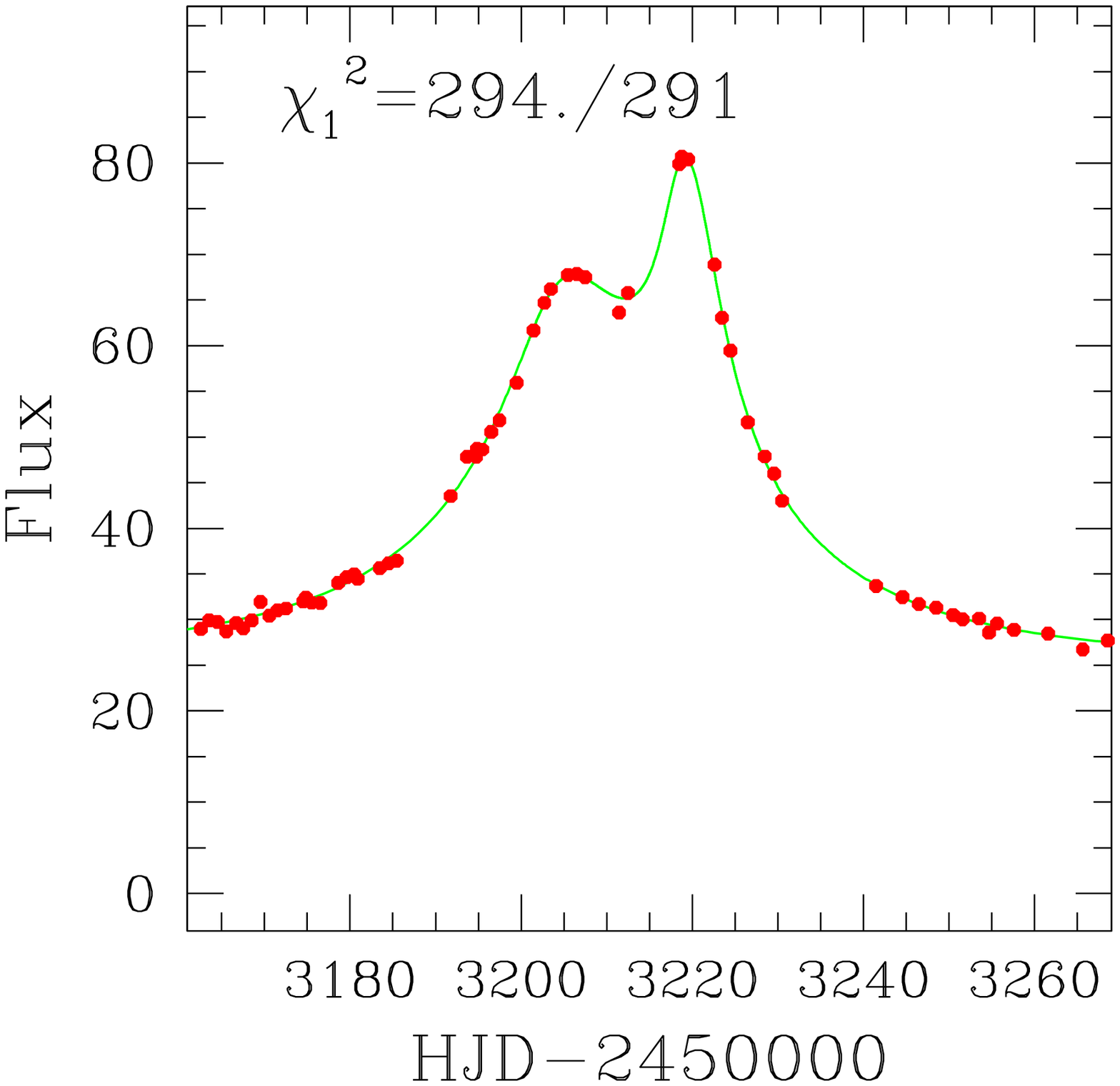}%
 \includegraphics[height=63mm,width=62mm]{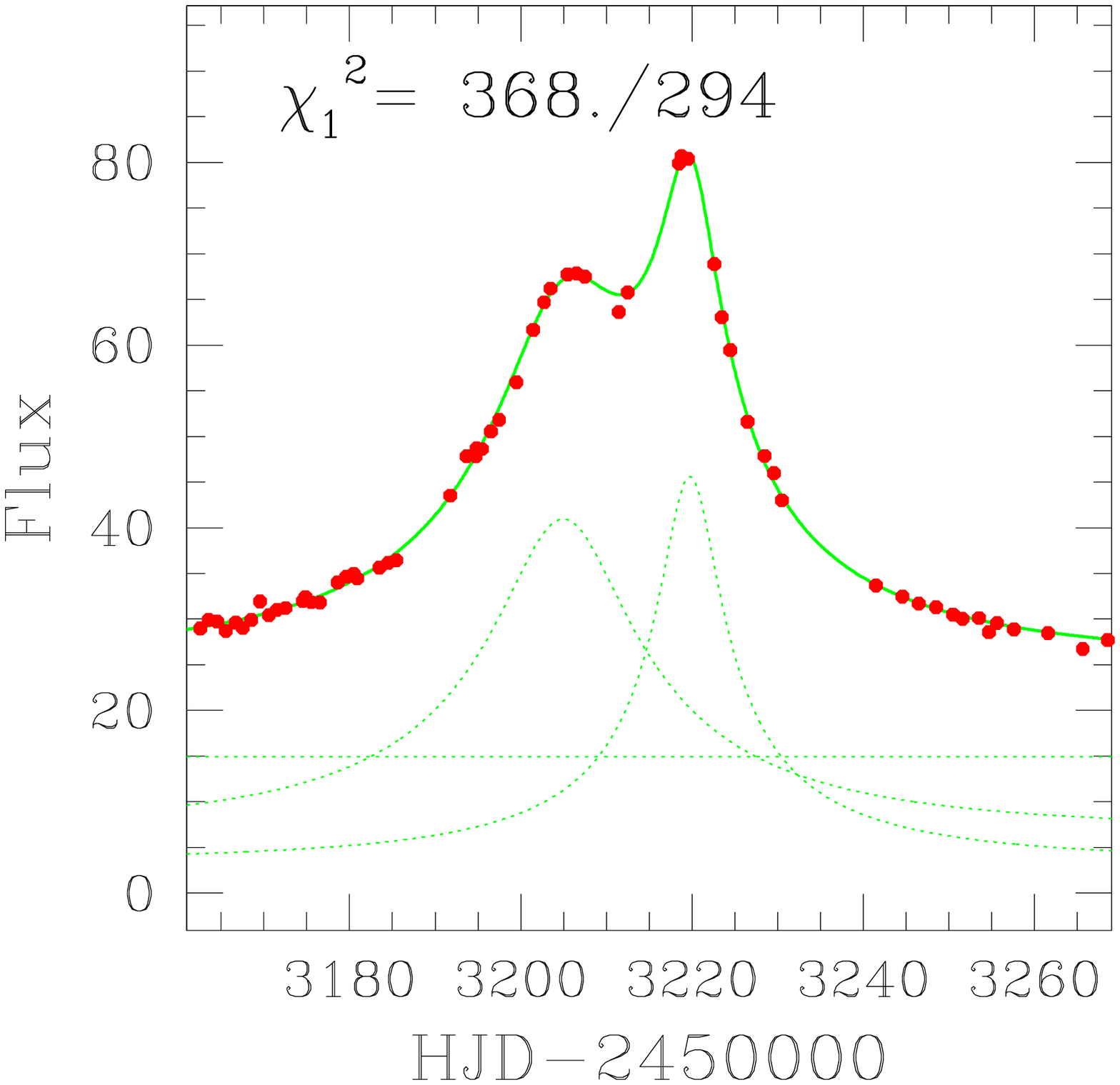}%

}

\noindent\parbox{12.75cm}{
\leftline {\bf OGLE 2004-BLG-354} 

 \includegraphics[height=63mm,width=62mm]{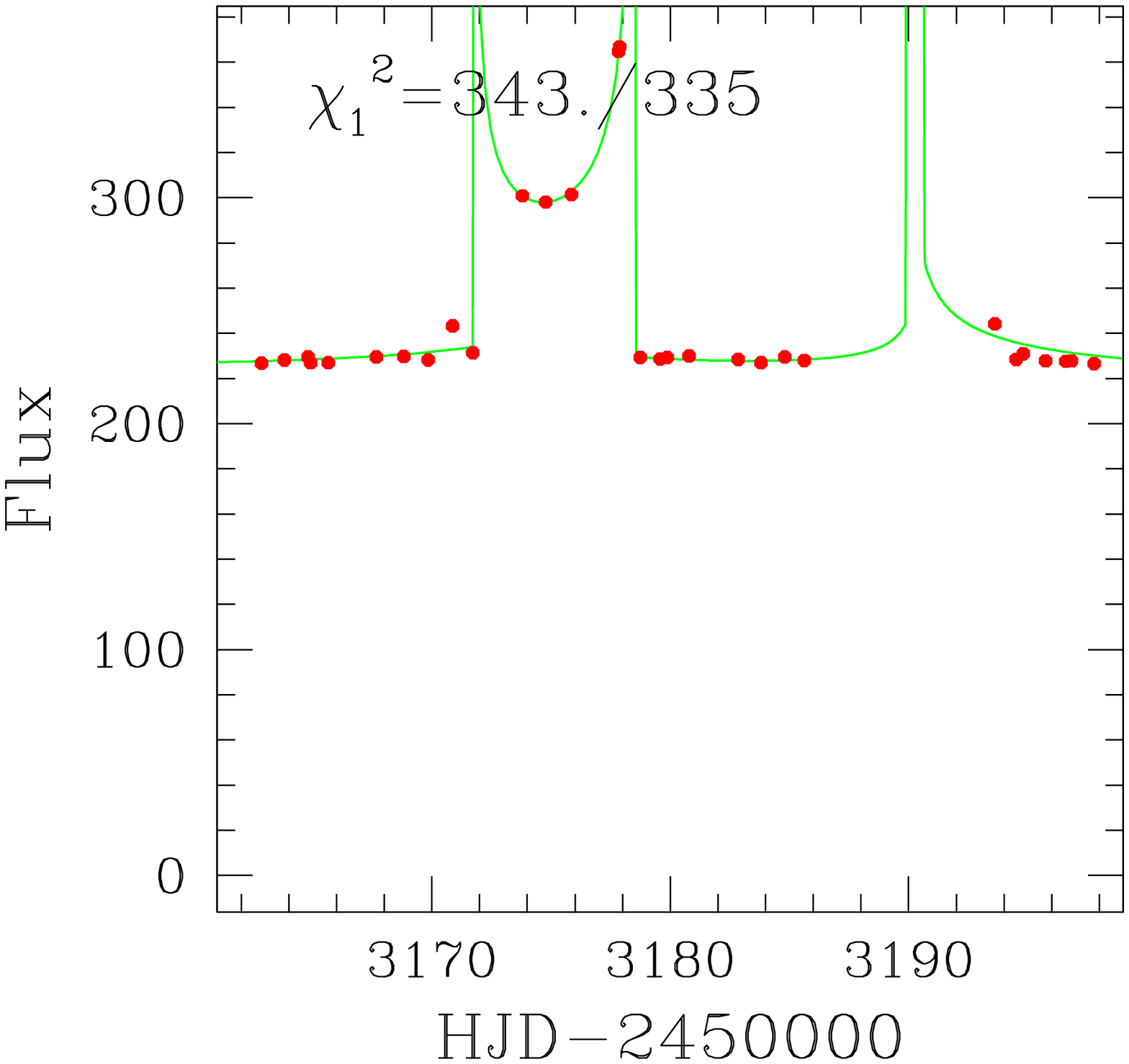}%
 \includegraphics[height=63mm,width=62mm]{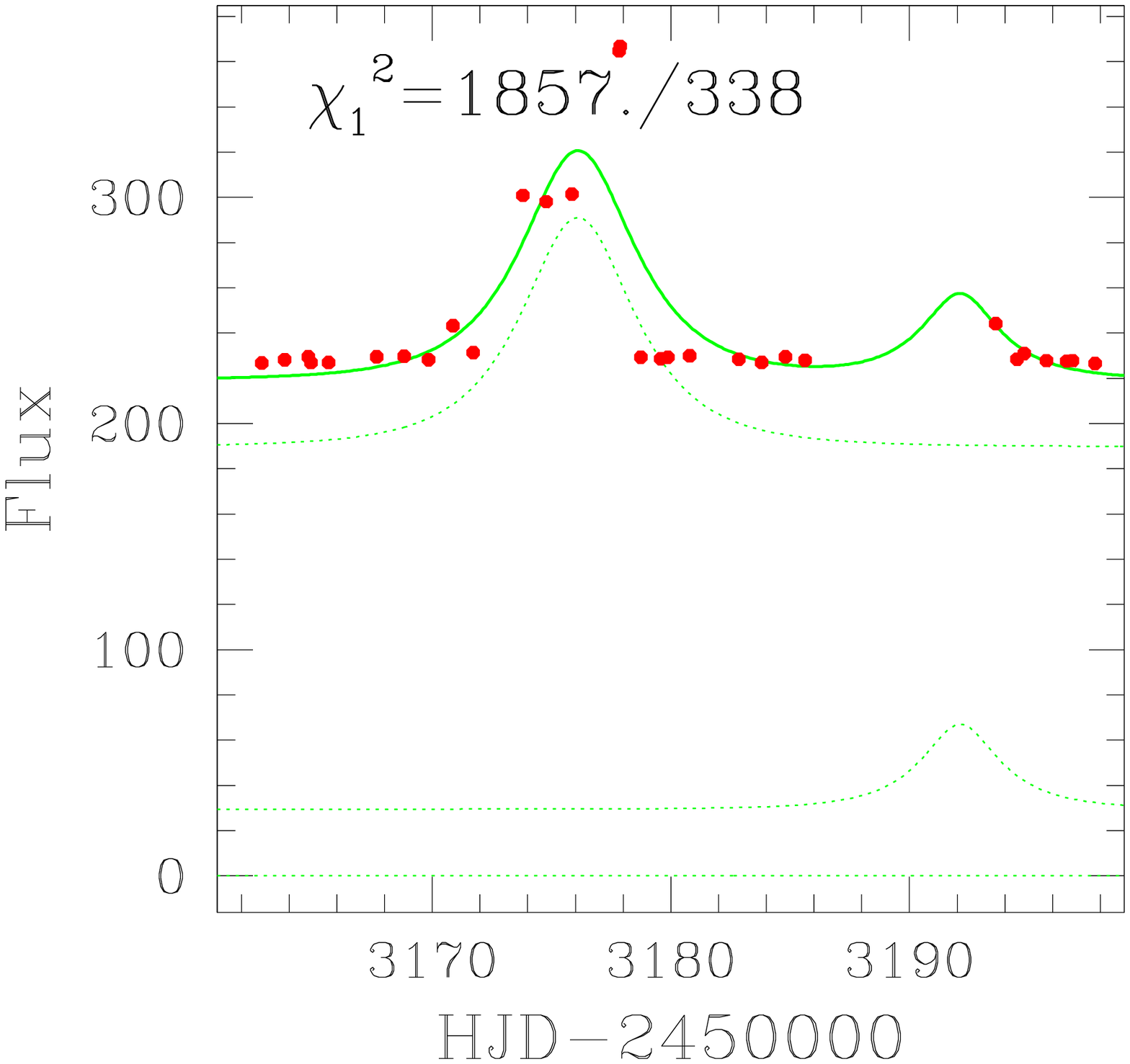}%

}

\noindent\parbox{12.75cm}{
\leftline {\bf OGLE 2004-BLG-362} 

 \includegraphics[height=63mm,width=62mm]{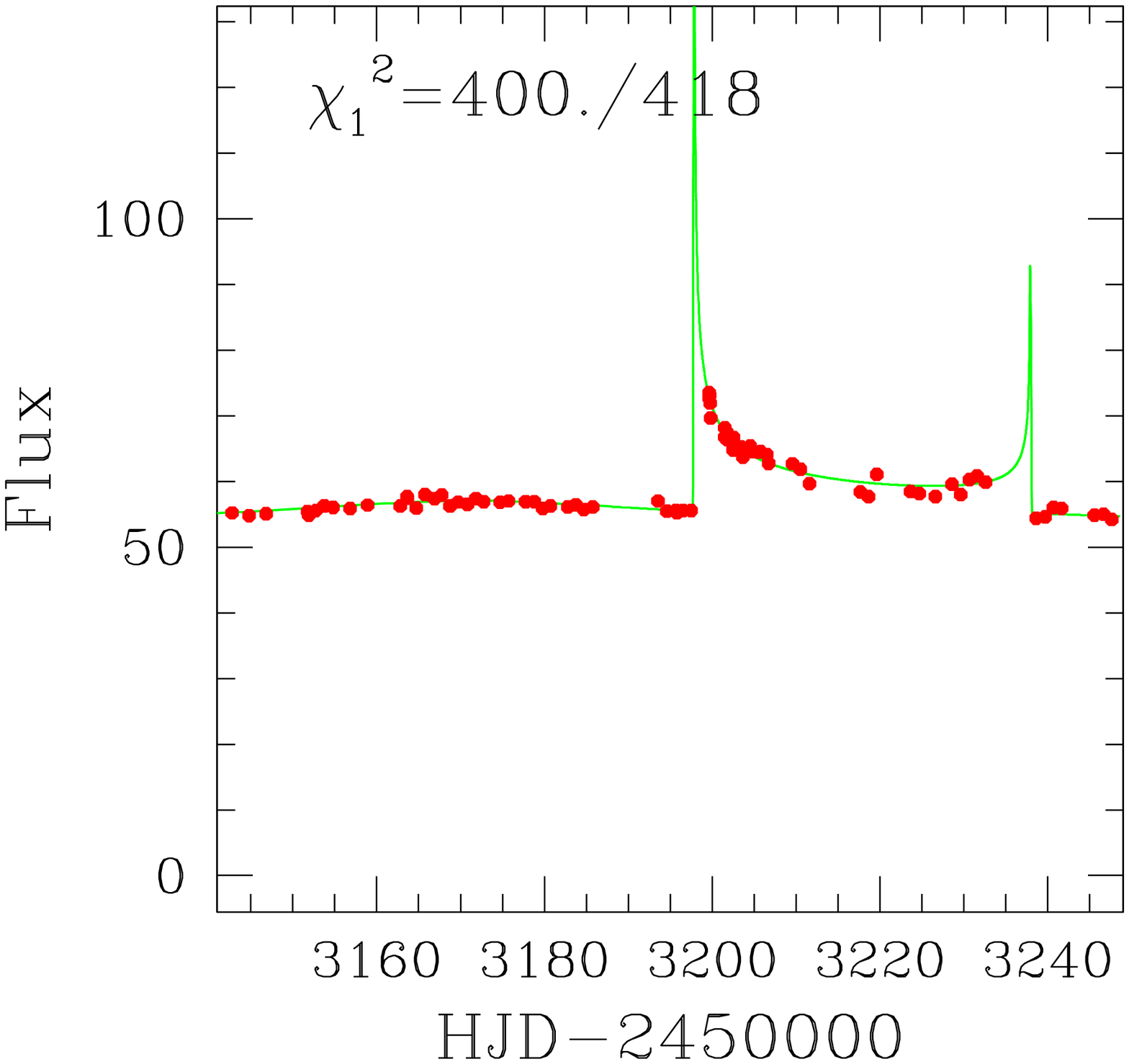}%
 \includegraphics[height=63mm,width=62mm]{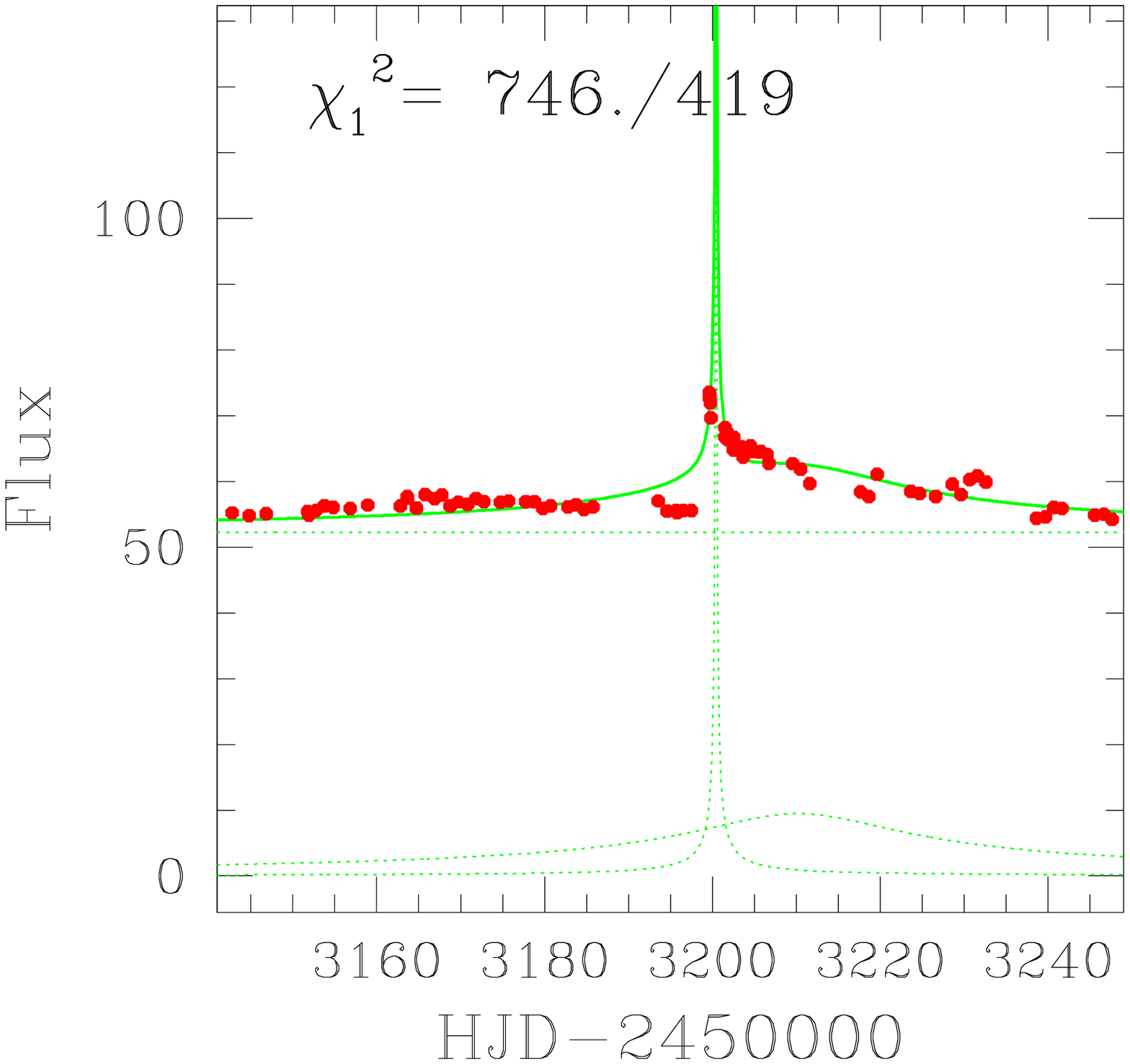}%

}
 
\noindent\parbox{12.75cm}{
\leftline {\bf OGLE 2004-BLG-366} 

 \includegraphics[height=63mm,width=62mm]{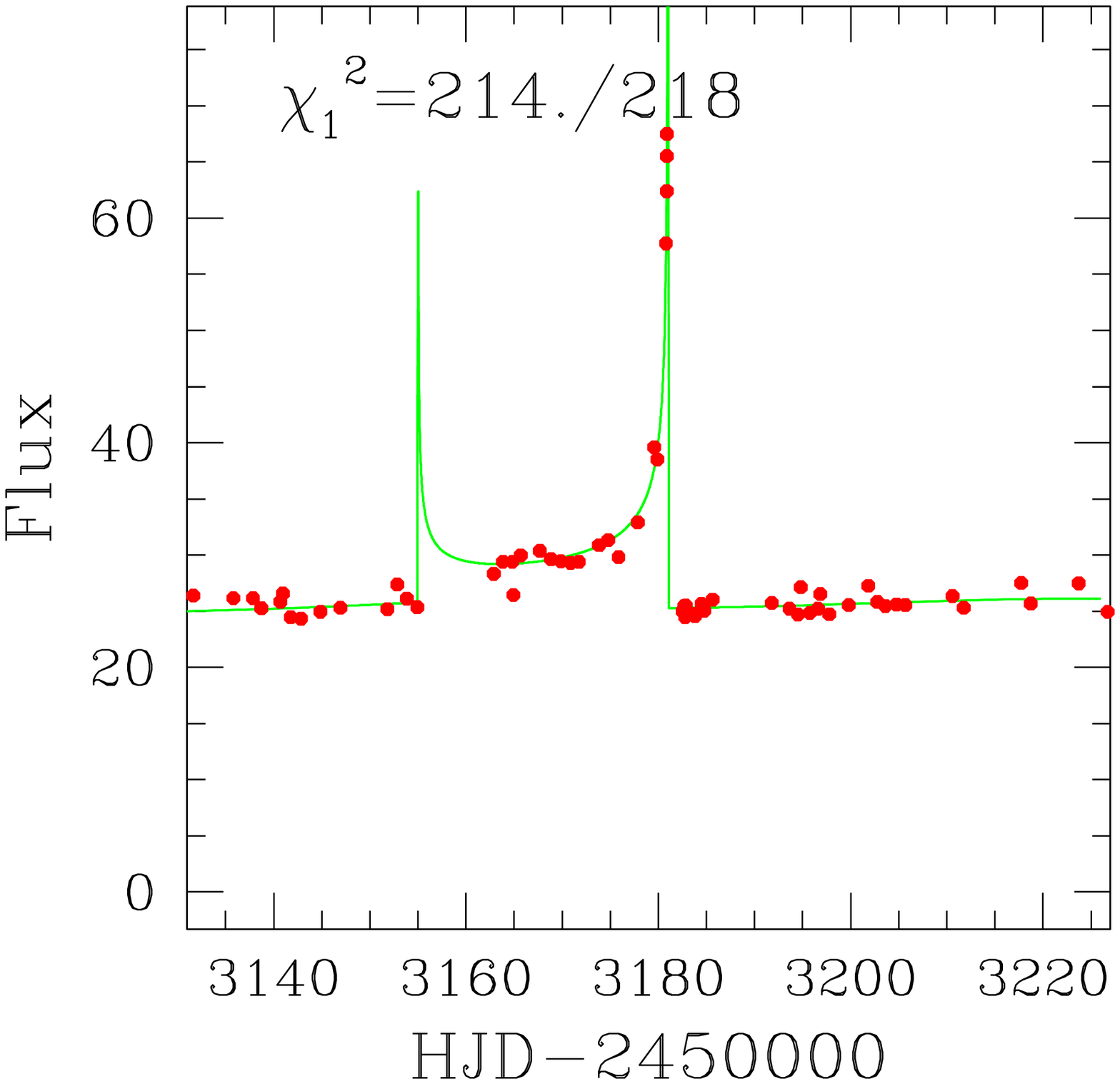}%
 \includegraphics[height=63mm,width=62mm]{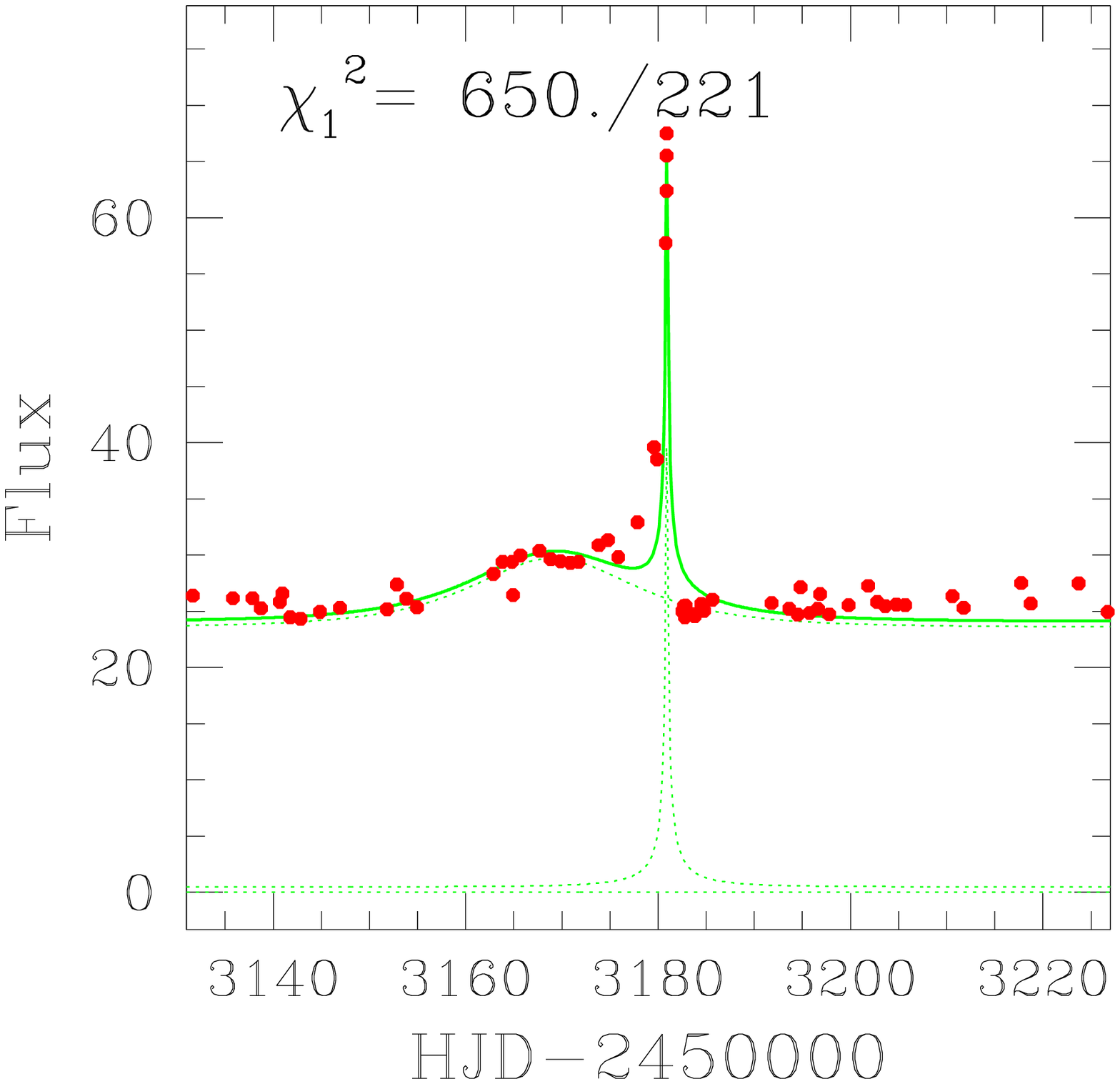}%

}

\noindent\parbox{12.75cm}{
\leftline {\bf OGLE 2004-BLG-367} 

 \includegraphics[height=63mm,width=62mm]{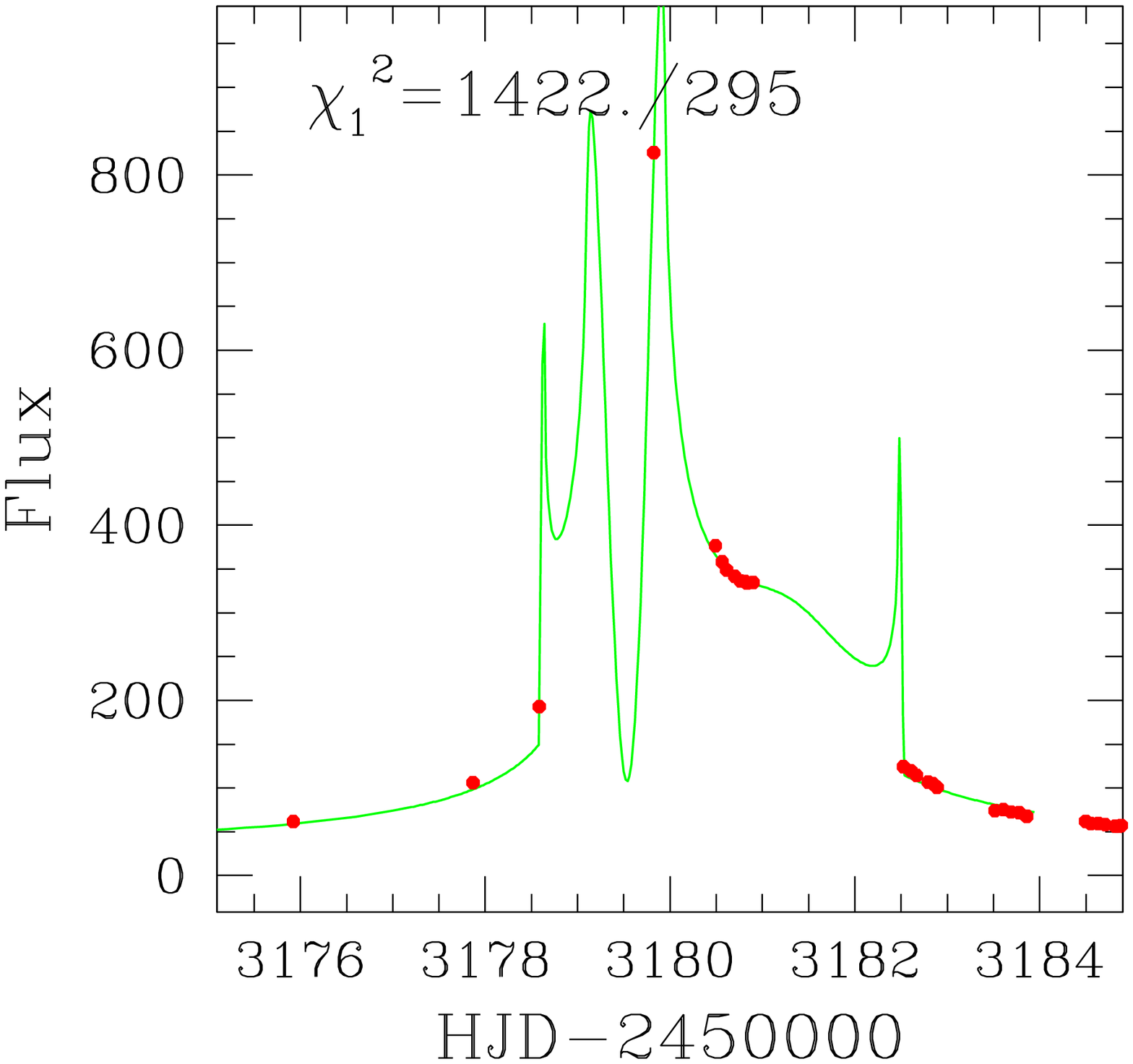}%
 \includegraphics[height=63mm,width=62mm]{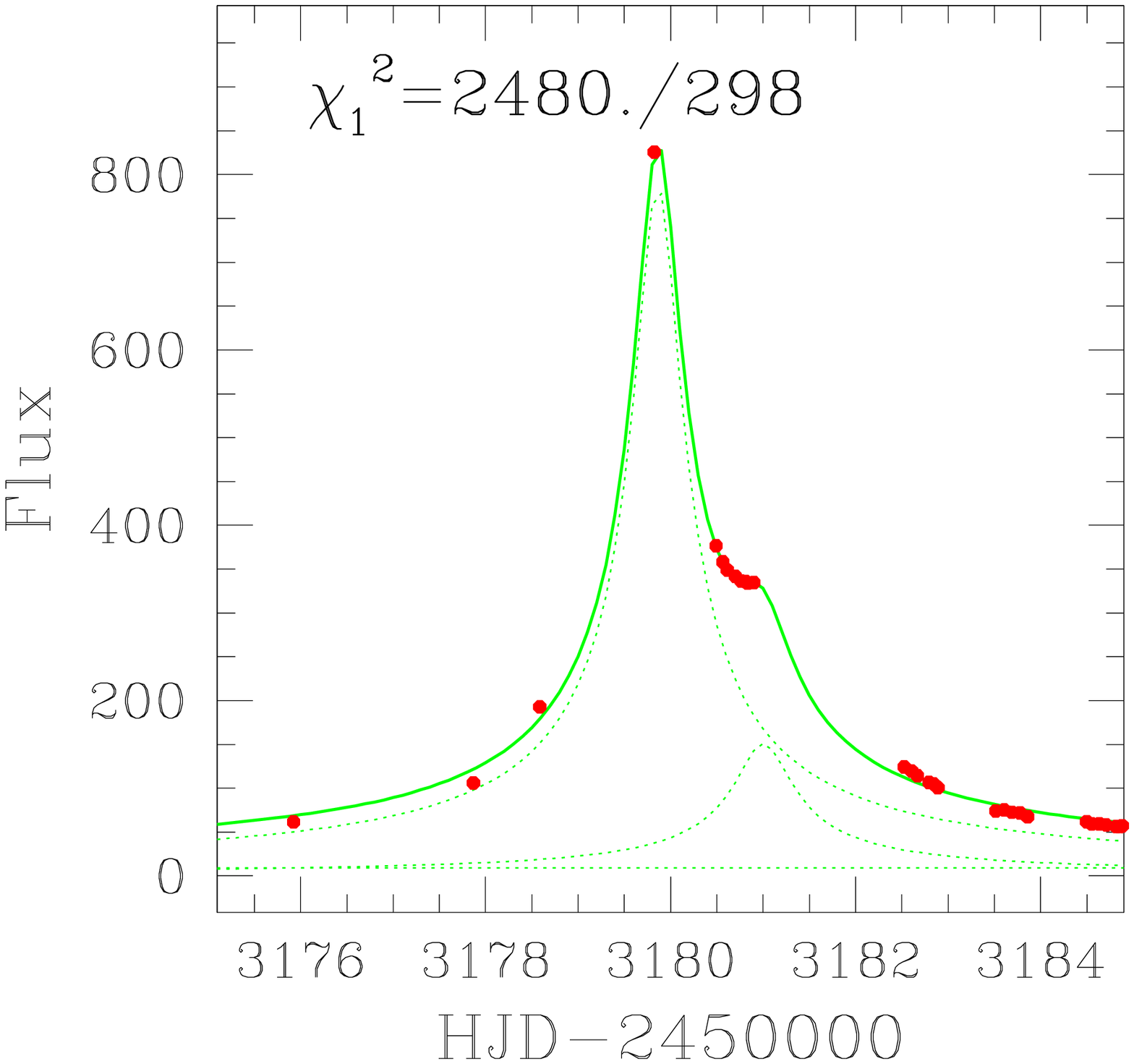}%

}

\noindent\parbox{12.75cm}{
\leftline {\bf OGLE 2004-BLG-444} 

 \includegraphics[height=63mm,width=62mm]{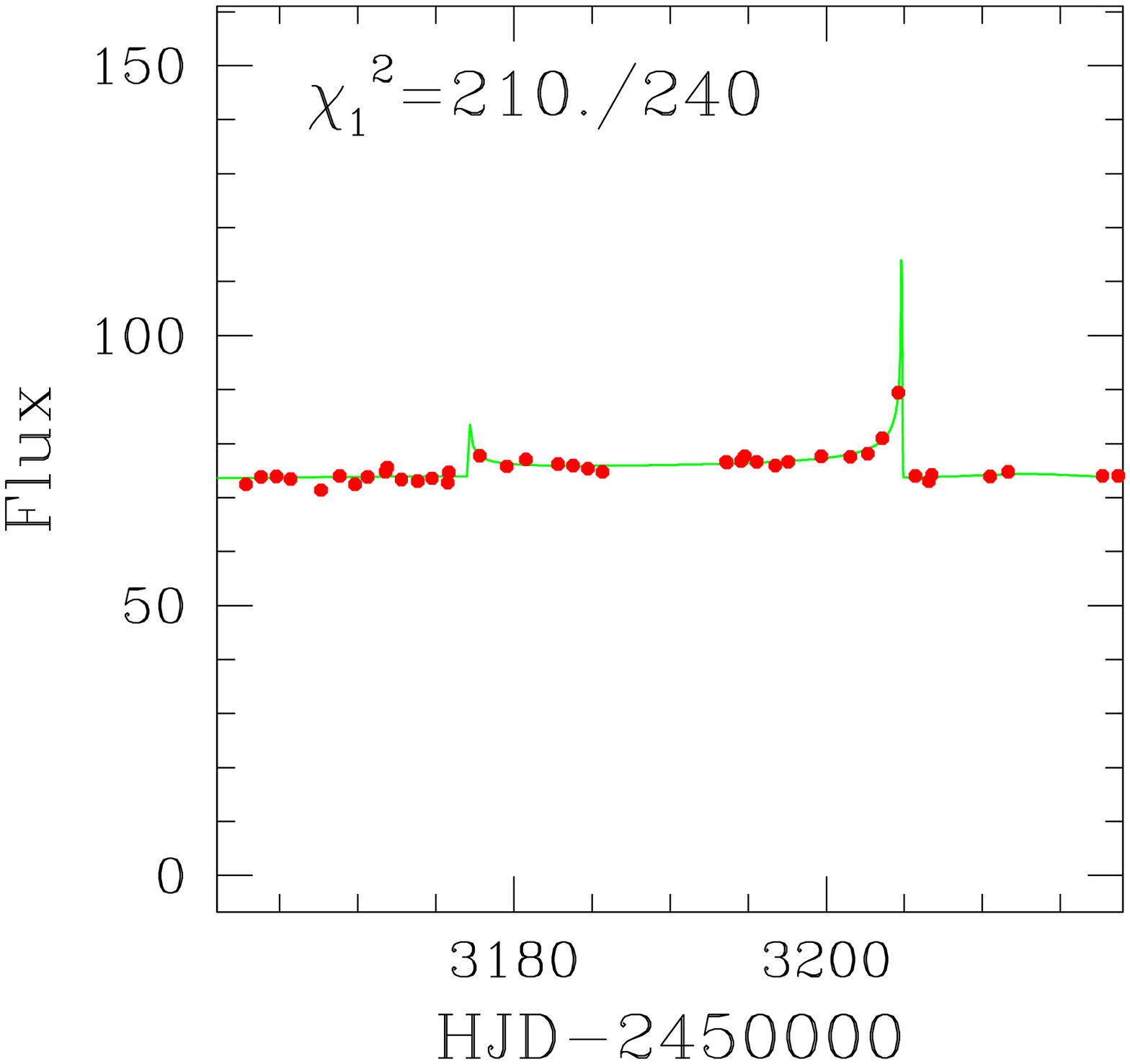}%
 \includegraphics[height=63mm,width=62mm]{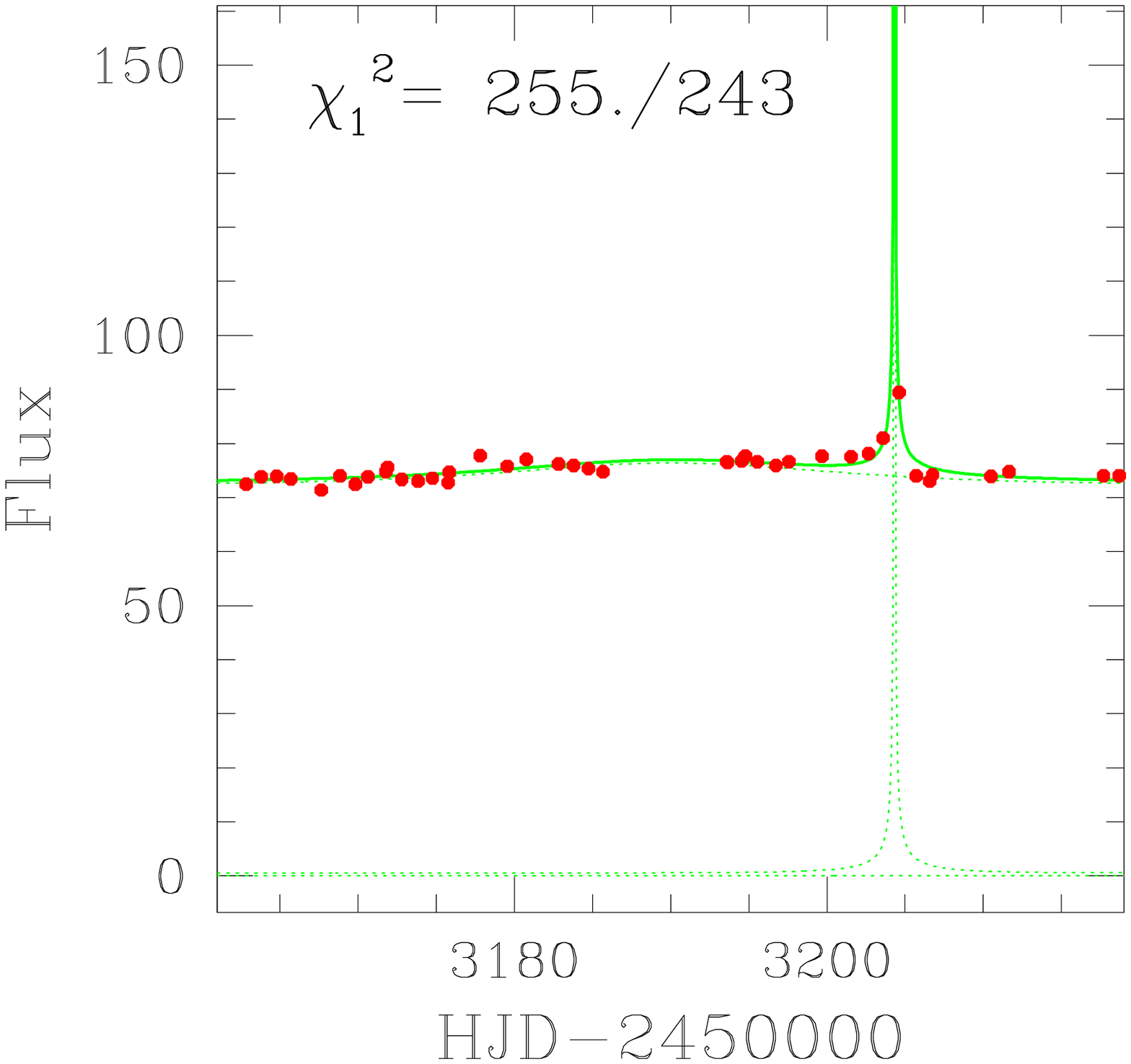}%

}

\noindent\parbox{12.75cm}{
\leftline {\bf OGLE 2004-BLG-451} 

 \includegraphics[height=63mm,width=62mm]{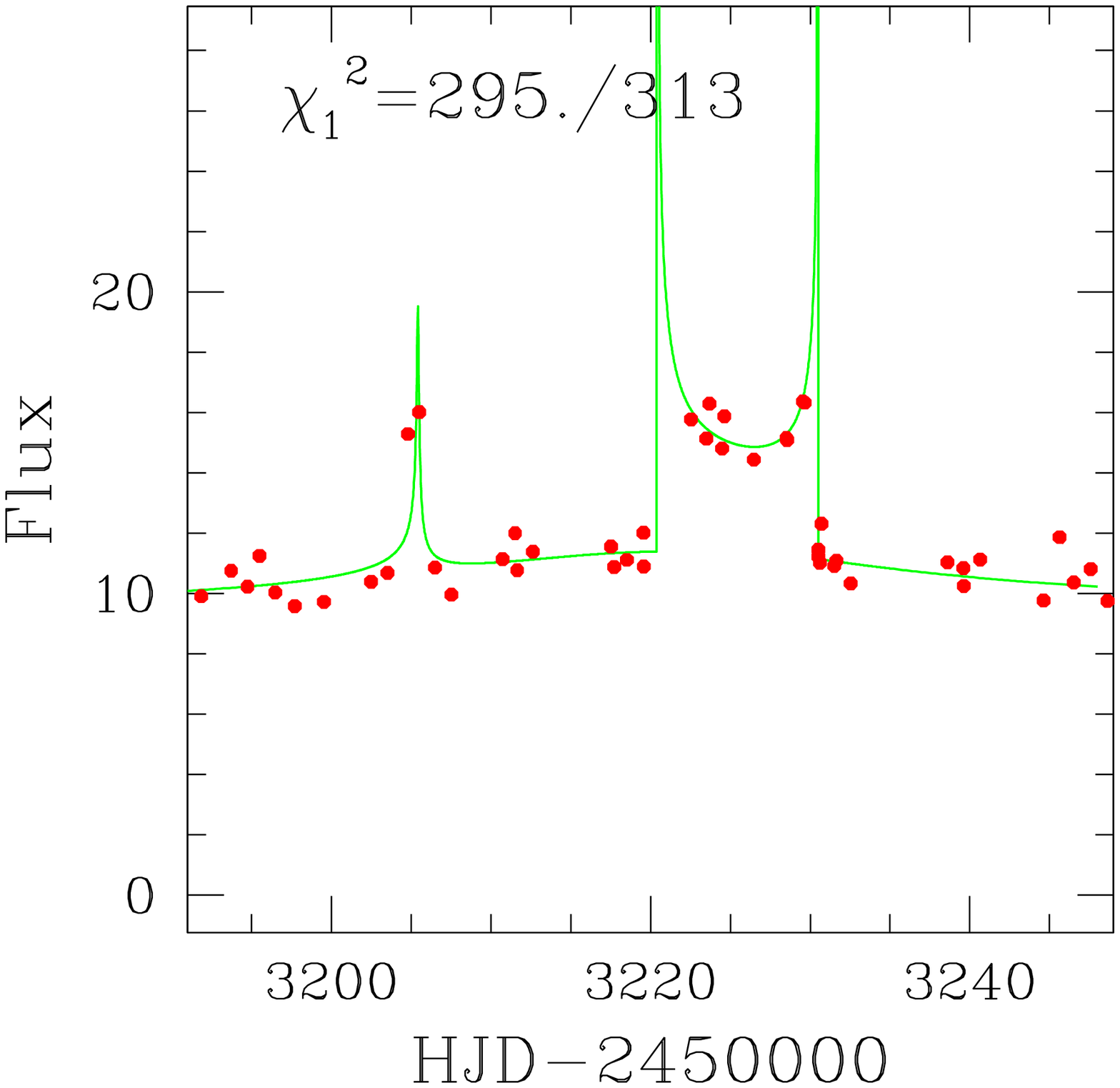}%
 \includegraphics[height=63mm,width=62mm]{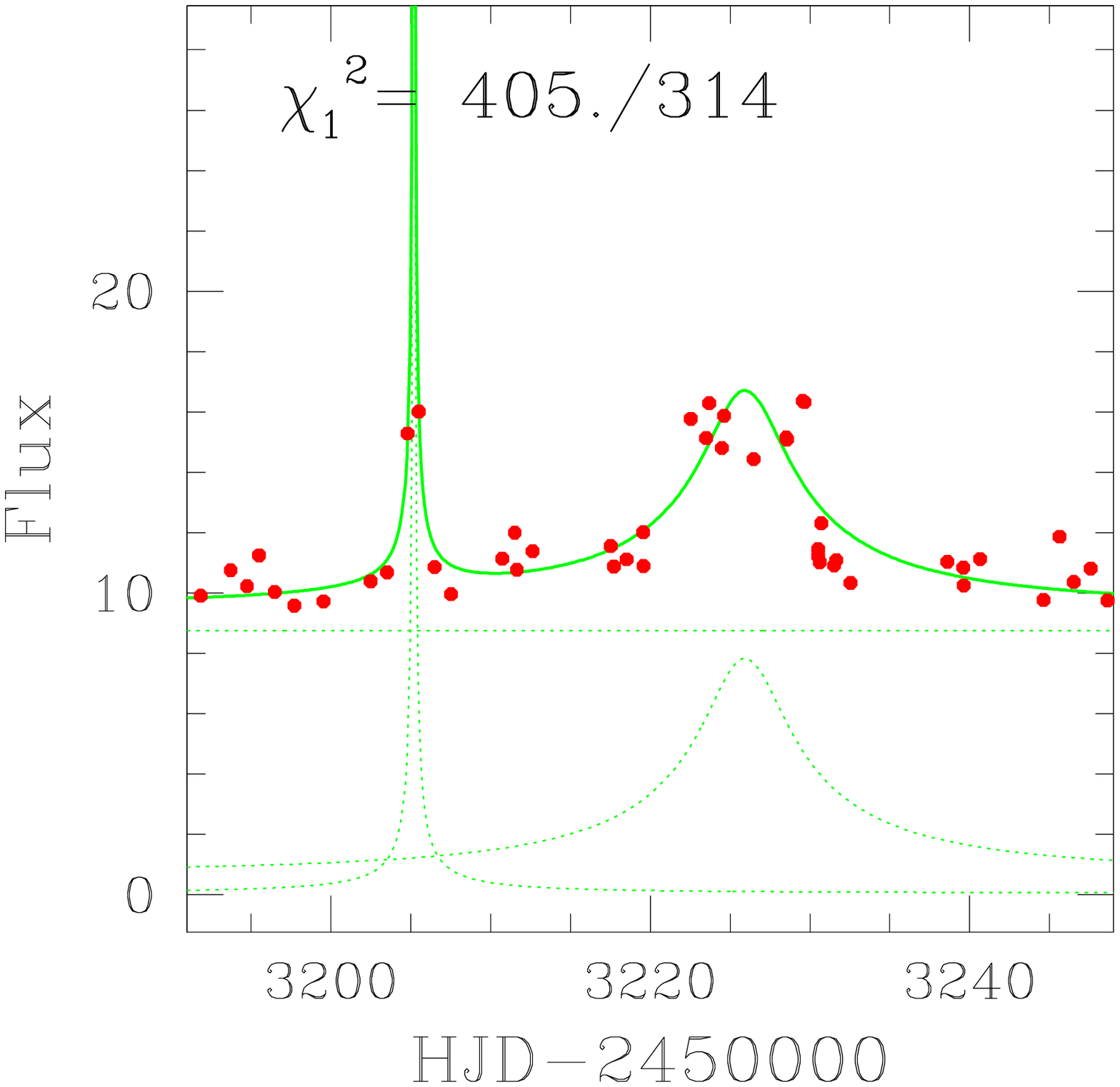}%

}

\noindent\parbox{12.75cm}{
\leftline {\bf OGLE 2004-BLG-460} 

 \includegraphics[height=63mm,width=62mm]{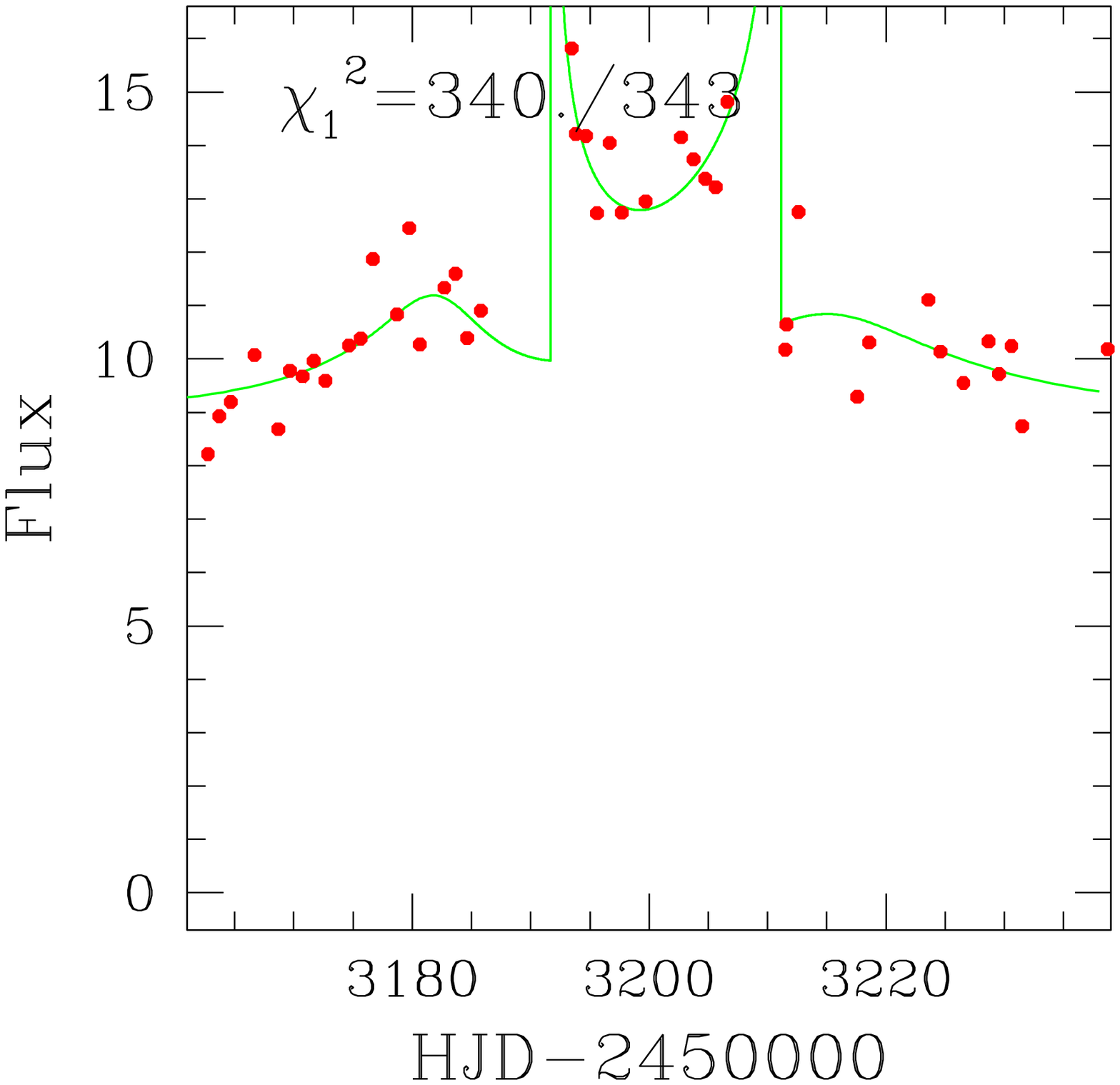}%
 \includegraphics[height=63mm,width=62mm]{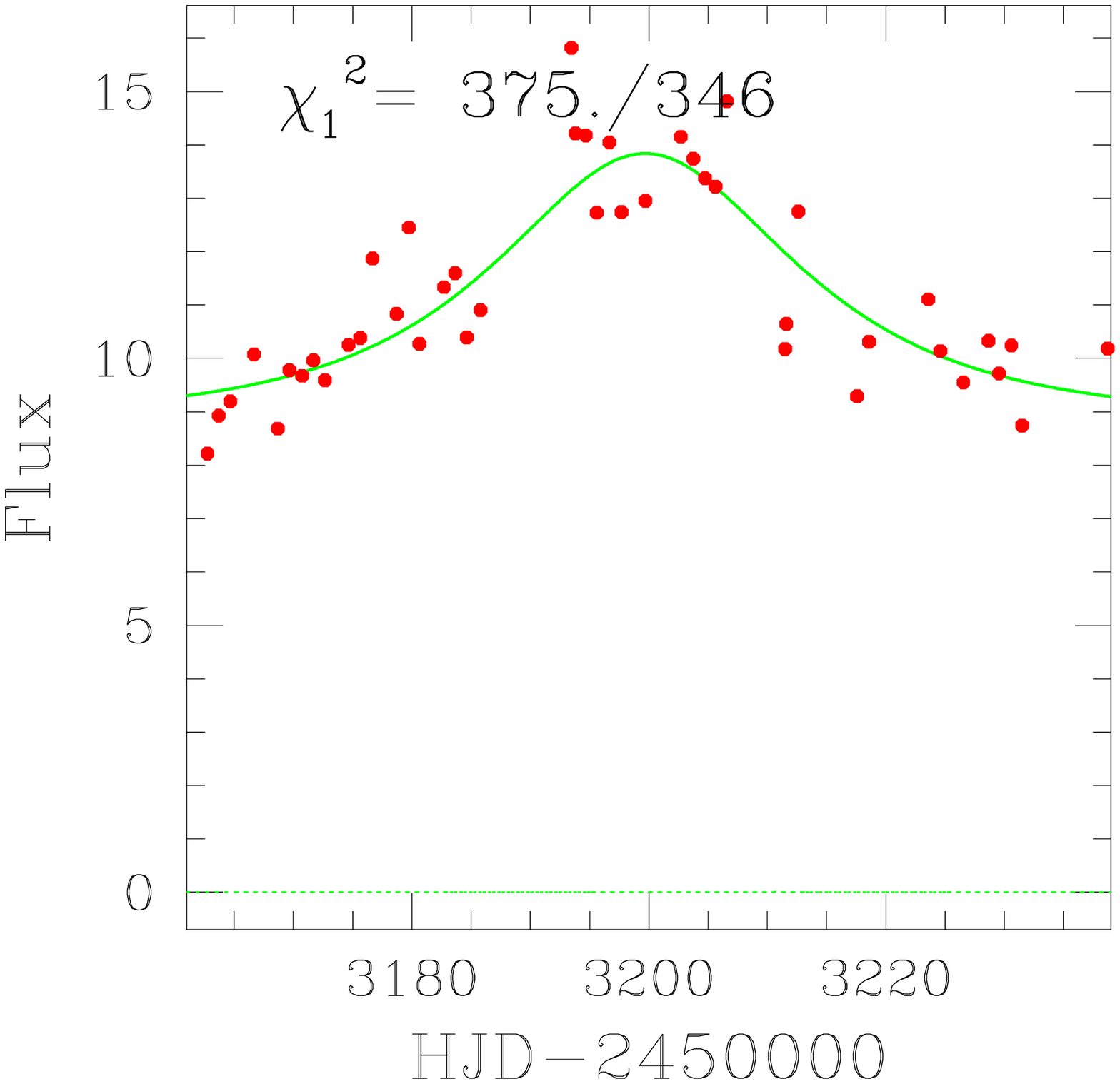}%

}

\noindent\parbox{12.75cm}{
\leftline {\bf OGLE 2004-BLG-480} 

 \includegraphics[height=63mm,width=62mm]{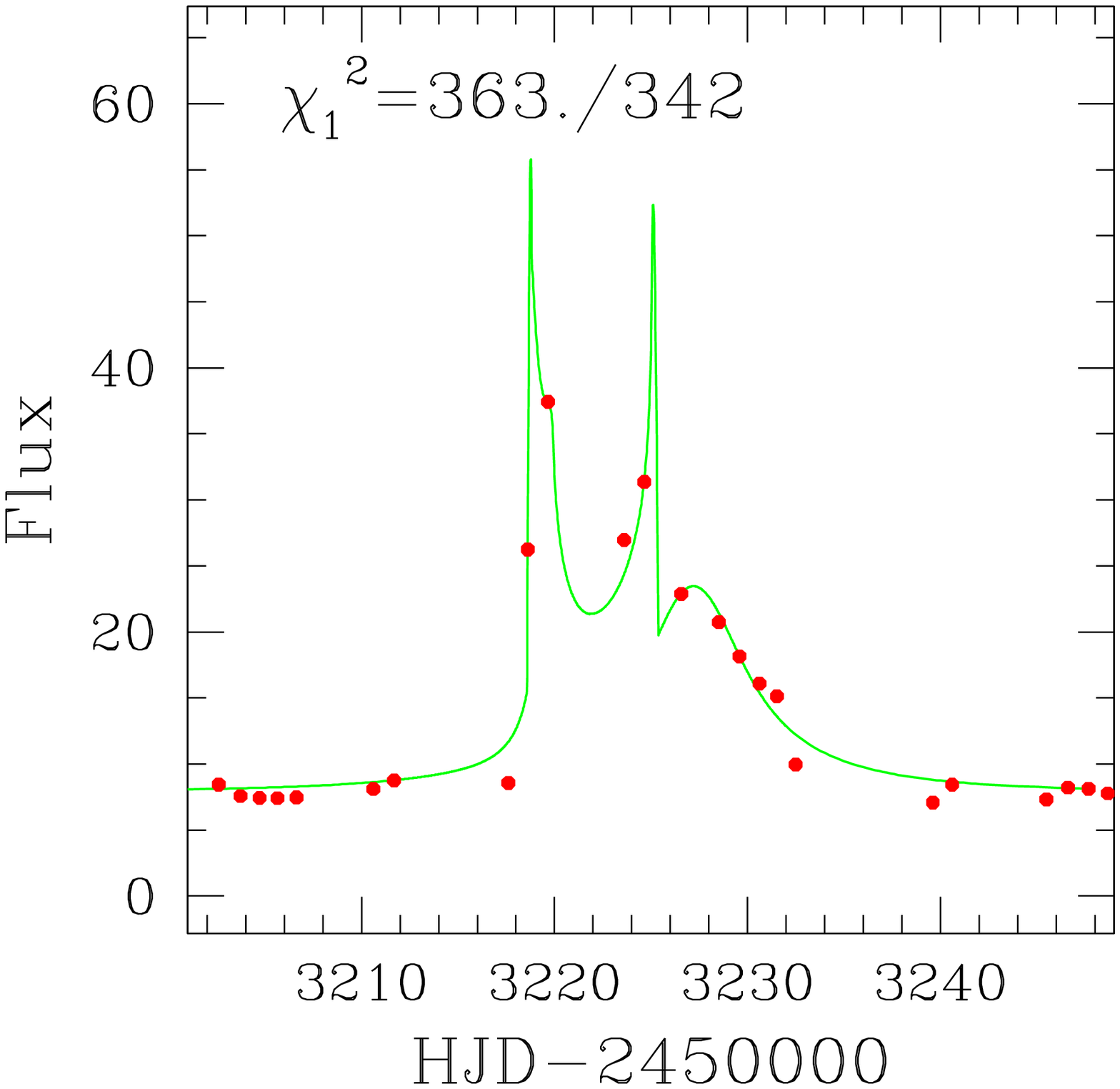}%
 \includegraphics[height=63mm,width=62mm]{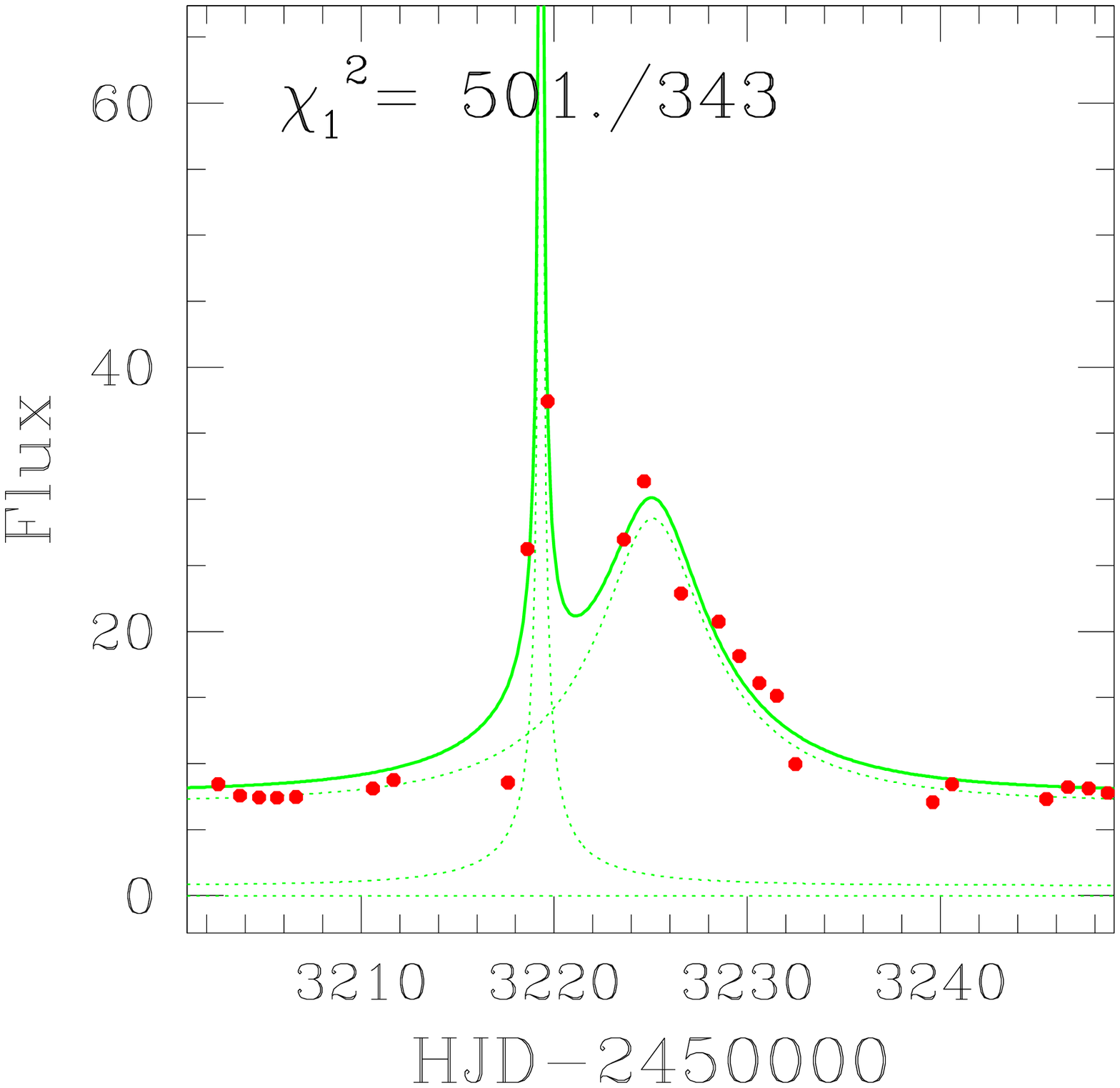}%

}

\noindent\parbox{12.75cm}{
\leftline {\bf OGLE 2004-BLG-490} 

 \includegraphics[height=63mm,width=62mm]{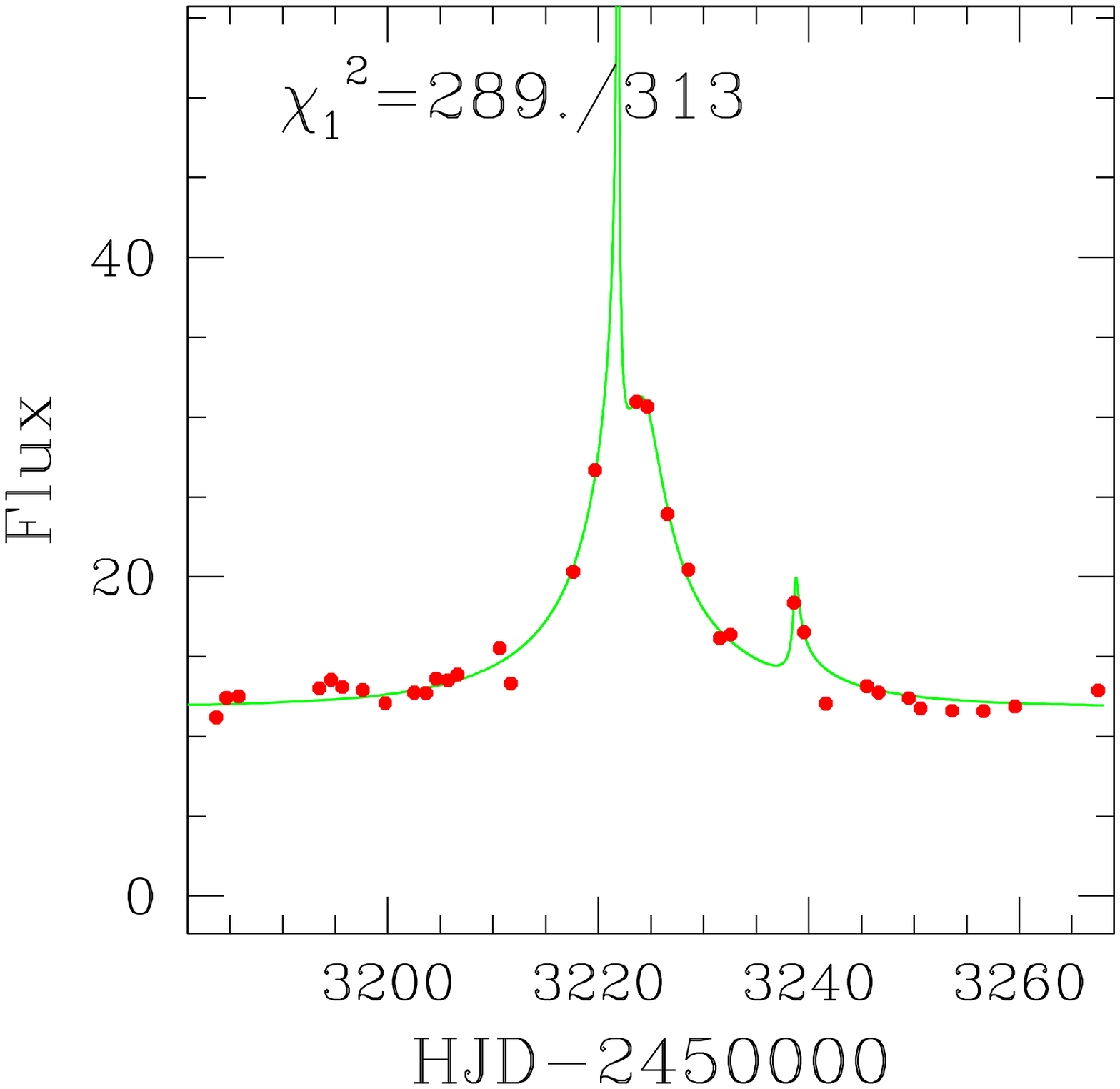}%
 \includegraphics[height=63mm,width=62mm]{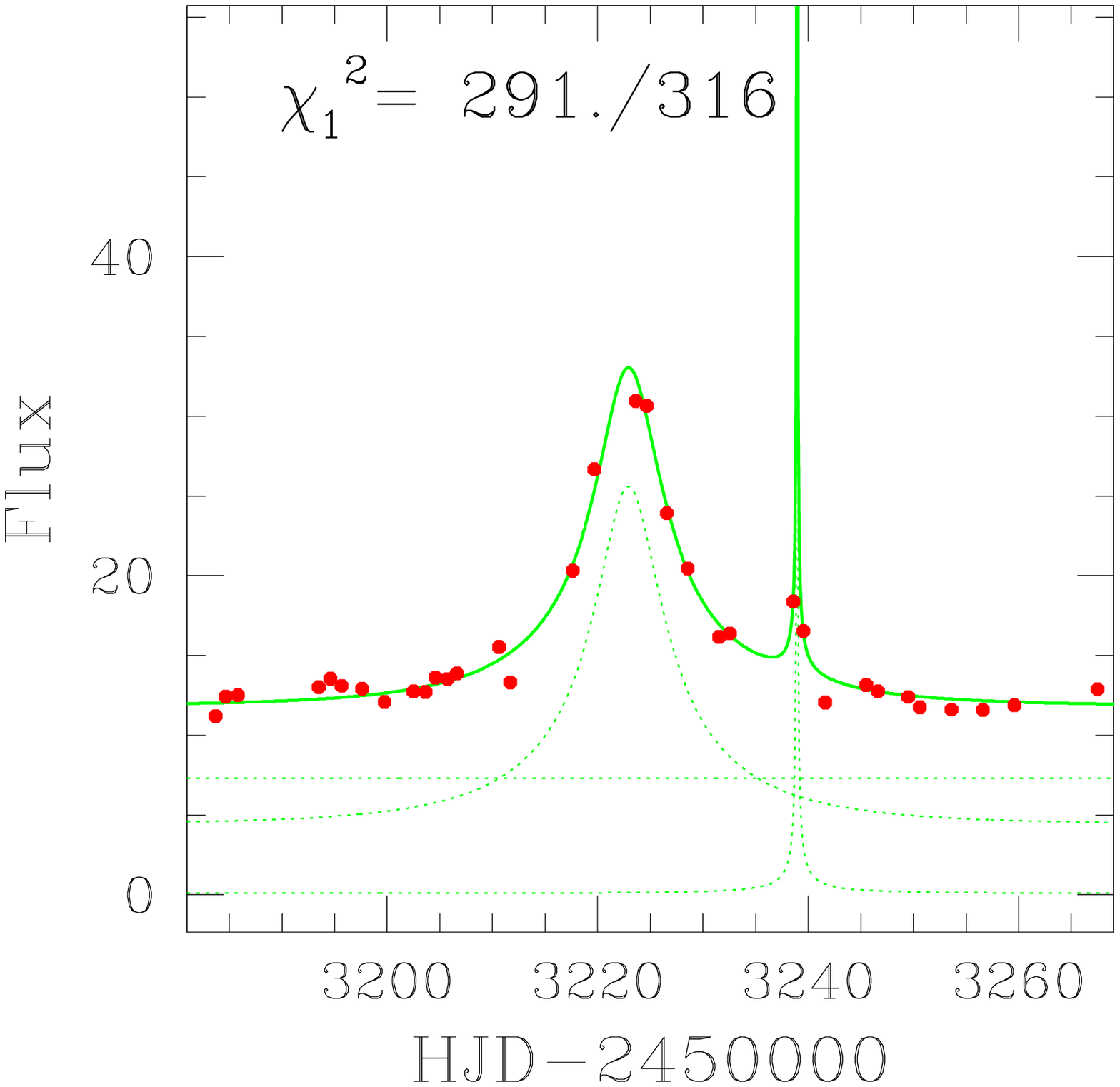}%

}

\noindent\parbox{12.75cm}{
\leftline {\bf OGLE 2004-BLG-559} 

 \includegraphics[height=63mm,width=62mm]{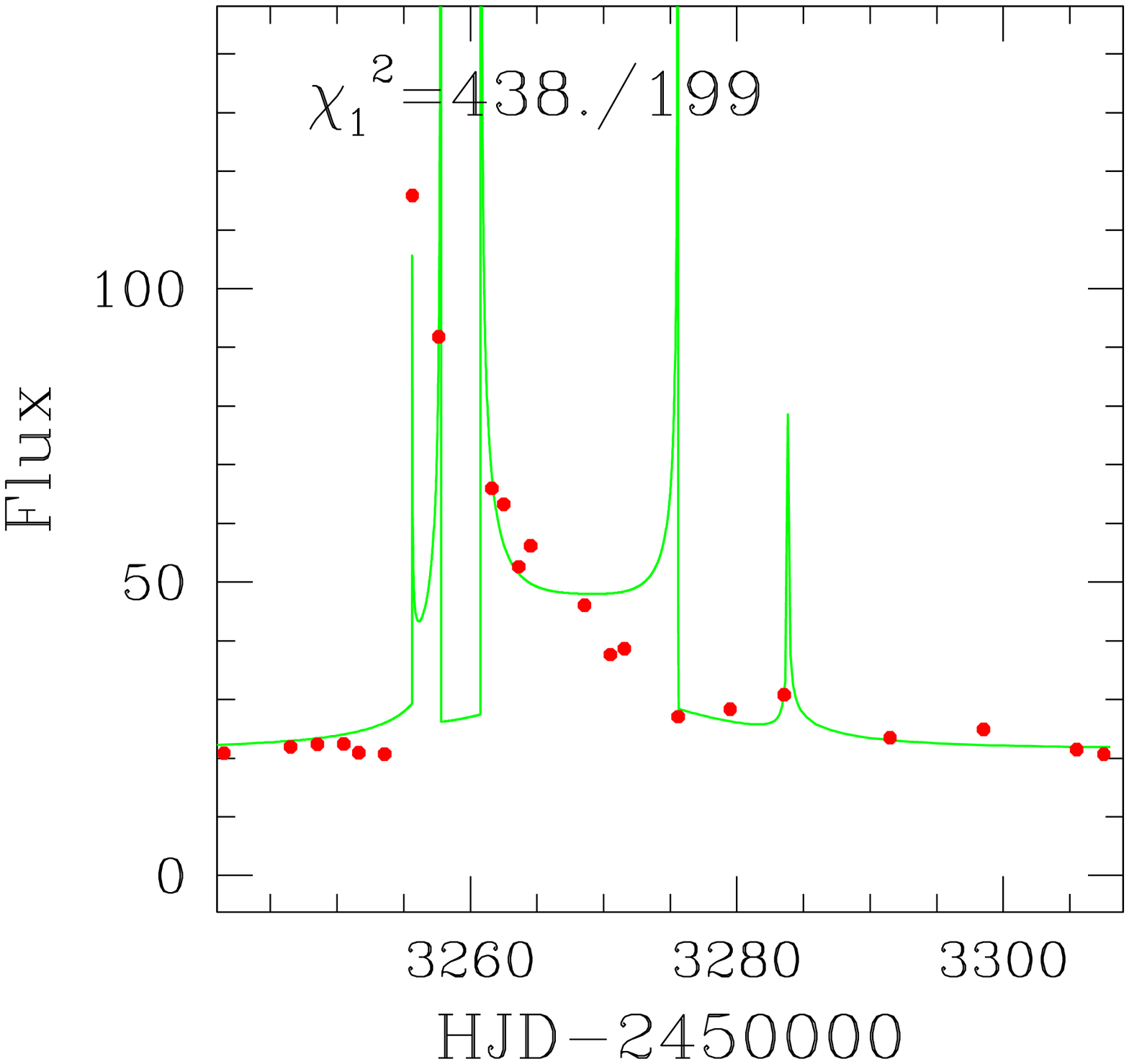}%
 \includegraphics[height=63mm,width=62mm]{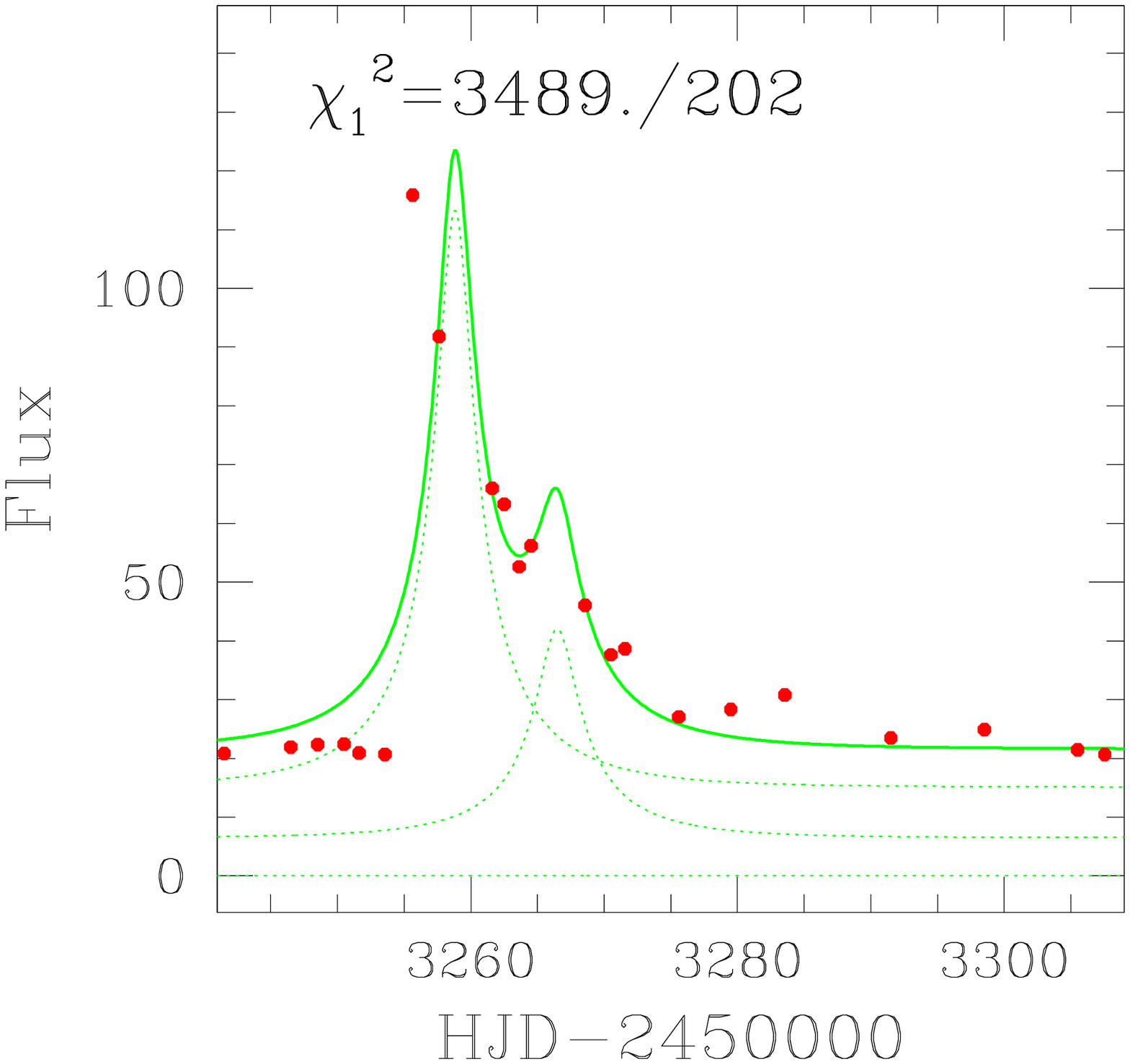}%

}

\noindent\parbox{12.75cm}{
\leftline {\bf OGLE 2004-BLG-572} 

 \includegraphics[height=60mm,width=62mm]{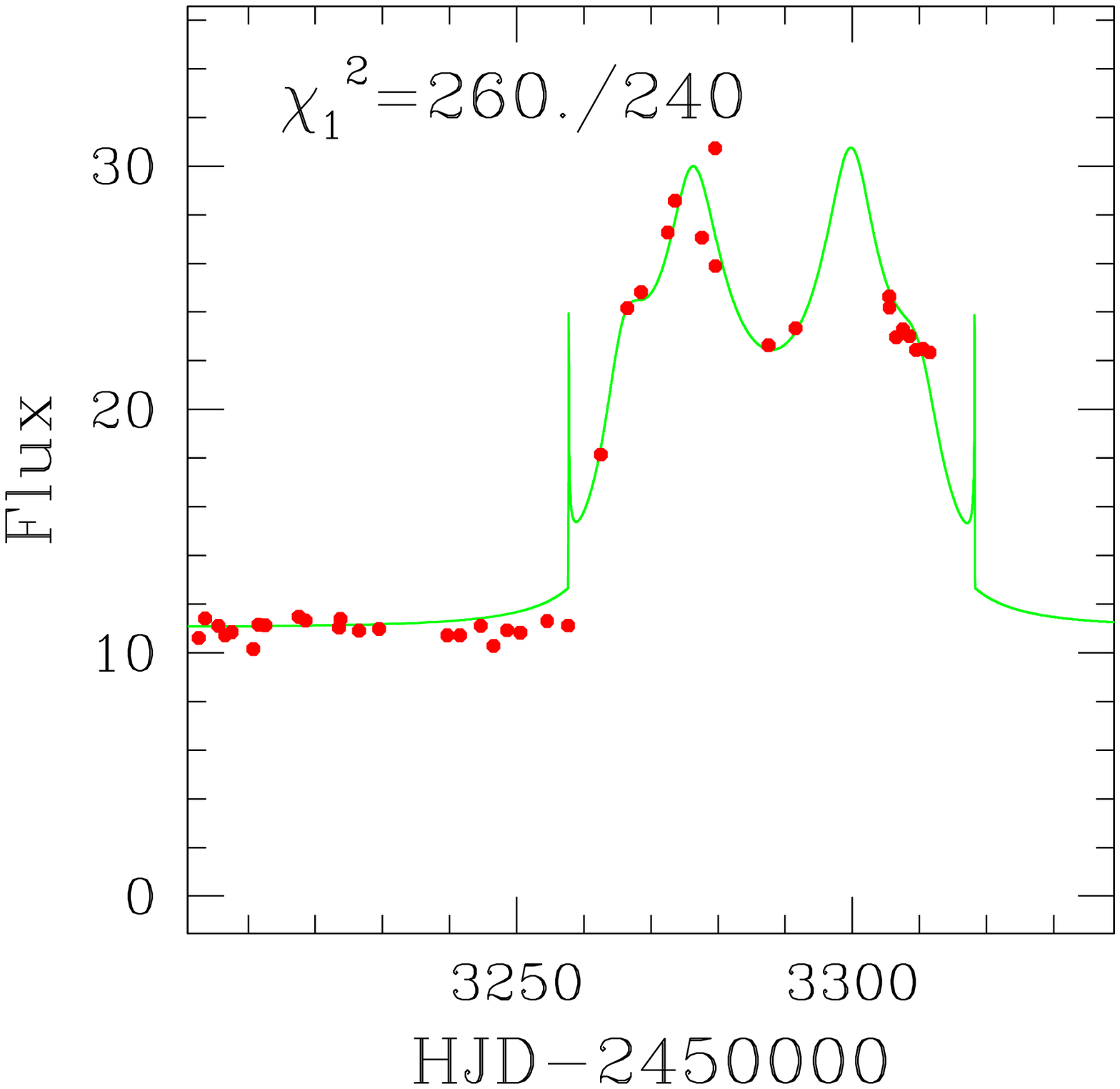}%
 \includegraphics[height=60mm,width=62mm]{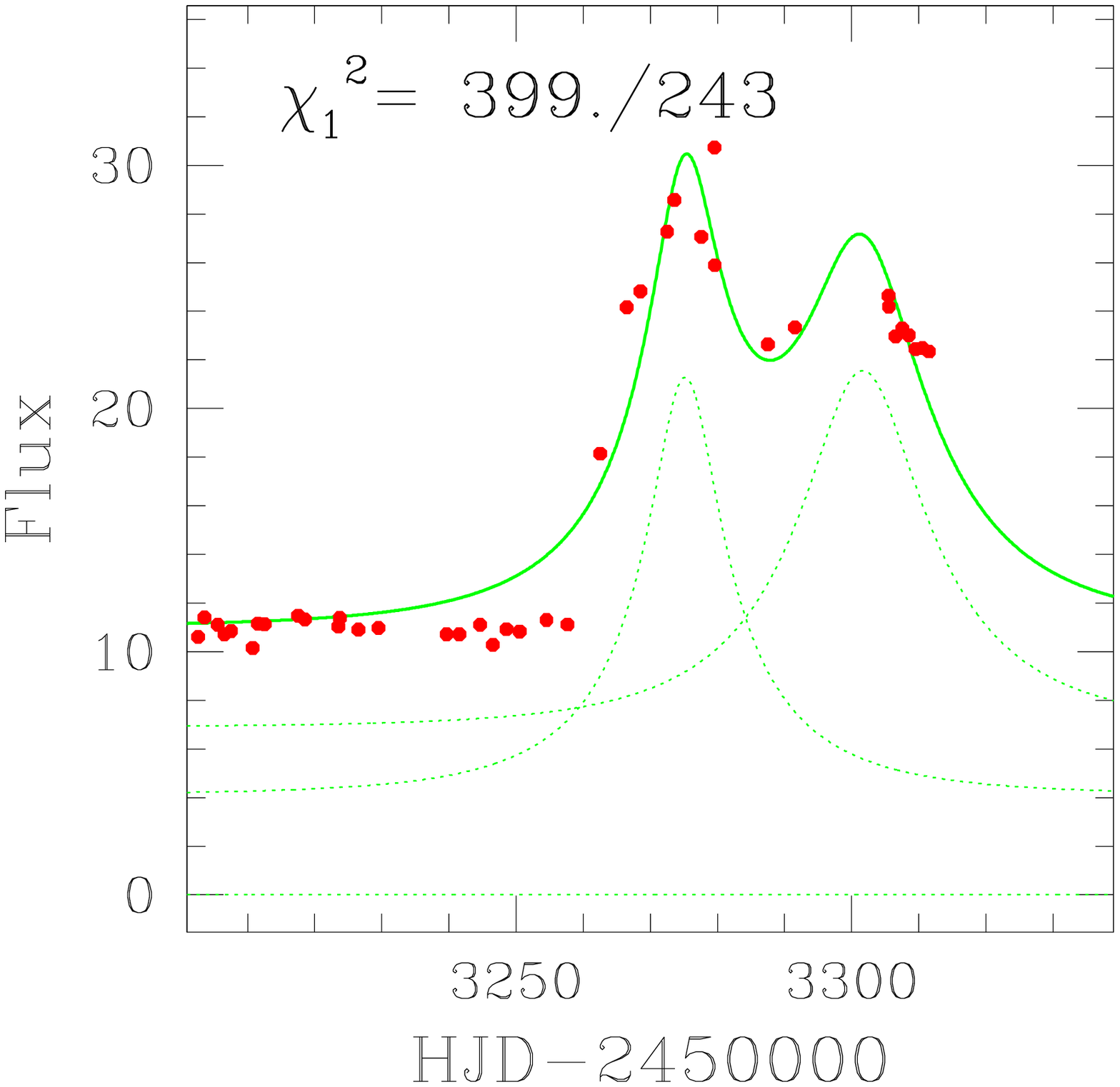}%

}

\noindent\parbox{12.75cm}{
\leftline {\bf OGLE 2004-BLG-605} 

 \includegraphics[height=60mm,width=62mm]{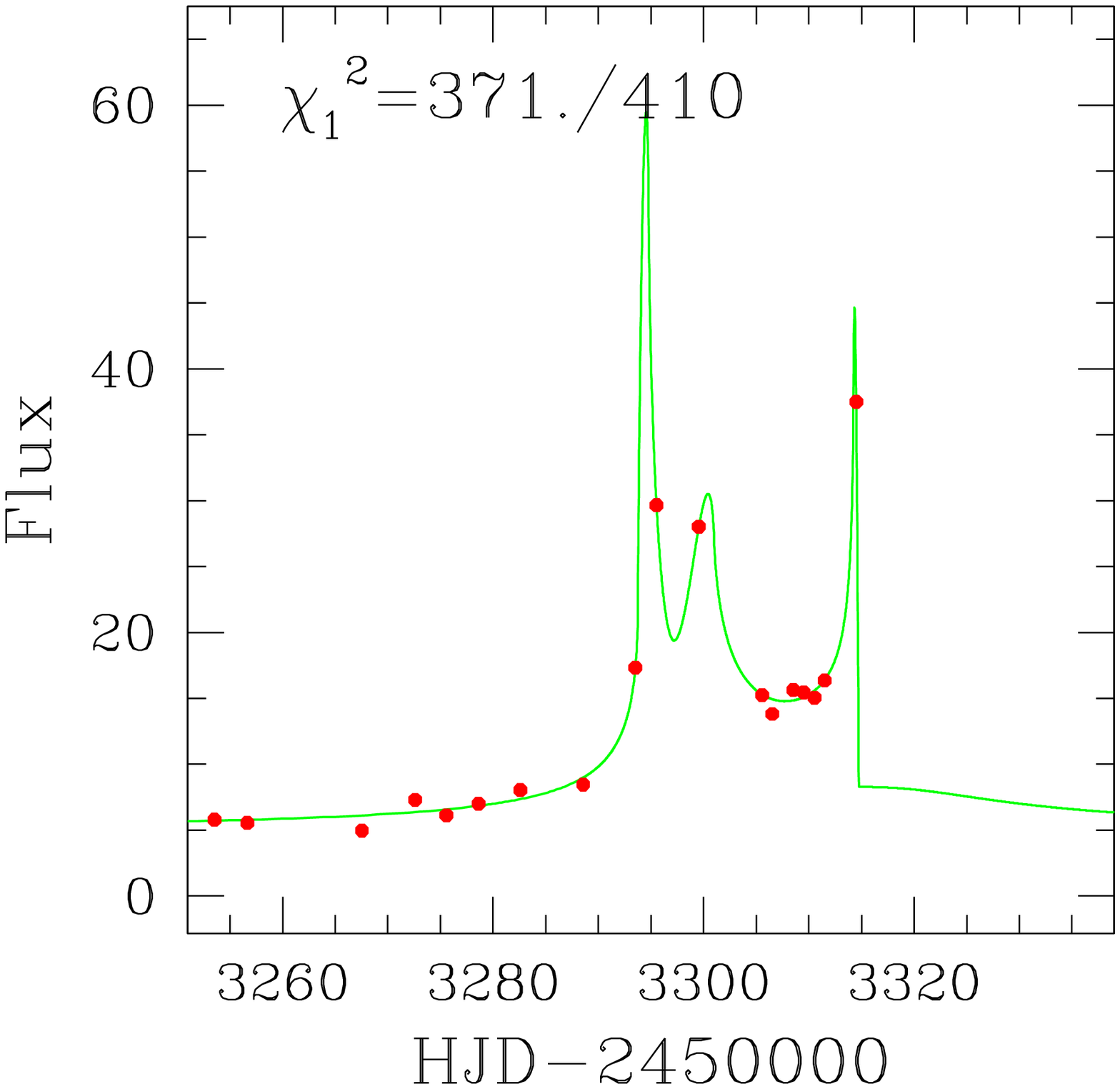}%
 \includegraphics[height=60mm,width=62mm]{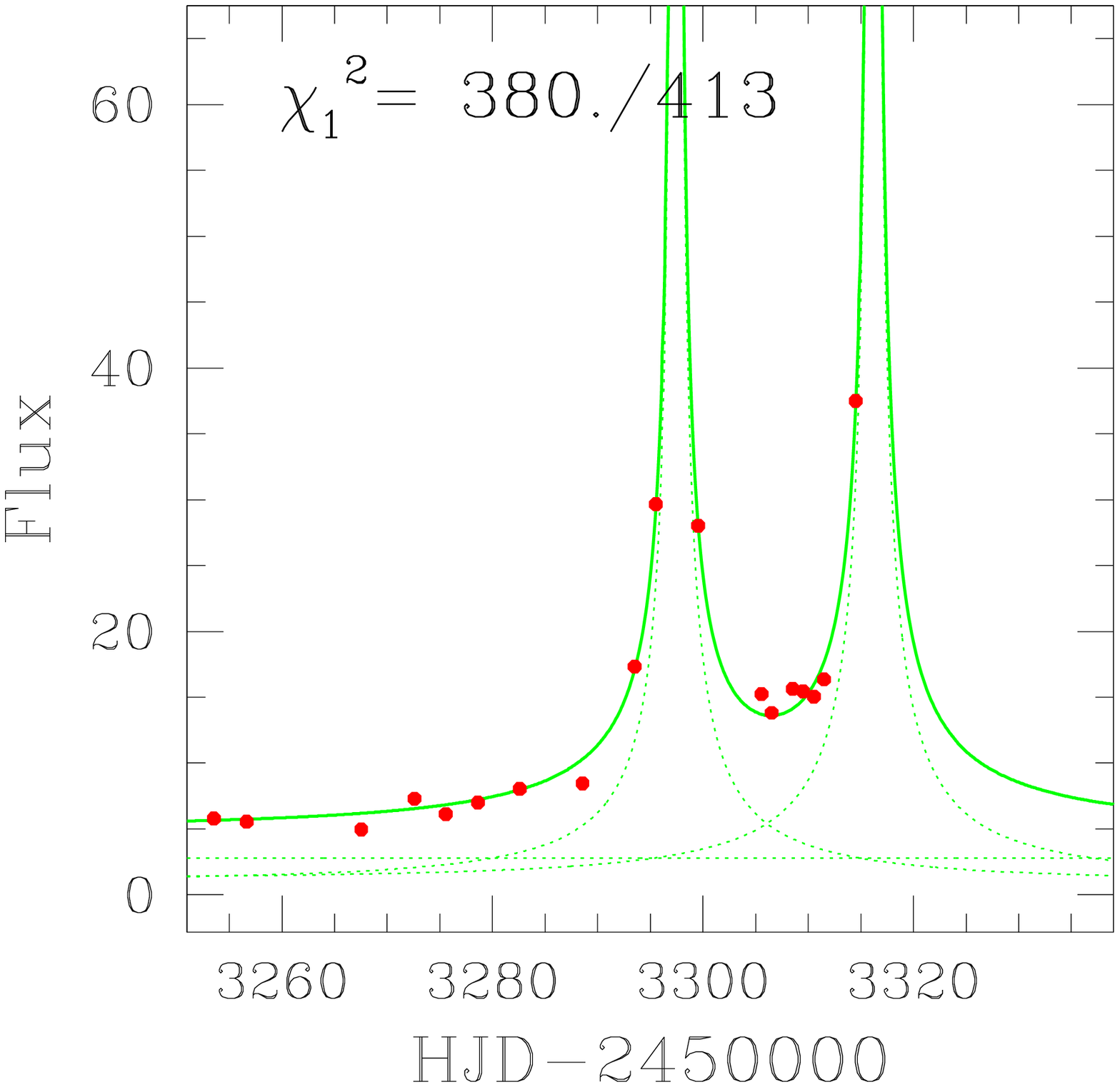}%

}
 
\vskip12pt
\noindent 
{\bf  Double source events}

\vskip 0.5cm
\vskip12pt
\noindent\parbox{12.75cm}{
\leftline{ {\bf OGLE 2004-BLG-004}\hskip3.5cm {\bf OGLE 2004-BLG-328}}

 \includegraphics[height=60mm,width=62mm]{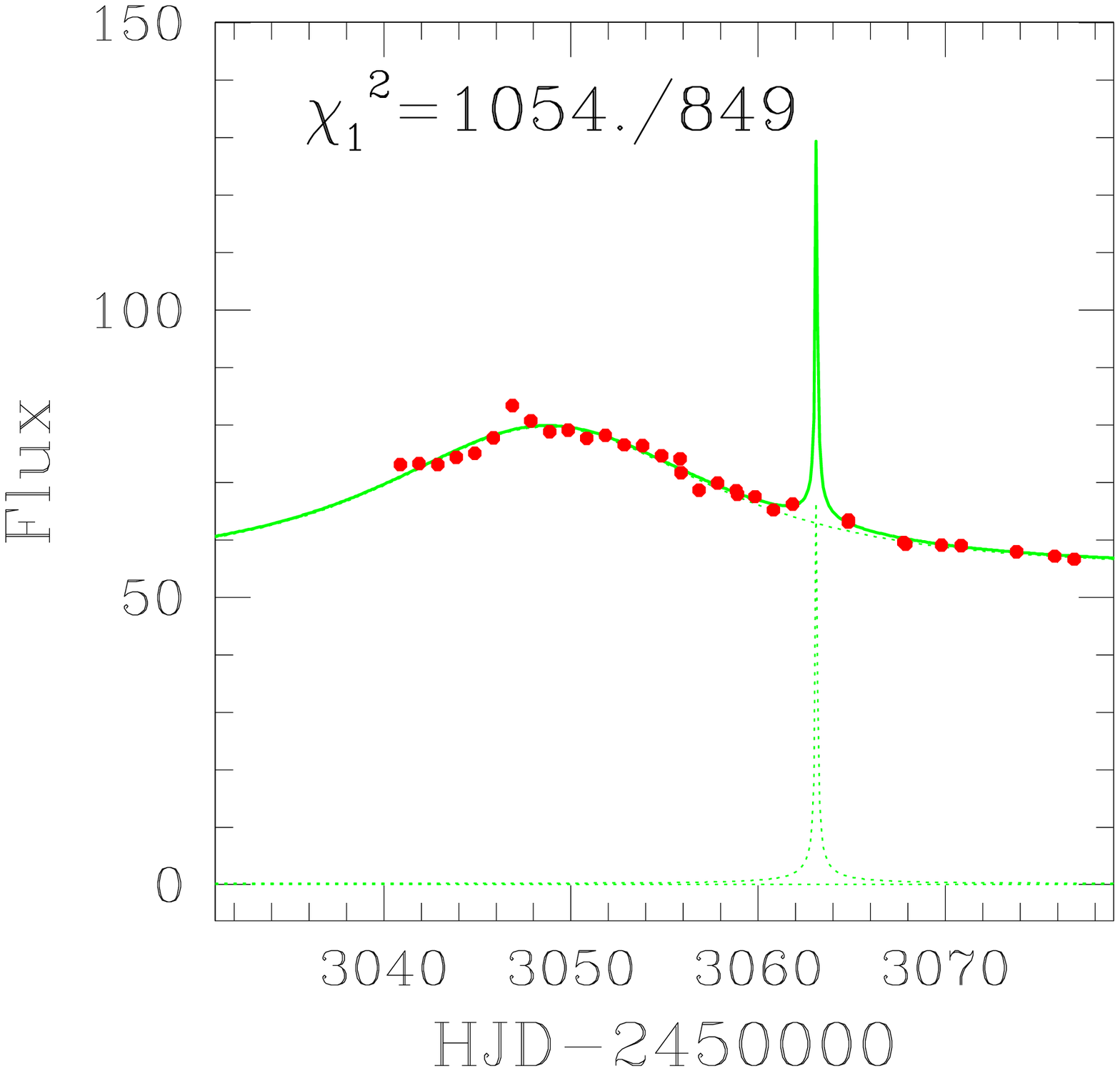}%
 \includegraphics[height=60mm,width=62mm]{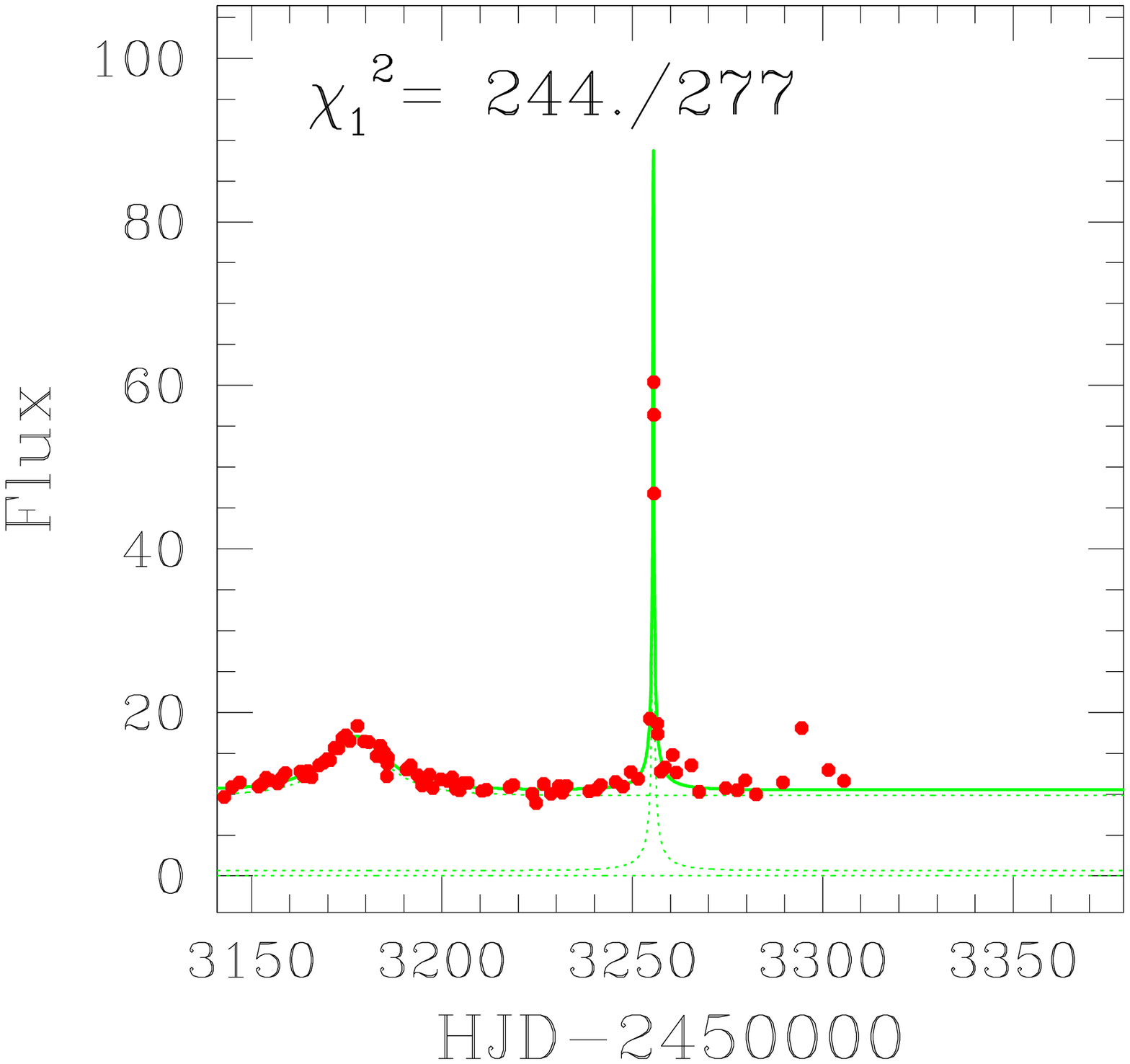}%

}
\end{document}